\pdfoutput=1 
\documentclass{elsart}
\usepackage{natbib}
\usepackage{epsfig}
\usepackage{dsfont}
\usepackage{hyperref}
\usepackage[normalem]{ulem}
\usepackage{color}
\journal{Journal of Fluids and Structures}
\graphicspath{{./FIGURES/}}
\begin{document}
\begin{frontmatter}
\title{Experimental investigation of dynamic pressure loads during dam break}
\author[label1]{L. Lobovsk\'{y}\corauthref{cor}},
\corauth[cor]{Corresponding author.} \ead{lobo@kme.zcu.cz}
\author[label2]{E. Botia-Vera},
\author[label3]{F. Castellana},
\author[label2]{J. Mas-Soler},
\author[label2]{A. Souto-Iglesias}
\address[label1]{NTIS - New Technologies for Information Society, Faculty of Applied Sciences,\\
University of West Bohemia, Univerzitn\'{i} 22, 30614 Plze\v{n}, Czech Republic.}
\address[label2]{Naval Architecture Department (ETSIN),\\
Technical University of Madrid (UPM), 28040 Madrid, Spain.}
\address[label3]{DYNATECH-Naval Architecture and Marine Engineering Group,\\
University of Genoa, 16145 Genoa, Italy.}
\begin{abstract}
The objectives of this work are to revisit the experimental measurements on dam break flow over a dry horizontal bed and to provide
a detailed insight into the dynamics of the dam break wave impacting a vertical wall downstream the dam, with emphasis on the pressure loads. The measured data are statistically analyzed and critically discussed. As a result, an extensive set of data for validation of computational tools is provided.
\end{abstract}
\end{frontmatter}

\section{Introduction}


The first studies analyzing the dam break flow date back to the 19th century. In 1892, Ritter \citep{Ritter_1892} published a theoretical solution of the free surface profile evolution for a collapsing rectangular column of fluid over a dry horizontal downstream bed based on Barr\'{e} de Saint-Venant's shallow water theory \citep{SaintVenant_1871a}\citep{SaintVenant_1871b}. In his approach, Ritter neglected friction over the horizontal bed and turbulence effects.
The effect of bed friction was investigated by Dressler \citep{Dressler_1952}. In general, Dressler's solution of the free surface profile agrees well with Ritter's one except for the retarded initial downstream wave propagation and a non-zero wave front depth. These results were also confirmed by other independent theoretical studies, e.g. \citep{Pohle_1950}\citep{Whitham_1955}.
%
%
Dressler also published an experimental study \citep{Dressler_1954} that confirmed his theoretical solution of the early stages of the dam break flow over an initially dry bed with friction and refered to the experimental results by Schoklitsch \citep{Schoklitsch_1917} and Eguiazaroff \citep{Eguiazaroff_1935} which were one of the first experimental works on dam break flow published in the first half of 20th century.

In 1950's, Martin and Moyce performed series of tests of dam break flow over an initially dry horizontal bed \citep{Martin_1952a} and provided a complete set of data on the collapse of a two-dimensional dam with rectangular or semi-circular initial profile and data on the three-dimensional axial collapse of circular cylinder. The main attention was paid to kinematics of the flow. The wave front velocity was found to be proportional to the root of original column height. This was in agreement with theoretical solution of Ritter \citep{Ritter_1892}.


A rigorous experimental and theoretical study on dam break flow over both dry and wet horizontal bed was published by Stansby et al. \citep{Stansby_1998}. They were the first who identified the mushroom-like jets that appear right after the dam release for wet downstream beds and provided data confirming that the experimental free surface profiles for two different initial water depths scaled approximately according to Froude's criteria.


Few studies addressing the viscous effects and influence of surface tension were performed recently, e.g. \citep{Nsom_2002}\citep{Janosi_2004}.
Although there were numerous experimental studies on dam break flow performed to date, e.g. \citep{Estrade_1967}\citep{Lauber_1998}, 
some of them being designed for validation of novel numerical schemes, e.g. \citep{Cruchaga_2007}\citep{koshizuka1996}, most of the published studies were focused purely on kinematics of the collapsing fluid column. That included analysis of the propagation of the positive wave downstream and of the negative wave upstream, the shape of the wave front, water level measurements and the overall free surface profile.
There were also many studies focused on a dam break evolution over a sloped bed, e.g. \citep{Chervet_1970}\citep{USCorps_1960a}, and recently a number of authors focused on kinematics of a fully three-dimensional flow after a dam failure in well defined laboratory conditions, e.g.
\citep{Bellos_2004}\citep{soaresFrazao_1999}.
Despite the fact that an extensive discussion of dam break wave kinematics can be found in the literature, there is a lack of data describing its dynamics.


In 1999, Zhou et al. \citep{Zhou_1999} validated their numerical scheme using an experimental work performed at Maritime Research Institute Netherlands (MARIN) that provided a description of dam break wave kinematics as well as data on wave impact on a solid vertical wall downstream the dam. Measurements of impact pressure at several locations were performed by force transducers with circular impact panels of large diameter. The details on experimental setup and on applied force transducers were published in the work of Buchner \citep{Buchner_2002PhD}. The same experimental data were applied for validation in the work of Lee et al. \citep{Lee_2002}.

The MARIN experimental setup was also used in the work of Wemmenhove et al. \citep{Wemmenhove_2010} and  Kleefsmann et al. \citep{kleefsman_etal_jcp_2005}.
\footnote{Experimental setups published in \citep{Janosi_2004}\citep{kleefsman_etal_jcp_2005} were approved as two benchmark tests for validation of numerical codes by Smoothed Particle Hydrodynamics European Research Interest Community
\citep{SPHERIC}.}
The former repeated and slightly altered experiments of Zhou \citep{Zhou_1999}, the latter presented a fully three-dimensional dam break problem.
From the presented figures in these publications, it can be deduced that the new pressure measurements in both studies were performed using pressure transducers of smaller diamater than the transducers used in \citep{Zhou_1999}. Thus more localised data could be measured and the transducers could be located closer to the horizontal bed.
However, description of applied pressure transducers was missing in both \citep{kleefsman_etal_jcp_2005}\citep{Wemmenhove_2010}.

An elementary research on dam break flow dynamics was also conducted by Bukreev et al. who studied the overall forces exerted by the dam break wave on the vertical structures downstream the dam \citep{Bukreev_2009,Bukreev_2008}.
A similar test case was studied by G\'{o}mez-Gesteira et al. \citep{GomezGesteira_2004}.
Besides applications in describing and predicting the flood events, understanding the dynamics of dam break flows is also useful when assessing certain types of impact flows like those found in slamming and green water events \citep{Greco2004251}\citep{Greco2012148}.


Except for few studies, such as \citep{Martin_1952a} or \citep{USCorps_1960a}, that provide a complete set of data on kinematics of series of several tests, there is a lack of a thorough discussion on repeatibility of the dam break experiments in the literature. However, such an analysis is crucial when assessing the fluid dynamics. The statistical analysis and the probability ranges of the measured nominal values of pressures are missing and little information on setup and precision of the pressure measurements is provided in the related dam break studies cited above.


This paper aims to provide a detailed insight into the dynamics of the dam break flow over a dry horizontal bed under well controlled laboratory conditions.
In order to do so, a similar experimental tank setup to Zhou et al. \citep{Zhou_1999} and Buchner \citep{Buchner_2002PhD} is constructed and an extensive experimental campaign is performed.
That provides a large set of data on both the kinematics of the free surface evolution after the dam burst and the dynamics of the downstream wave.
Similarity between the experimental setups enables a direct comparison of the measured data and the data from literature.

A special attention is paid to measurements of the impact pressure of the downstream wave on the flat vertical wall and the propagation of the pressure along this wall in vertical direction.
Unlike in the previous dam break studies, the impact pressure is recorded right above the horizontal bed as the pressure probes of a small diameter are used.
This enables measurements of the peak impact pressure.
Although the experimental setup is designed so that the studied case can be idealized as a two-dimensional problem, a three-dimensionality of the dam break flow is also investigated.

In order to address the repeatability of the experiments, a large set of measurements is performed under the same experimental conditions.
As a result, a statistically relevant sample of data on impact pressure at the downstream wall as well as on kinematics of the dam gate motion is recorded.
A special care is taken so that all tests are performed for a completely dry horizontal bed as well as the tank walls downstream the dam gate. That is achieved due to a novel design of the dam gate and its guiding system which features a leak-proof sealing and enables rapid removal of the gate in the vertical direction.

The obtained results from a thorough statistical analysis of the experimental data provide a novel information about the dam break wave dynamics and complement the knowledge of the previously published studies discussed above.
Thus this study aims to serve as a basis for computational fluid dynamics (CFD) validations for which
information about median and dispersion values and confidence intervals is also provided. To help with this task,
video movies associated with all the flow images presented in the paper and the MATLAB figure files
can be downloaded from \href{http://canal.etsin.upm.es/papers/lobovskyetaljfs2013/}{http://canal.etsin.upm.es/papers/lobovskyetaljfs2013/}.

\section{Experimental setup}
\label{sec:exp_setup}

\subsection{General}
\begin{figure}
\centering
\includegraphics[width=0.9\textwidth]{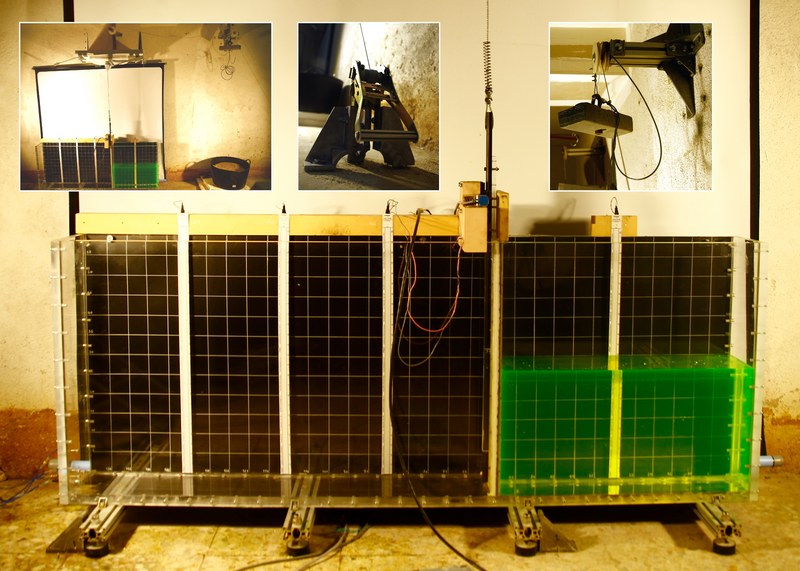}
\caption{Experimental setup of the dam-break tests (data acquisition system is not displayed). The imbedded figures display from left to right: the overall setup, the release mechanism, the weight used to induce gate's motion.}
\label{fig:setup}
\end{figure}

\begin{figure}
\centering
\includegraphics[width=0.99\textwidth]{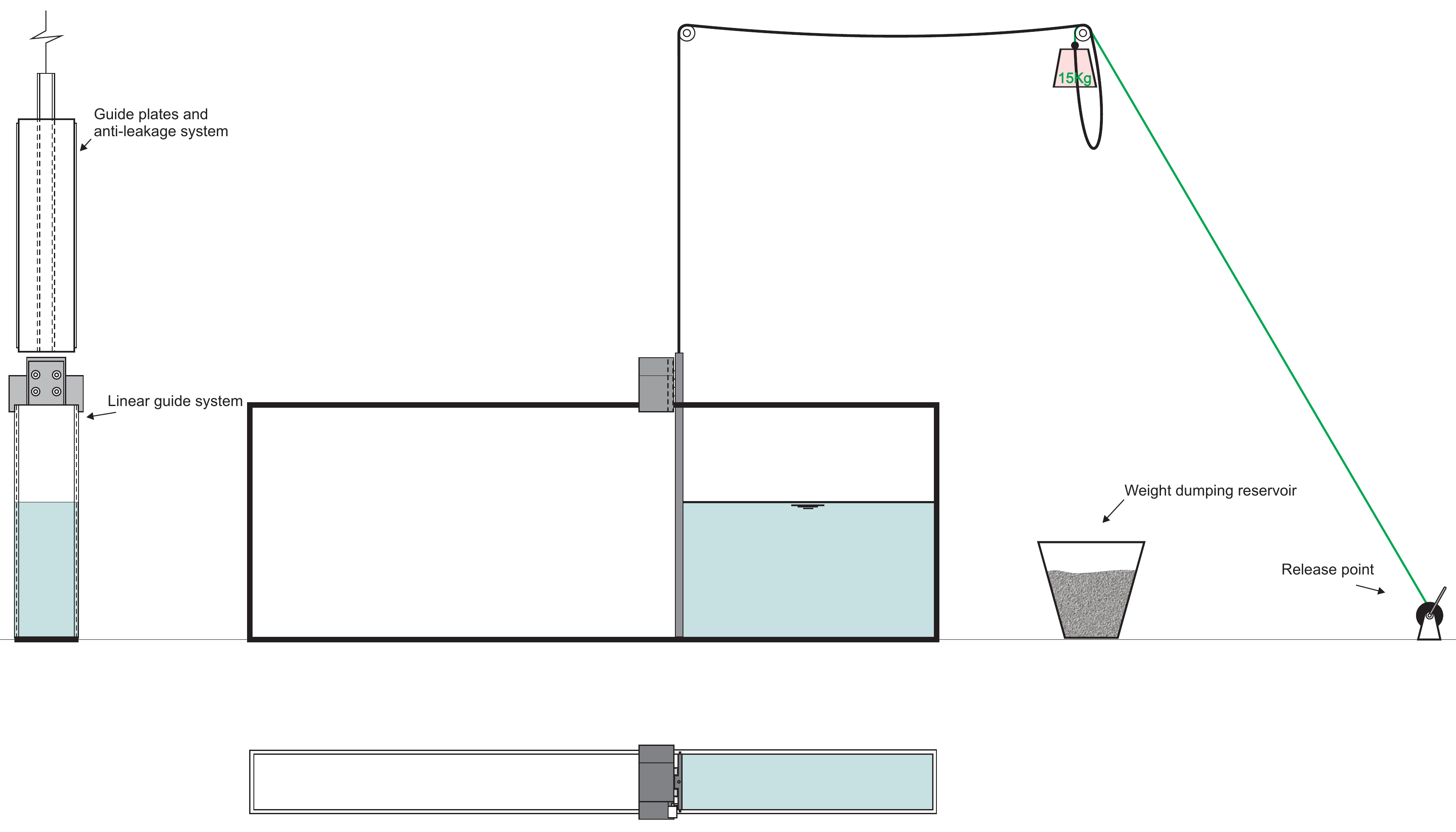}
\caption{Schematic view of the experimental setup.}
\label{fig:experimental_setup}
\end{figure}

The dam-break experimental setup was built and installed at the facilities of the Technical University of Madrid (UPM) where several experimental campaigns dealing with sloshing flows had been carried out in the past (see e.g. \citep{botia_etal_spheric10}\citep{souto_botia_martin_perezarribas_part0_oe2011}).
For the dam-break experiments, a dedicated tank setup was designed and assembled.
It consisted of a prismatic tank that could be divided into two separate parts by a removable gate, a release system with a sliding mechanism, a weight inducing the gate motion and a damping system, as shown in Figs.~\ref{fig:setup} and \ref{fig:experimental_setup}.

The experimental measurements were performed using the instrumentation, control and data acquisition systems that had been well-proven during the sloshing tests.
Some of the previously published data gathered by these systems served well for CFD validation in a number of recent papers, e.g. in \citep{leonardi_etal_spheric2012}\citep{Khayyer_gotoh_ijope09}\citep{bulian_etal_jhr09}\citep{idelsohn_etal_09}\citep{degroote_etal_cmame_2010}\citep{souto2006}.

All parts of the experimental setup are explained in detail hereafter.

\subsection{Tank}
\begin{figure}
\centering
\includegraphics[width=0.595\textwidth]{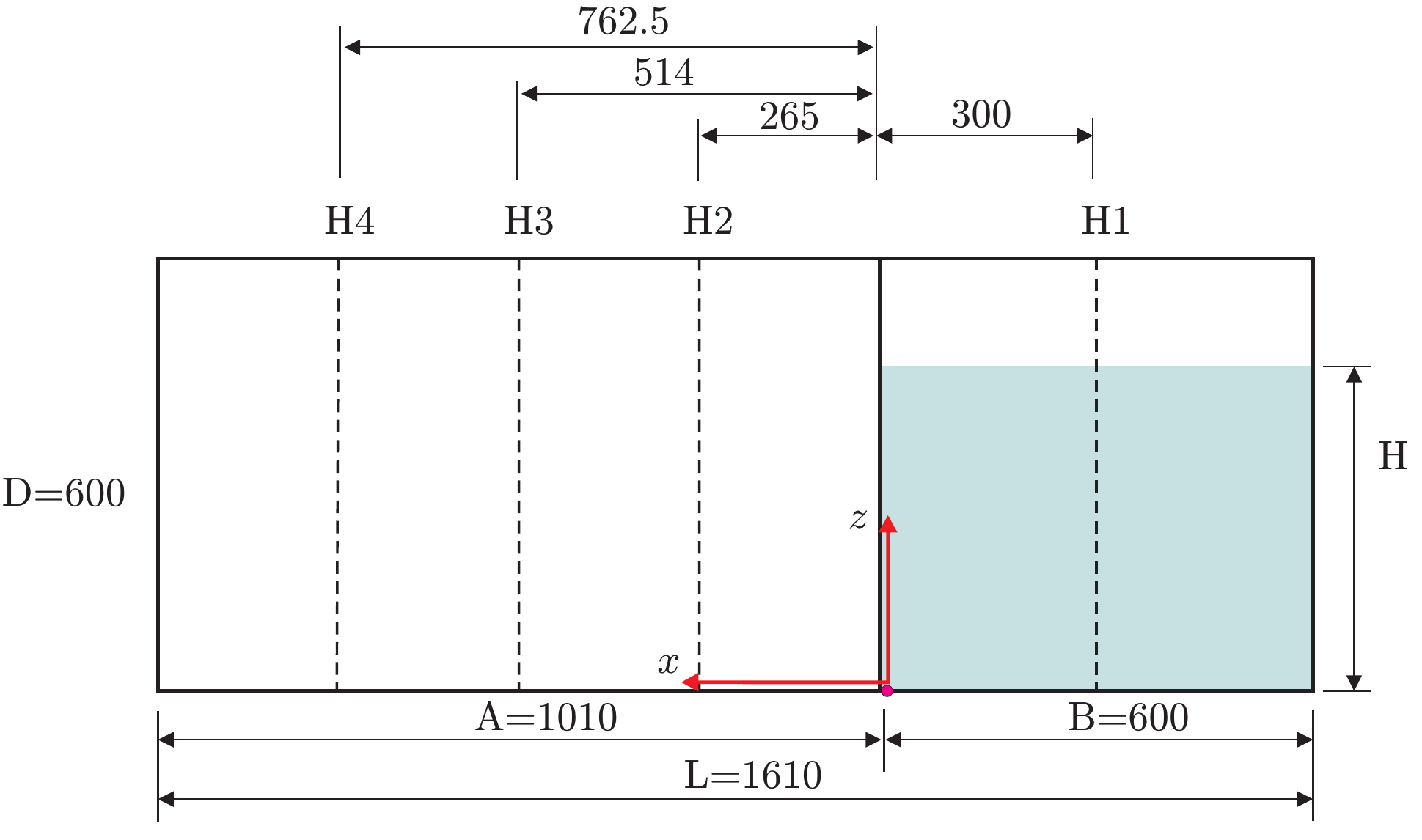}
\includegraphics[width=0.395\textwidth]{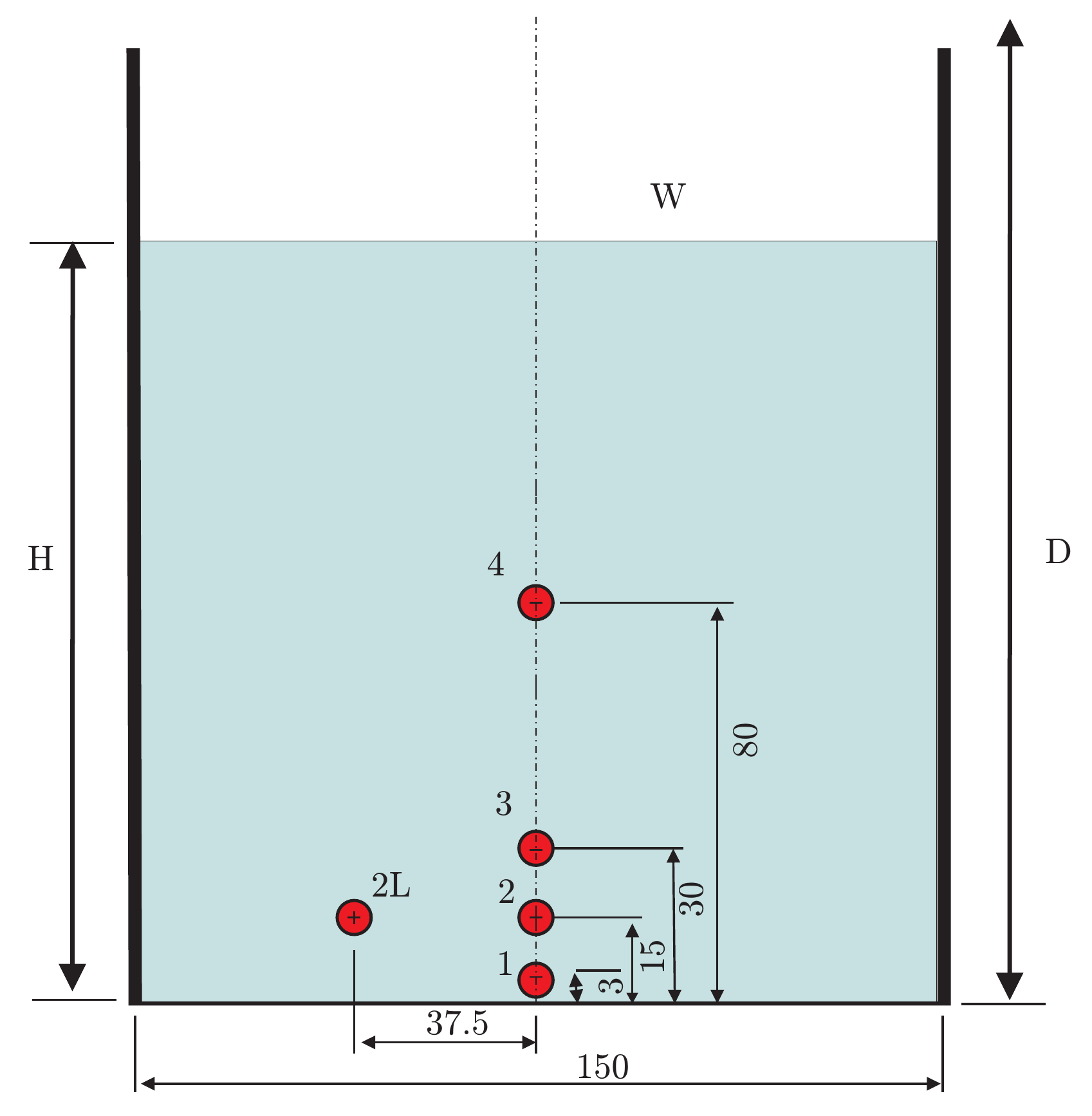}
\caption{A side view of plexiglass tank with locations of water level measuring positions (left) and a front view showing the locations of pressure sensors at the impact wall downstream the dam (right). Dimensions in millimeters.}
\label{fig:dam_scheme}
\end{figure}

A prismatic tank with inner dimensions of $1610\times600\times150\,{\rm mm}$ was made of polymethyl methacrylate (PMMA) (see Fig. \ref{fig:dam_scheme}). With regards to the tank dimensions and PMMA mechanical properties, the tank walls were made of $20\,{\rm mm}$ thick PMMA sheets so that the walls could be considered rigid enough to avoid hydroelastic effects \citep{choi_etal_omae12_sloshing_hydroelasticity}\citep{abrahamsen_faltinsen_2011} while keeping a reasonable overall weight of the tank.
The PMMA walls were transparent and the tank was assembled with a precision of $0.1\,{\rm mm}$.\footnote{Precision of tank manufacturing at UPM is reflected in the cooperative study \citep{Loysel_Chollet_Gervaise_etal_gtt_isope2012} where UPM tank construction was evaluated as the most accurate one.}

The breadth of the tank was chosen with regards to the study on a two-dimensionality of breaking waves in prismatic tanks \citep{souto_etal_isope2012}
so that the resulting dam break flow experiments could be idealized as a two-dimensional phenomenon and wall effects could be considered as not affecting the main flow dynamics. Nevertheless, a discussion on a three-dimensionality of the dam-break flow is provided below.

In order to avoid spurious tank motion the tank was fixed to the ground and levelers were used to guarantee that the tank bottom is leveled.
The deviation from the horizontal plane was bounded by $0.5\,{\rm mm}$ measured in the corners of the tank and therefore the maximum angular deviation of the free surface along the full tank attained is $\pm 0.018\,deg$ in $x$ direction and  $\pm 0.18\,deg$ in $y$ direction.
\subsection{Dam gate and gate release system}
The dam gate was made of $10\,{\rm mm}$ thick PMMA and was placed $600\,{\rm mm}$ from the lateral side of the tank as shown in Fig.~\ref{fig:experimental_setup}.
This defined the length of the dam reservoir region and left $1000\,{\rm mm}$ of unobstructed horizontal bed downstream the dam gate.

\begin{figure}
\centering
\includegraphics[width=0.6\textwidth]{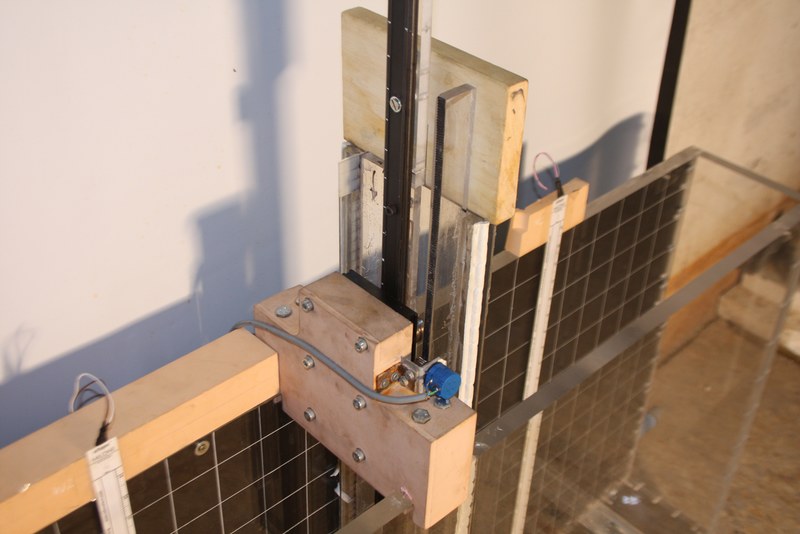}
\caption{Gate guiding system}
\label{fig:gate_rail}
\end{figure}

The wet side of the dam gate was kept flat while there was a robust steel guiding rail mounted in the center of its dry side, Fig.~\ref{fig:gate_rail}.
This rail together with the guiding system mounted on the top of the tank was used to assure that the dam gate is released straight in a vertical direction.
The smooth vertical motion of the rail was guided by four roller-bearing wheels.
The guiding rail itself also increased the rigidity of the gate and served as a support along the entire gate length.
Furthermore, $3\,{\rm mm}$ thick steel side rails were mounted to the gate edges.
These side rails fitted into $3.5\,{\rm mm}$ wide and $4\,{\rm mm}$ deep grooves that were grooved into the frontal and distal tank wall.
This provided an additional support for the dam gate and prevented its displacement in horizontal direction which could be caused by the mass of the liquid column inside the dam reservoir.

There was a $1\,{\rm mm}$ gap between the edge of PMMA gate and the tank's frontal and distal walls. This was necessary in order to provide a smooth motion of the gate. Dam reservoir watertightness was implemented by a soft rubber band placed around gate's edge and smeared by vaseline in order to decrease the friction.

The experimental setup was constructed so that the dam gate motion had one degree of freedom and the gate could be moved only in the vertical direction as long as the side rails were sliding in the grooves and the guiding rail was moving inside the guiding system, Fig.~\ref{fig:gate_rail}.

The vertical motion of the gate was induced by a similar release mechanism as used by Stansby \citep{Stansby_1998} or Koshizuka \citep{koshizuza1995}.
The mechanism incorporated a $15\,{\rm kg}$ weight
that was connected to the top of the guiding rail using a $5\,{\rm mm}$ steel braided wire
and a system of pulleys, Fig.~\ref{fig:experimental_setup}.
In order to release the dam gate, the $15\,{\rm kg}$ weight was dropped under gravity while it pulled the gate upwards.
The steel wire was loose in the initial configuration, Fig.~\ref{fig:experimental_setup}.
Thus, when the wire got stretched, the weight already had a non-zero velocity and the gate's motion initiated rapidly.
To ensure the integrity of the experimental setup, there was a dedicated damping system above the tank which stopped the gate's motion and there was a damping reservoir filled with sand which stopped the falling weight.

At the beginning of each test, the initial vertical position of the weight was set using the ratchet mechanism. In order to release the weight, the gate respectively, the ratchet mechanism was set free to move.
The free falling height before the steel wire got stretched was $1100\,{\rm mm}$ which corresponded to theoretically predicted velocity of the falling weight equal to $4.65\,{\rm ms^{-1}}$.

The non-zero velocity of the falling weight at the moment when the wire got stretched caused a strong shock which resulted in rapid acceleration of the dam gate.
However, this shock also induced a high tension in the steel wire and non-negligible plastic deformations of the wire appeared.
Therefore the cable was frequently adjusted so that the cable length stayed constant.
This way, a reasonable repeatability of the dam gate's motion was achieved.

\subsection{Tested fluid and preparation of each run}
During the entire experimental campaign presented in this paper, the dam-break flow was investigated for fresh water, whose temperature was controlled by standard thermometer and was preheated to $25\,^{\circ}{\rm C}$ with an uncertainity of $\pm 0.1\,^{\circ}{\rm C}$ before each run.
Under the given testing conditions, the fresh water could be considered Newtonian with the density $997\,{\rm kg\,m^{-3}}$, the kinematic viscosity $8.9\cdot10^{-7}\,{\rm m^2s^{-1}}$ and surface tension $0.072\,{\rm N\,m^{-1}}$.

Before each experimental run, the bed and the walls of the tank downstream the dam gate were dried and watertightness of the dam reservoir was secured so that the dam gate prevented any leaks. Thus the downstream part of the tank, i.e. the tank walls and especially the horizontal bed, was kept completely dry before each run.
The aforementioned drying process of the tank was performed by using a pump and traditional drying accessories.
A leak-proof system of the dam gate enabled a precise adjustment of the initial water level in the dam reservoir.

For the experimental runs when digital images and videos were recorded, the fluid was dyed using a small amount of fluorescein in order to increase fluid's contrast. The presence of fluorescin dye in the tested fluid did not exhibit any notable influence on the studied fluid dynamics.

\subsection{Data acquisition}
\subsubsection{General}
Data acquisition system comprises all devices necessary to assess both the kinematics and the dynamics of the dam break flow.
It is used to measure the velocity of the dam gate motion, the pressure at the downstream impact wall and to record the video data that are used for analysis of the free surface evolution.
The elements of the applied apparatus are explained in detail hereafter and further
information about some of them in sloshing context can be found in
reference \cite{souto_botia_martin_perezarribas_part0_oe2011}.
\subsubsection{Pressure sensors}
\label{sss:presssensors}
Five pressure sensors have been used. The sensor matrix arrangement is shown in Fig.~\ref{fig:dam_scheme}.
The first center-line sensor (sensor 1) is placed exactly at the bottom of the tank in order to measure the impact pressure in the tank corner, as shown in Fig. \ref{fig:dam_scheme} (right). In practice it means that its center is located $3\,{\rm mm}$ above the bed.
The other three pressure sensors at the center-line are positioned corresponding to locations of selected sensors from the literature, i.e. their centers are $15\,{\rm mm}$ \cite{Wemmenhove_2010} (sensor 2), $30\,{\rm mm}$ \cite{kleefsman_etal_jcp_2005} (sensor 3) and $80\,{\rm mm}$ \cite{Zhou_1999} (sensor 4) above the tank bed. The off-centered sensor (sensor 2L) is located at the same height like sensor 2 but half way to the back wall.

The piezo-resistive type of pressure sensors (KULITE XTL-190 series) is applied.
The sensing diameter of the sensors is $4.2\,{\rm mm}$ and their pressure signal is amplified before the A/D conversion using a NI-PCI6221 card. The digital signal is recorded at sampling rate of $20\,{\rm kHz}$.
The effective full scale output (FSO) of the sensors is $400\,{\rm mb}$.
With our experimental capabilities, the bias uncertainty in the pressure measurements is
about $0.5\,{\rm mb}$ (a detailed uncertainty analysis for a previous experimental campaign with similar equipment can be found in \cite{souto_botia_martin_perezarribas_part0_oe2011}).
Taking into account the pressure ranges, these measurements are comparable in accuracy
to those of the state of art laboratories, as discussed
in \cite{Loysel_Chollet_Gervaise_etal_gtt_isope2012}.
\subsubsection{Gate velocity sensor}
\label{sss:gatesensor}
The dam gate removal velocity is monitored with a multiturn potentiometer using a gear wheel at the potentiometer axis and a gear rack attached to the gate dry side, Fig.~\ref{fig:gate_rail}. The information from the gate sensor is used to set time zero for all measurements.
A range of $240\,{\rm mm}$ of gate's displacement is used to define the gate removal duration as the time the gate
needs to cover this displacement. The corresponding mean velocity is taken as the representative gate velocity for each experiment.
Analysis of this data is provided later in the paper.
\subsubsection{Video recording}
\label{sss:video}
In parallel to pressure measurements, the evolution of the free surface profile and the impact on the vertical wall downstream the dam gate is captured by a digital camera (Casio EXILIM F1) that enables recording of 300 frames per second (fps) at resolution $512\times384$ pixels.
In order to get a complex information on the free surface evolution of the breaking wave, lateral, frontal and top views of the flow evolution are taken. In addition, a close investigation of the surge front shape development is carried out using a camera that moves along the tank on a linear guide.

The time history of water elevation at several locations and the propagation of the downstream
wave front is analyzed from the video images captured during randomly selected experimental runs.
Since a confrontation of the new data with previously published studies is desirable, the water level measurements are conducted at locations as specified in works of \cite{Buchner_2002PhD}\cite{Lee_2002}\cite{Zhou_1999}.
These correspond to a location $300\,{\rm mm}$ upstream the dam gate, i.e. in the center of the dam reservoir, and to locations at $265\,{\rm mm}$, $514\,{\rm mm}$ and $762.5\,{\rm mm}$ downstream the gate in the present experimental setup, see Fig.~\ref{fig:dam_scheme}. The water levels at these locations are denoted as $H1$, $H2$, $H3$ and $H4$ respectively.
\subsubsection{Synchronization}
Within the performed dam-break experiments, the zero time is assigned to the moment when the gate starts moving upwards.
Synchronization of the video recordings, the water level measurements, the pressure signals and
the data describing the dam gate movement is enabled using a potentiometer that is activated
when the dam gate starts to move (see section \ref{sss:gatesensor}).
The precision uncertainty related to this synchronization process is of the order of $0.01\,{\rm s}$.
For further details on the experimental setup and the data acquisition system a kind reader may refer to \cite{souto_botia_martin_perezarribas_part0_oe2011}.

\section{Test matrix}
\label{sec:test_matrix}

The tank dimensions and the complete experimental setup is designed so that a direct comparison with the data from MARIN measurements can be made. Particularly a confrontation of the new results with data of Buchner \cite{Buchner_2002PhD}, Lee et al. \cite{lee_zhou_cao_2002_jfe_dambreak} and Zhou et al. \cite{Zhou_1999} is of interest.

The presented experimental campaign consists of two parts. Within each part a set of $100$ experimental runs is performed using the hardware setup described in section \ref{sec:exp_setup}.
The key and only setup parameter that differs between the two sets is the initial water depth $H$ in the dam reservoir.
The two tested initial water depths $H$ are $300\,{\rm mm}$ and $600\,{\rm mm}$ with an
uncertainty in the filling levels of $\pm 0.5\,{\rm mm}$.

When the reservoir fluid depth of $300\,{\rm mm}$ is applied, the experimental setup resembles the experiments of \cite{Buchner_2002PhD}
\cite{lee_zhou_cao_2002_jfe_dambreak}\cite{Zhou_1999} with all geometrical dimensions scaled by factor $\lambda=0.5$. This enables a direct comparison of the recorded measurements with the published data using appropriate scaling laws.
However, the setup that involves a fluid level of $600\,{\rm mm}$ uses a full capacity of the developed tank and enables recording of higher impact pressures at the vertical wall downstream the dam while fully utilizing FSO of the applied pressure sensors.

For this type of gravity current flow over a horizontal bed that is initiated from the rest, the Froude number $Fr$ is technically equal to unity regardless of the initial water depth in the dam reservoir. Furthermore the Froude scaling applies, i.e. assuming the fluid density is identical in the two models and the model geometrical dimensions are scaled by $\lambda$, the pressure also scales with $\lambda$ and the time scales with $\sqrt{\lambda}$. The latter was experimentally confirmed by Stansby et al. \cite{Stansby_1998}.

The Reynolds number $Re$ of a dam break flow for the initial water level $H=300\,{\rm mm}$, $H=600\,{\rm mm}$ respectively, equals to $3.8\cdot10^6$, $5.5\cdot10^6$ respectively, based on distance $A$ between the gate and the wall and on the theoretically predicted velocity $v$ of the wave front \cite{Ritter_1892} ($v=2\sqrt{gH}$, where $g$ is gravity). Considering the transitional values
in flows across flat plates (see e.g. \cite{VanDyke1982}), it is expected to have a certain amount of turbulence developing on the boundary layer prior to the impact. The theoretically predicted Weber number $We$ based on the same magnitudes equals to $1.64\cdot10^5$, $3.29\cdot10^5$ respectively, large enough to hypothesize that surface tension effects may not be relevant.

The peak pressure of the wave front impact on the vertical wall downstream is treated in this paper as in the literature on experimental sloshing flows where impact pressure is considered a random variable \cite{graczyk_moan_oe2008,bulian_etal_pof2012}.
In order to characterize its confidence intervals, many repetitions are needed.
Within this study, 100 repetitions have been carried out for each of the two filling levels.
More repetitions would be desirable but with regards to the complexity of the experimental setup it was not suitable for conducting more repetitions with the means available.

\section{Results - Flow Pattern}

\subsection{Free surface profile}


In the mathematical solution of the kinematics of the evolving free surface flow over a dry horizontal bed after the dam burst by Ritter \cite{Ritter_1892}, three regions of interest are distinguished: the undisturbed upstream region with constant water depth, the region with concave parabolic decay of water level that connects the upstream region with the wave front and the region of dry downstream bed with zero water level. This solution features a smooth transition between the downstream dry bed region and the wave front.
However, when the bed friction is considered, a non-zero wave front depth is expected, as shown by Dressler \cite{Dressler_1952}.



\begin{figure}
\centering
\includegraphics[width=0.325\textwidth]{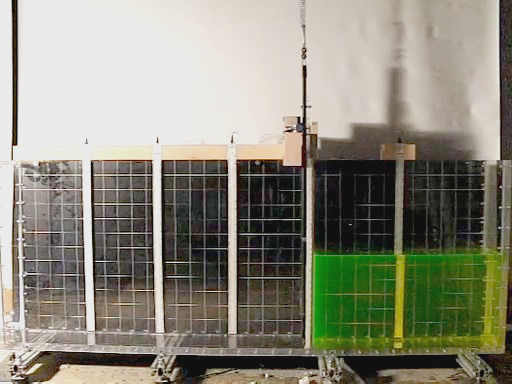}
\includegraphics[width=0.325\textwidth]{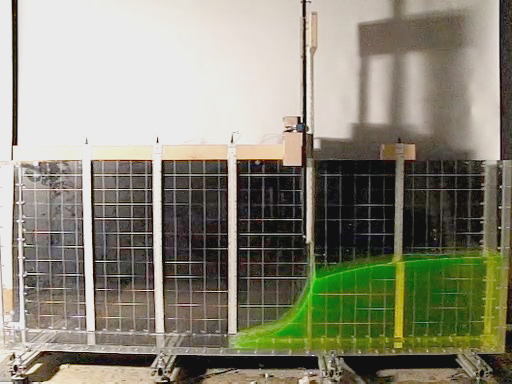}
\includegraphics[width=0.325\textwidth]{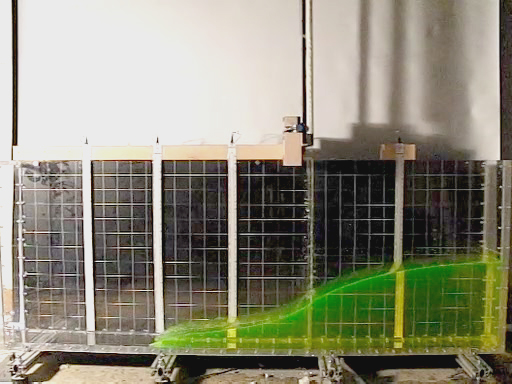}
\includegraphics[width=0.325\textwidth]{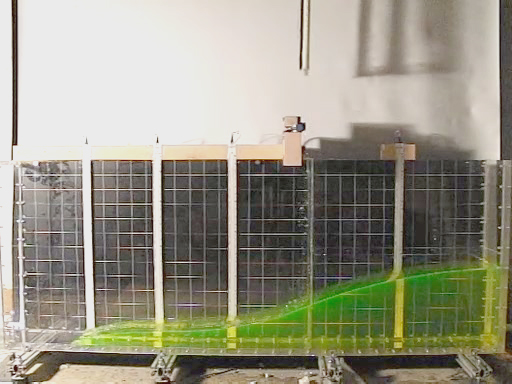}
\includegraphics[width=0.325\textwidth]{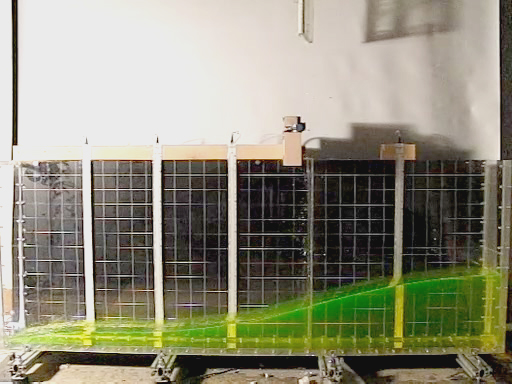}
\includegraphics[width=0.325\textwidth]{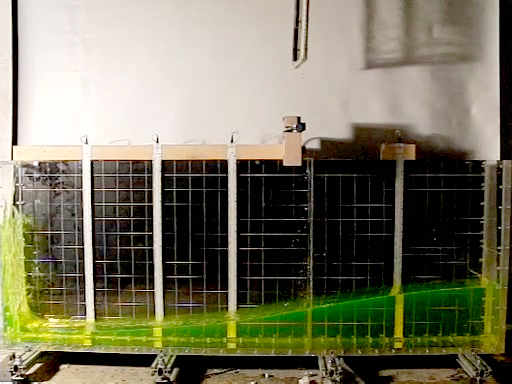}
\includegraphics[width=0.325\textwidth]{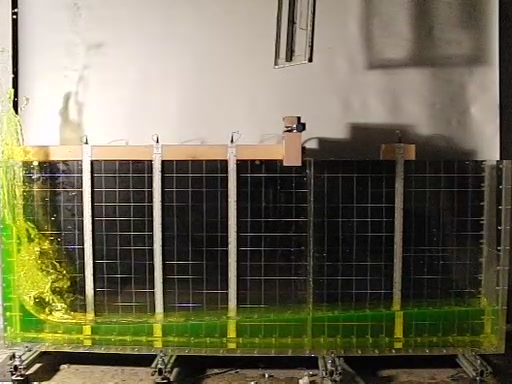}
\includegraphics[width=0.325\textwidth]{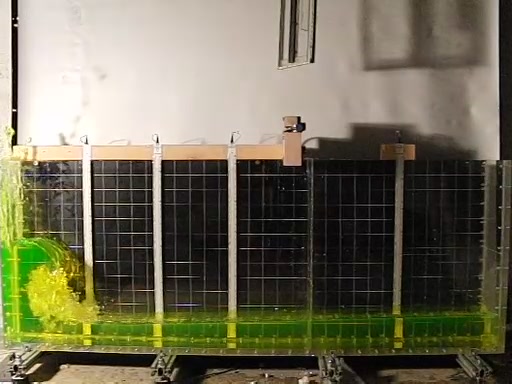}
\includegraphics[width=0.325\textwidth]{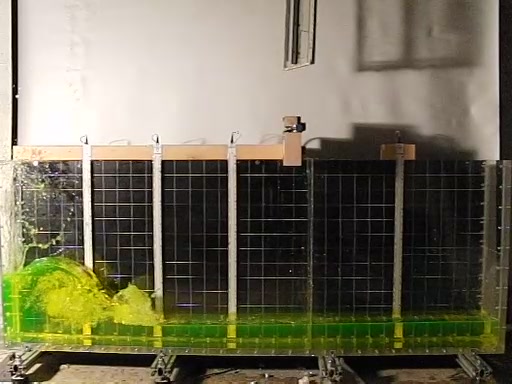}
\includegraphics[width=0.325\textwidth]{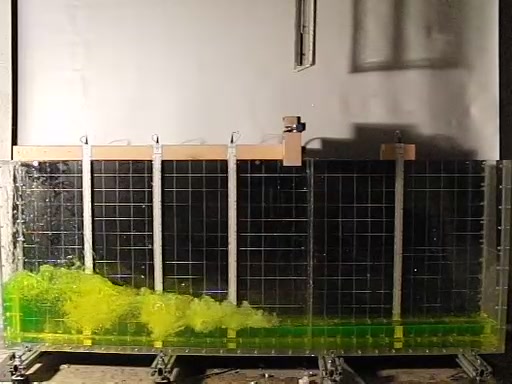}
\includegraphics[width=0.325\textwidth]{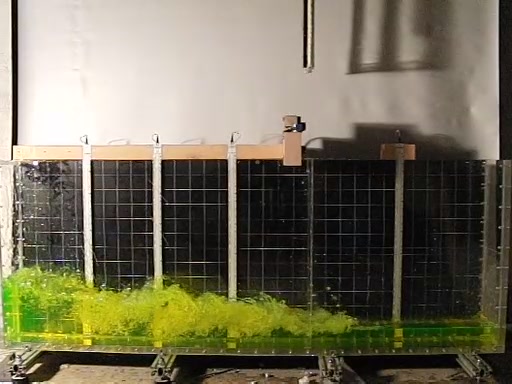}
\includegraphics[width=0.325\textwidth]{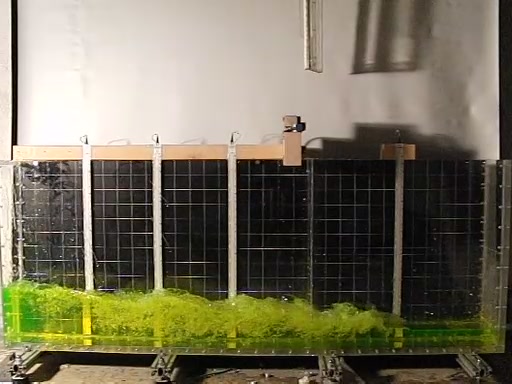}
\caption{$H=300\,\mathrm{mm}$; evolution of free surface profile (0.0, 159.9, 276.6, 373.3, 449.9, 573.3, 862.3, 1023.3, 1166.6, 1320.0, 1473.3, 1626.7, $\pm 3.3$ ms.). See supplementary materials at \href{http://canal.etsin.upm.es/papers/lobovskyetaljfs2013/}{http://canal.etsin.upm.es/papers/lobovskyetaljfs2013/} for a complete movie.}
\label{fig:H30_fs_evolution}
\end{figure}

A side view of a free surface profile evolution in time for an experiment with $300\,{\rm mm}$ filling height is presented in Fig.~\ref{fig:H30_fs_evolution}. Several stages of the advancing downstream wave front are displayed followed by the the wave impact on the downstream wall and the consequent run-up. After that a plunging breaker is formed in the back flow which develops in a combination of a bore propagation and a mixing layer on top of the laminar layer of fluid yet advancing towards the wall. As a result, a lot of vorticity is generated. This corresponds to findings published by Landrini et al. \cite{landrini_etal_jfm2007} who analyzed this phenomenon numerically.

\begin{figure}
\centering
\includegraphics[width=0.3\textwidth]{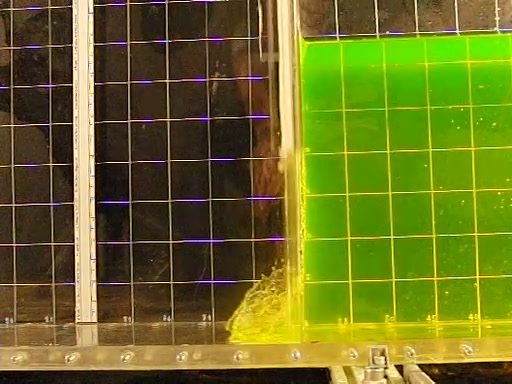}
\includegraphics[width=0.3\textwidth]{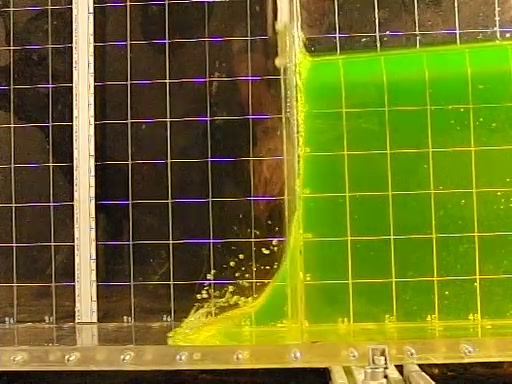}
\includegraphics[width=0.3\textwidth]{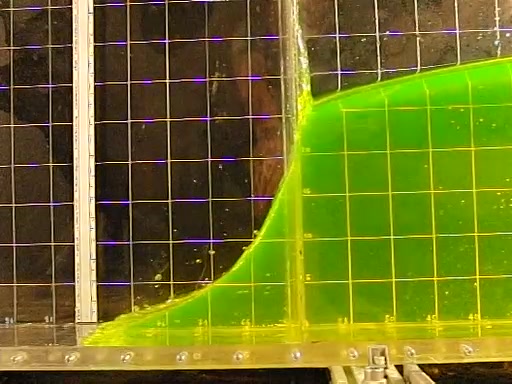}
\caption{$H=300\,\mathrm{mm}$; free surface after dam removal (50.0, 90.0, 146.7 $\pm 3.3$ ms.). See supplementary materials for a complete movie.}
\label{fig:H30_tongue_after_dam_removal}
\end{figure}
\begin{figure}
\centering
\includegraphics[width=0.3\textwidth]{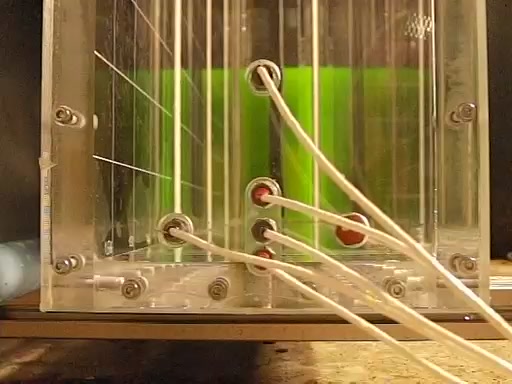}
\includegraphics[width=0.3\textwidth]{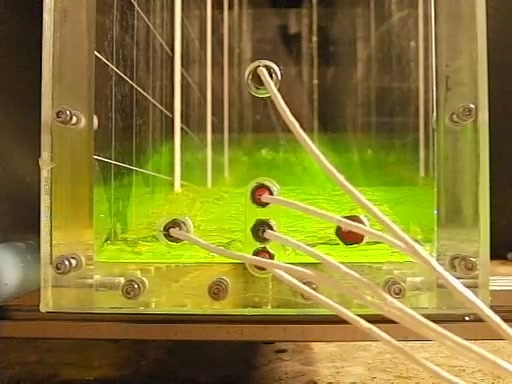}
\includegraphics[width=0.3\textwidth]{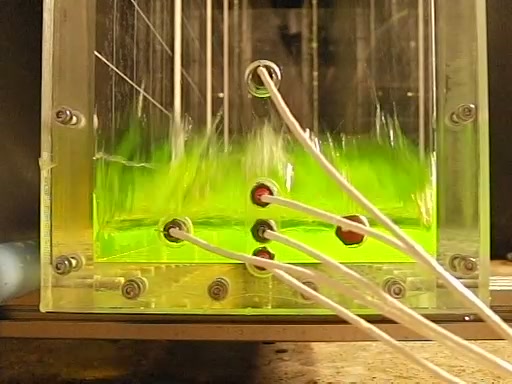}
\includegraphics[width=0.3\textwidth]{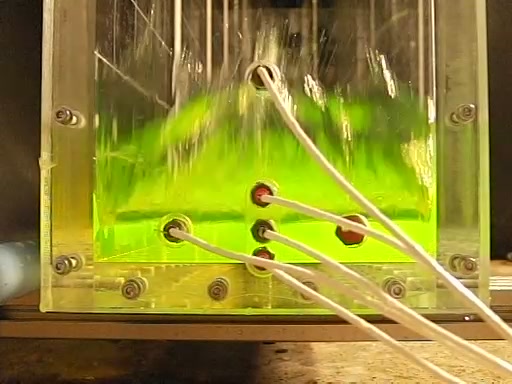}
\includegraphics[width=0.3\textwidth]{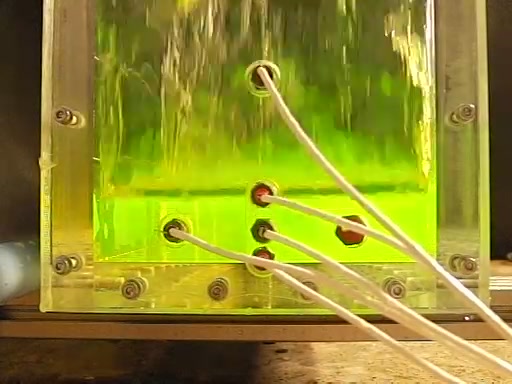}
\includegraphics[width=0.3\textwidth]{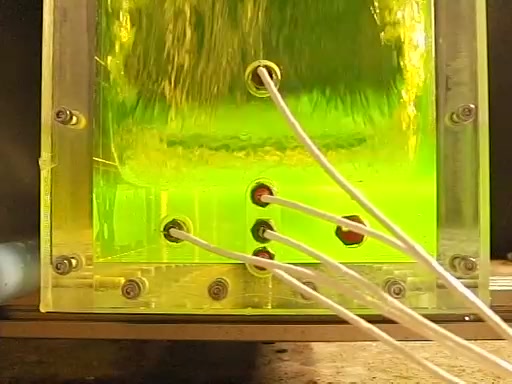}
\includegraphics[width=0.3\textwidth]{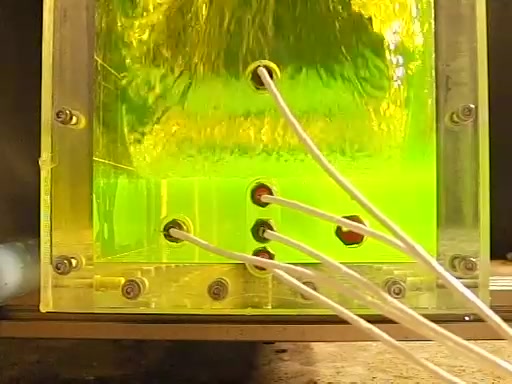}
\includegraphics[width=0.3\textwidth]{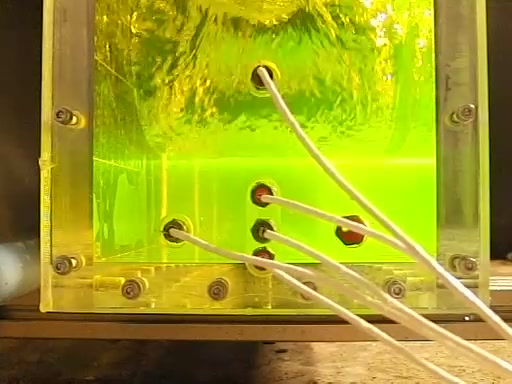}
\includegraphics[width=0.3\textwidth]{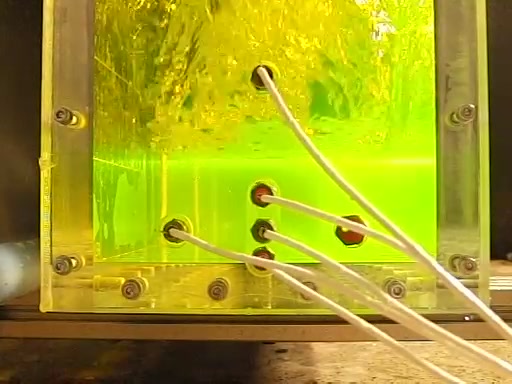}
\caption{$H=300\,{\rm mm}$; free surface during impact and run-up (front view) (0.0, 450.0, 463.3, 470.0, 486.7, 540.1, 646.7, 853.3, 916.7, $\pm 3.3$ ms.). See supplementary materials for a complete movie.}
\label{fig:H30_frontview}
\end{figure}

When inspecting the free surface profile right after the dam gate removal, it can be observed that the rapid removal of the gate, which moves with a finite speed, induces strong shear and a vertical jet that affects the shape of the free surface close to the gate, Fig.~\ref{fig:H30_tongue_after_dam_removal}.
This jet disintegrates into droplets that splash onto and are reabsorbed by the main volume of the fluid flow. After that the free surface becomes smooth and the entire phenomenon does not induce significant waves of any kind.
Although two critical moments can be appreciated, the very start of the gate's motion Fig.~\ref{fig:H30_tongue_after_dam_removal} (left) and the moment when the gate's lower edge exceeds the initial dam filling hight Fig.~\ref{fig:H30_tongue_after_dam_removal} (right), this jet was not proven to have a significant effect on the resulting flow.
The described phenomenon can be also appreciated in Fig.~\ref{fig:H30_frontview} which focuses on the evolution of the wave front during the downstream run.

\begin{figure}
\centering
\includegraphics[width=0.48\textwidth]{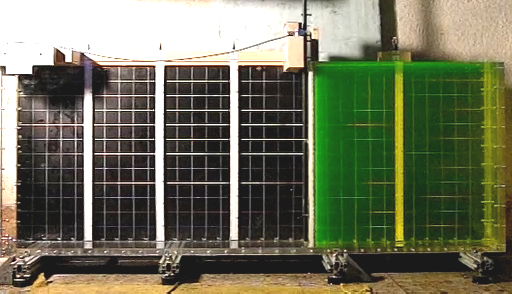}
\includegraphics[width=0.48\textwidth]{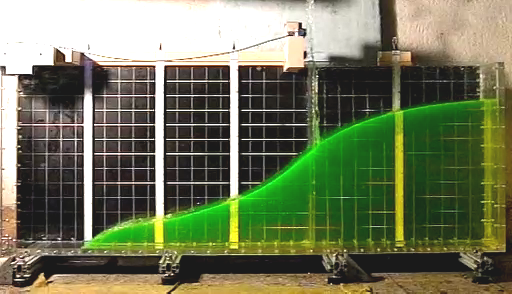}
\includegraphics[width=0.48\textwidth]{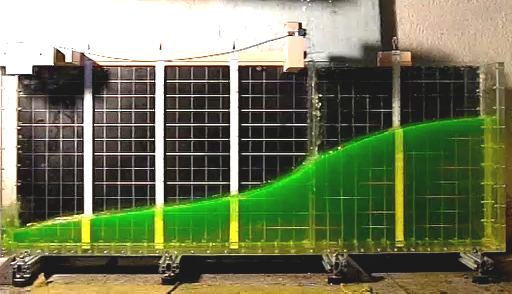}
\includegraphics[width=0.48\textwidth]{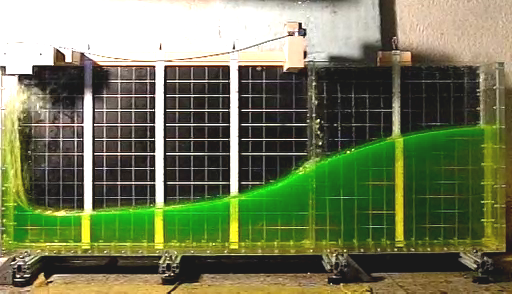}
\caption{$H=600\,\mathrm{mm}$; free surface evolution (0.0, 316.7, 413.4, 463.3, $\pm 3.3$ ms.)}
\label{fig:H60_fs_evolution}
\end{figure}
\begin{figure}
\centering
\includegraphics[width=0.325\textwidth]{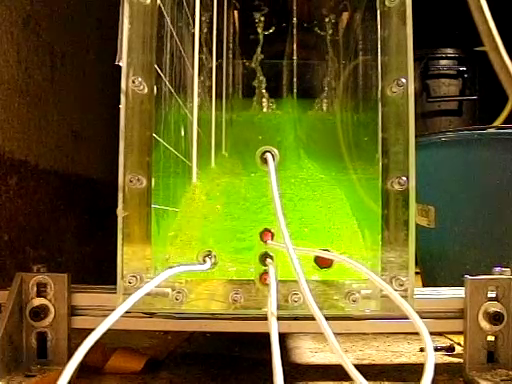}
\includegraphics[width=0.325\textwidth]{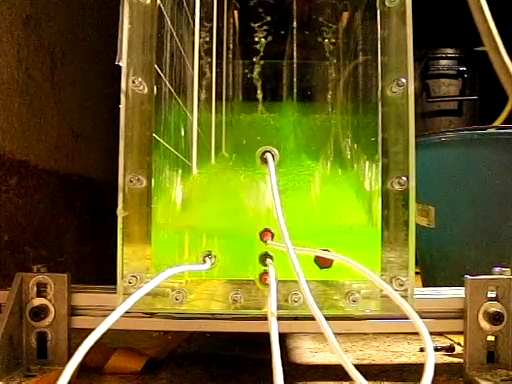}
\includegraphics[width=0.325\textwidth]{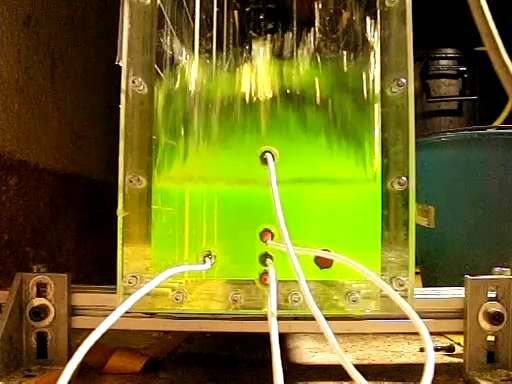}
\caption{$H=600\,\mathrm{mm}$; free surface during impact and run-up (front view) (390.0, 406.7, 420.0, $\pm 3.3$ ms.)}
\label{fig:H60_runup}
\end{figure}

The free surface profile evolution for experiments with $600\,{\rm mm}$ filling height shows qualitatively identical flow features as described above, Fig.~\ref{fig:H60_fs_evolution}, and a similar effect of the dam gate removal on the free surface shape is observed.
As the wave front advances towards the downstream vertical wall, small amplitude waves appear at the free surface and can be appreciated from the front views in Fig.~\ref{fig:H60_runup}. However, the run-up front remains homogeneous.

%


\subsection{Gate removal repeatability analysis}
\label{sss:gate_removal}

Since the velocity of the gate during its removal is the most important initial condition for
the dam-break problem, an analysis of the gate motion in absolute terms as well as its repeatability
from one experiment to another is carried out.

As discussed in section \ref{sec:exp_setup} the gate motion is induced by the free fall of a $15\,{\rm kg}$ weight.
This weight is connected to the dam by a steel wire that is guided through a pulley system and is loose initially.
The falling weight first stretches the steel wire before it actually causes a pull up of the dam gate.
The experimental time lapse is measured since the instant when the gate starts moving.

\begin{figure}
\centering
\includegraphics[width=0.495\textwidth]{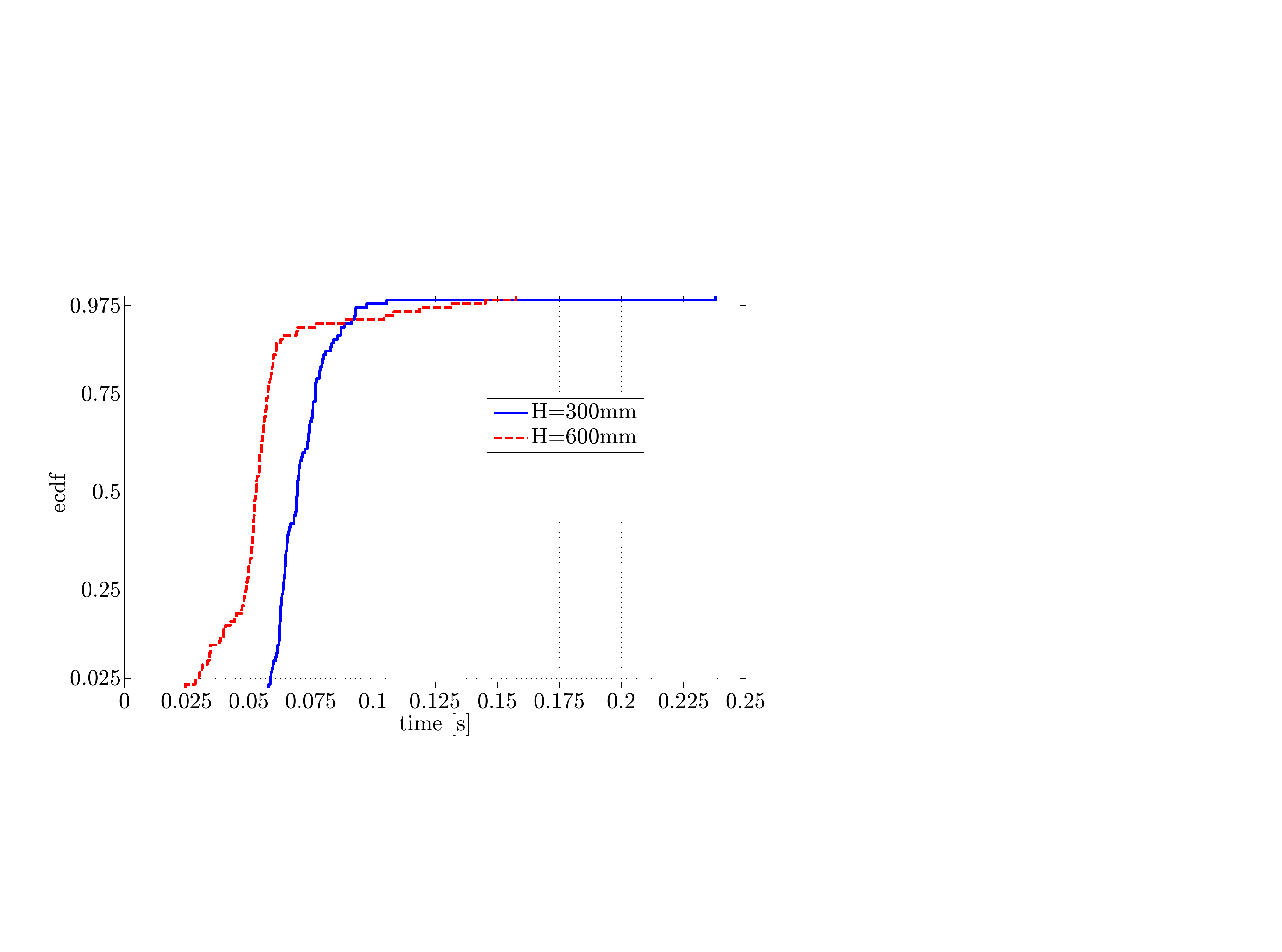}
\includegraphics[width=0.495\textwidth]{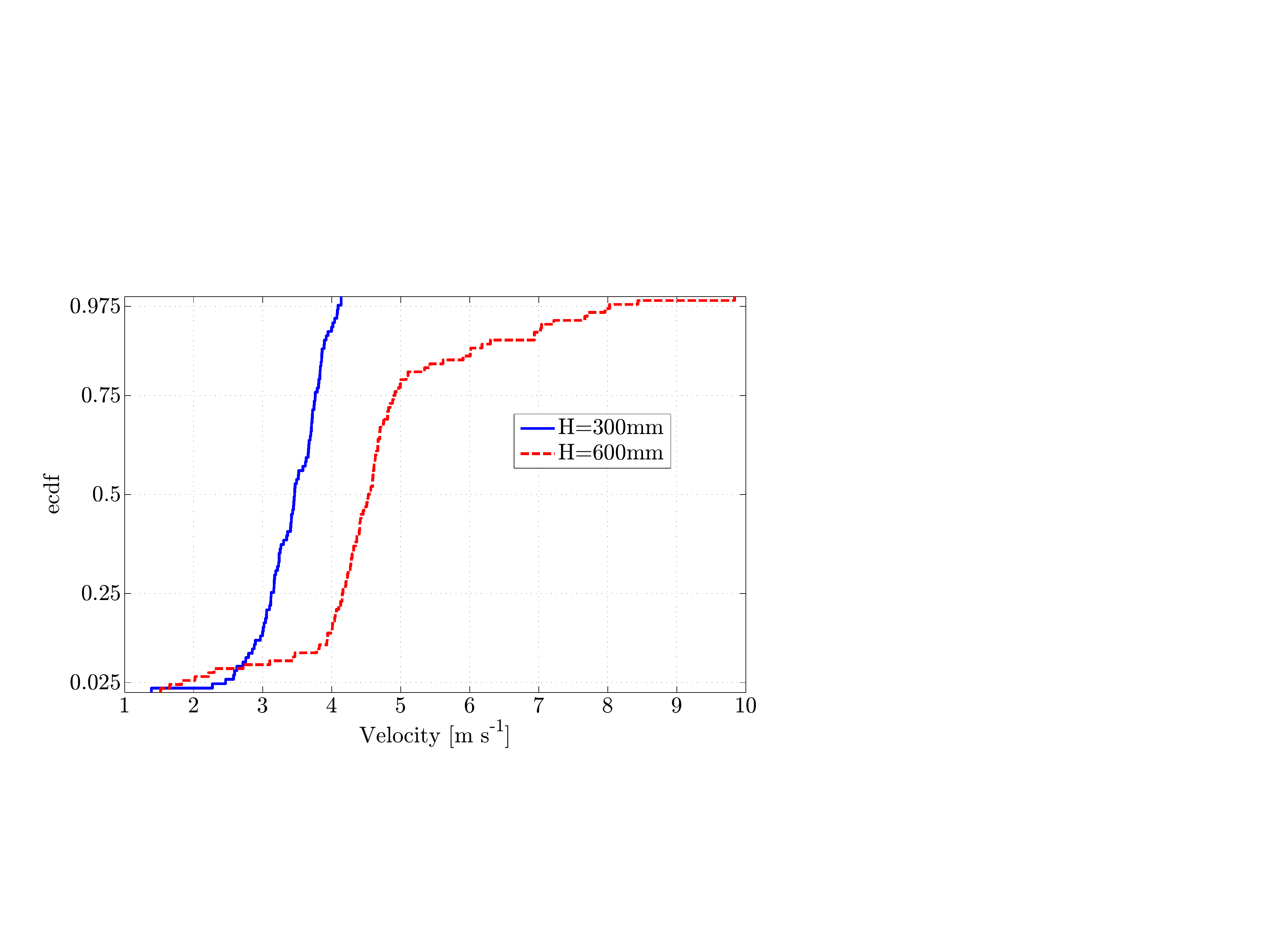}
\caption{Gate removal duration (left) and speed (right) empirical cumulative distribution function (ecdf) for both filling levels $H=300\,\mathrm{mm}$ and $H=600\,\mathrm{mm}$.}
\label{fig:CDF_GATE_REMOVAL_SPEED}
\end{figure}

Empirical cumulative distribution function (ecdf) graphs of the gate removal duration and speed (measured
as described in section \ref{sss:gatesensor})
for both $300\,{\rm mm}$ and $600\,{\rm mm}$ filling height tests are presented in Fig.~\ref{fig:CDF_GATE_REMOVAL_SPEED}.

For the tests with $300\,{\rm mm}$ filling height, the representative value (median) of the removal time
is $0.069\,{\rm s}$ and in $95\%$ of the cases the removal time falls within range $\left<0.06\,{\rm s}; 0.10\,{\rm s}\right>$.
The median value of measured removal time roughly corresponds to a theoretically predicted duration of the dam gate removal
based on the free fall of the weight. The gate velocity is obtained directly from the removal time.
Its median value is approximately $3.46\,{\rm ms^{-1}}$. Median value is recommended for setting up simulations, provided
gate velocity is decided to be relevant for the modeling; later in the paper the correlation between the pressures downstream and
the gate velocity is explored without finding a significant one. Regarding $H=600\,\mathrm{mm}$ filling height cases, the median value
of the removal time  is $0.053\,{\rm s}$ and median value of the velocity is $4.53\,{\rm ms^{-1}}$.

The difference in values between $H=300\,\mathrm{mm}$ and $H=600\,\mathrm{mm}$ test series may
be attributed to a minor adjustment of the steel wire which was required between the tests,
plus there might be a contribution of the extra static pressure due to the increased filling height.
However no thorough investigation of this phenomenon was performed.

\begin{figure}
\centering
\includegraphics[width=0.495\textwidth]{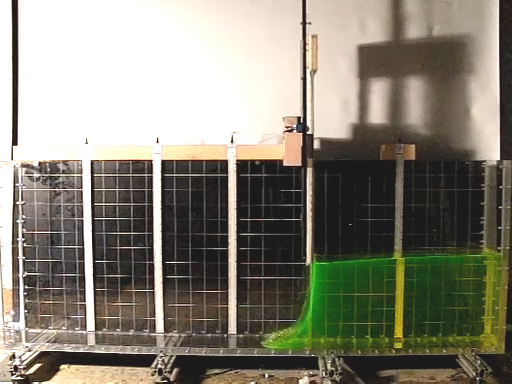}
\includegraphics[width=0.495\textwidth]{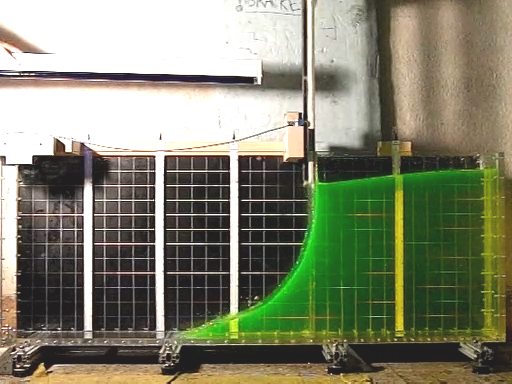}
\caption{Free surface profile and upstream wave at the instant of complete dam gate removal. $H=300\,{\rm mm}$ (left), $H=600\,{\rm mm}$ (right).}
\label{fig:upstream_wave}
\end{figure}

For the dam break flow analysis, it is also relevant to assess the kinematics of the upstream wave with regards to the dam gate removal.
Since the dam reservoir length is limited and the duration of the gate removal is finite, a comparison of the
two experimental setups is provided in Fig.~\ref{fig:upstream_wave}.
In both cases the front side of the liquid block is close to vertical when the gate stops touching the
fluid, which indicates that the gate rising speed is large enough so that its effect on
the liquid column collapse is minor, setting thus a good initial condition for the dam break experiment.

\subsection{Wave front and water level measurements}
%
In the following analysis, the displacement of the wave front from the gate and the water levels at the four locations are non-dimensionalized
with regards to the initial water level in the dam reservoir $H$. The recorded data are plotted versus the non-dimensional time $t^*=t\sqrt{g/H}$ (where $t$ is the dimensional time measured since the start of the gate's vertical motion).
%
%
\subsubsection{Wave front}
\label{sss:wavefront}
\begin{figure}
\centering
\includegraphics[width=0.495\textwidth]{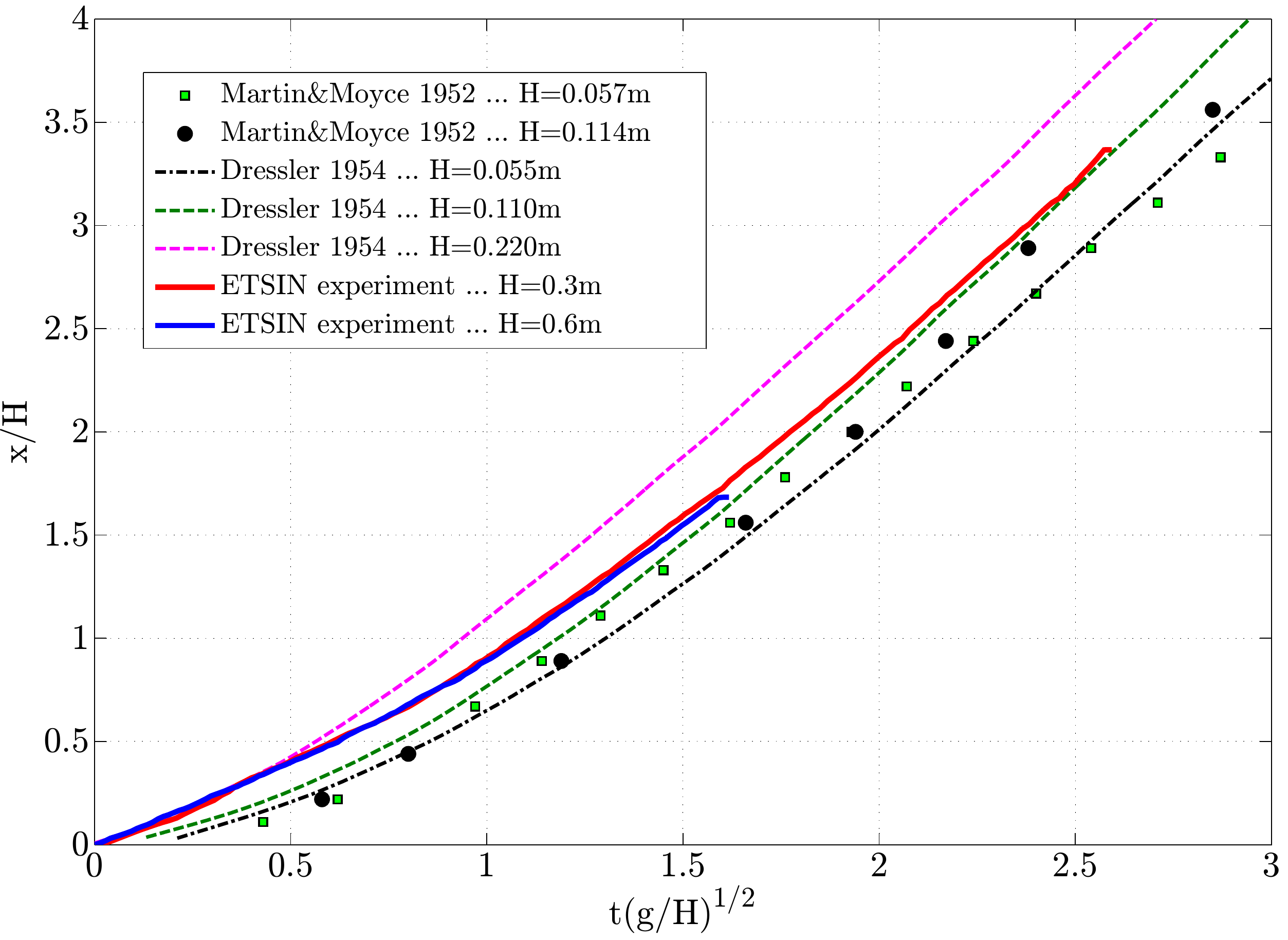}
\includegraphics[width=0.495\textwidth]{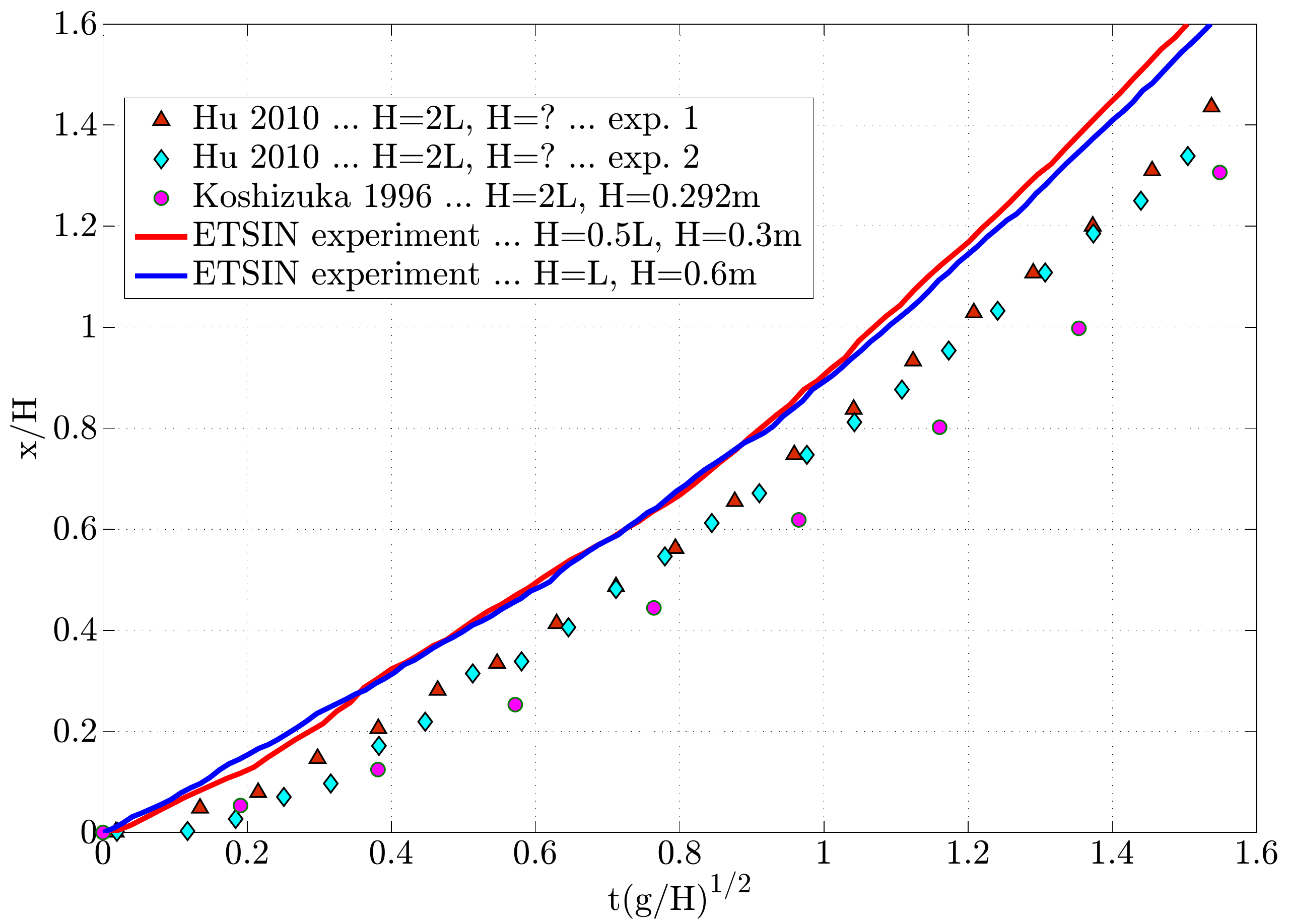}
\caption{Propagation of the surge front after dam gate removal compared to data from literature: \cite{Martin_1952a}\cite{Dressler_1954} (left) and \cite{Hu_2010}\cite{koshizuka1996} (right).}
\label{fig:SURGE_FRONT}
\end{figure}

The wave front propagation along the downstream horizontal dry bed after the gate release was read from the sequence of video
images using the digital image processing \cite{Parker_2010}, while the position of the wave tip was determined with an accuracy of $2.1\,{\rm mm}$.
The results for tests with filling heights $H=300\,{\rm mm}$ and $H=600\,{\rm mm}$ are compared in Fig.~\ref{fig:SURGE_FRONT} and show a very good agreement of the wave front propagation for both filling heights.
Furthermore, the recorded data correspond well to the previously published results in literature. The evolution of the displacement curves in time is qualitatively similar to findings of \cite{Martin_1952a}, \cite{Dressler_1954} or \cite{Hu_2010}.
The differences between the mentioned studies are pronounced the most for time $t^*<1$, i.e. in the initial stage of dam break wave evolution. This may be probably attributed to various techniques of dam gate removal which likely affects the way how the water column collapses from the dam reservoir onto the horizontal bed.
Nevertheless, a quantitative analysis of the recorded average velocity of the propagating downstream wave front after time $t^*>1$ shows reasonable results in comparison to the published experimental data, Tab.~\ref{tab:velo}.
All of these experimentally determined values are lower than the analytical solution by Ritter \cite{Ritter_1892}
who estimated the downstream wave front velocity equal to $v/\sqrt{gH}=2$.
Although the results indicate a reasonable agreement between the studies, a thorough analysis of the downstream wave front propagation would require a longer horizontal bed than the one provided in the presented experimental setup.

\begin{table*}[h]
	\centering
	\begin{tabular}{|l|c|}
		  \hline
		  Experimental run                         & Avg. velocity \\ \hline \hline
      ETSIN $H=300\,{\rm mm}$                  & $1.56$ \\ \hline
      ETSIN $H=600\,{\rm mm}$                  & $1.34$ \\ \hline
      Martin \& Moyce (1952) $H=57\,{\rm mm}$  & $1.48$ \\ \hline
      Martin \& Moyce (1952) $H=114\,{\rm mm}$ & $1.69$ \\ \hline
      Dressler (1954) $H=55\,{\rm mm}$         & $1.54$ \\ \hline
      Dressler (1954) $H=110\,{\rm mm}$        & $1.70$ \\ \hline
      Dressler (1954) $H=220\,{\rm mm}$        & $1.74$ \\ \hline
      Hu (2010) exp. 1                         & $1.21$ \\ \hline
      Hu (2010) exp. 2                         & $1.14$ \\ \hline
      Koshizuka (1996) $H=292\,{\rm mm}$       & $1.30$ \\ \hline
		\end{tabular}
	\caption{The non-dimensional average velocity $(v/\sqrt{gH})$ of the downstream wave front propagation along the dry horizontal bed for $t^*>1$.}
	\label{tab:velo}
\end{table*}

\begin{figure}
\centering
\includegraphics[width=0.325\textwidth]{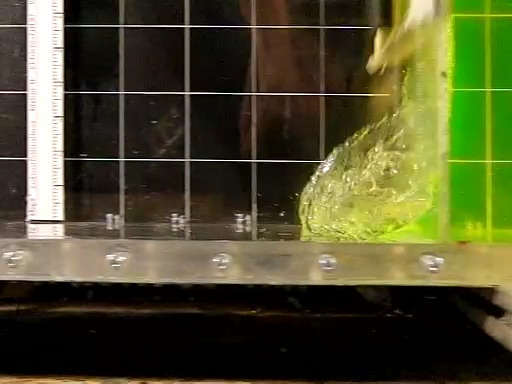}
\includegraphics[width=0.325\textwidth]{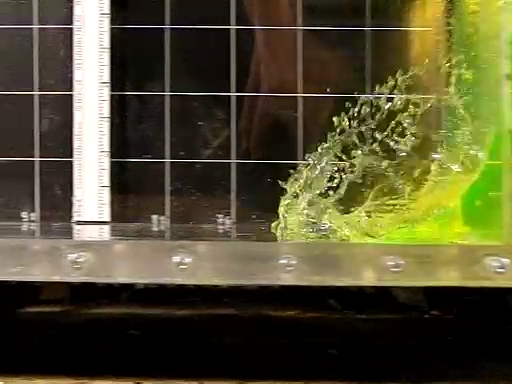}
\includegraphics[width=0.325\textwidth]{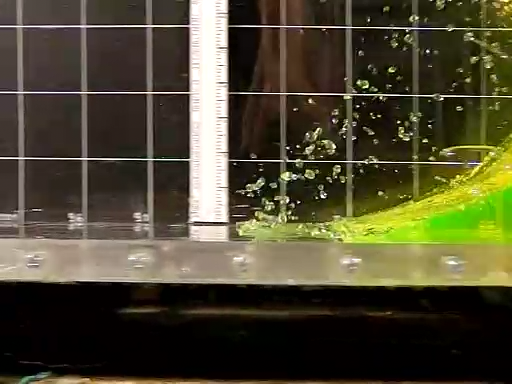}
\includegraphics[width=0.325\textwidth]{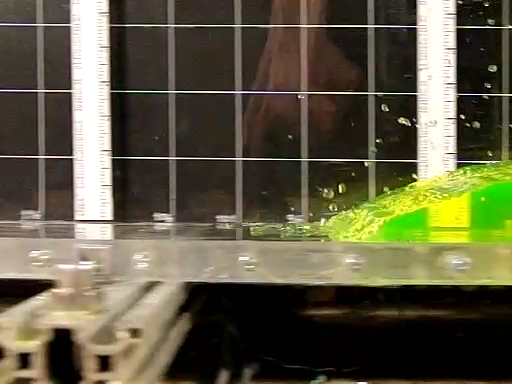}
\includegraphics[width=0.325\textwidth]{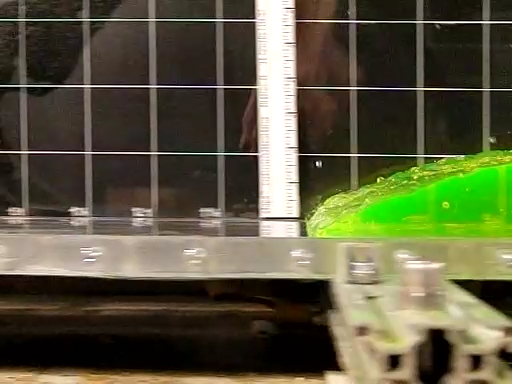}
\includegraphics[width=0.325\textwidth]{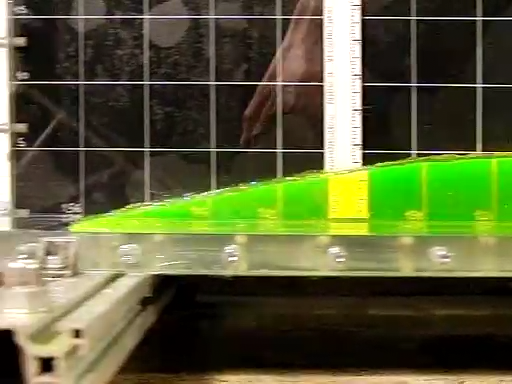}
\caption{Evolution of the downstream wave front after dam gate removal for $H=300\,{\rm mm}$ (33.4, 56.7, 110.1, 200.0, 266.7, 420.1, $\pm 3.3$ ms.). See supplementary materials at \href{http://canal.etsin.upm.es/papers/lobovskyetaljfs2013/}{http://canal.etsin.upm.es/papers/lobovskyetaljfs2013/} for a complete movie.}
\label{fig:H30_rail}
\end{figure}

In addition, a detailed high-speed video of the downstream wave front and its propagation in space was captured.
A close investigation of the surge front shape development was carried out using a camera that moved along the
tank on a linear guide, Fig.~\ref{fig:H30_rail}. 
When neglecting the friction at the horizontal bed, the theoretical solution of Ritter \cite{Ritter_1892}
shows that a tangent to the wave front free surface at its tip is horizontal. On the other hand, Dressler \cite{Dressler_1952}
in his theoretical solution considers a friction at the bed which results in vertical tangent to the free surface profile at the wave front tip.
This results in a convex shape of the wave front also confirmed by later experimental study of Dressler \cite{Dressler_1954}.
In general, the tangent direction is dependent on fluid's surface tension and wettability of the horizontal bed surface for given fluid.
For example in study by Nsom \cite{Nsom_2002}, a strongly convex shape of the wave front for highly viscous fluids was observed.

In the present experimental study, the initial phase of wave front evolution is rather complex due to the selected gate removal mechanism and materials applied. This is obvious from
Fig.~\ref{fig:H30_rail}, which displays several time instants after the gate removal.
First a rolling wave is created due to the vertical movement of the gate and adhesion between its surface and the tested fluid (water). This wave breaks down into droplets and only a thin horizontal jet remains travelling along the bed in front of the main bulk of the downstream wave.
Within a short time, this thin jet is absorbed by the main bulk of fluid traveling downstream and a convex parabolic shaped profile, as reported by Dressler
\cite{Dressler_1954}, can be identified. The tip of this parabolic shaped front becomes sharper with an increased distance from the dam reservoir.

As discussed in Janosi et al. \cite{Janosi_2004} and Stansby et al. \cite{Stansby_1998},
the wave front propagation and a complexity of the flow may be significantly affected by isolated drops or a layer of fluid covering the horizontal bed donwstream the gate.
However, such problems are avoided in this study by proper sealing of the dam gate and careful drying process before each run.
As a result, a linear flow downstream the dam gate with a smooth free surface profile is developed.

\begin{figure}
\centering
\includegraphics[width=4.33cm]{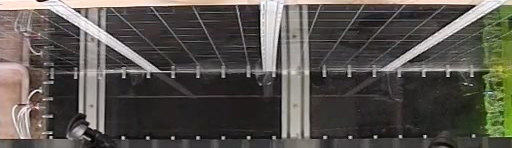}\\
\includegraphics[width=4.33cm]{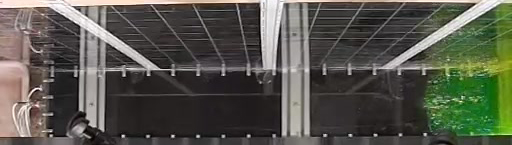}\\
\includegraphics[width=4.33cm]{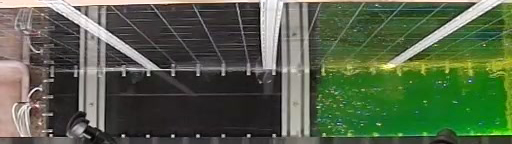}\\
\includegraphics[width=4.33cm]{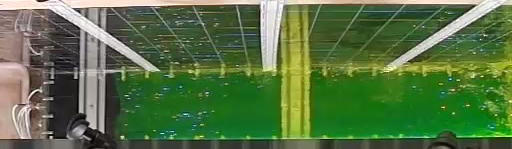}
\caption{$H=300\,{\rm mm}$; free surface (top view) (26.7, 106.7, 233.4, 393.4, $\pm 3.3$ ms.). See supplementary materials for a complete movie.}
\label{fig:H30_fstopview}
\end{figure}
\begin{figure}
\centering
\includegraphics[width=4.33cm]{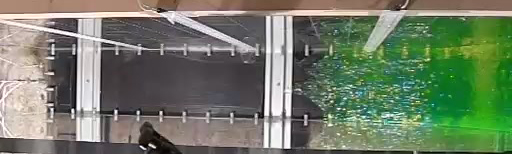}\\
\includegraphics[width=4.33cm]{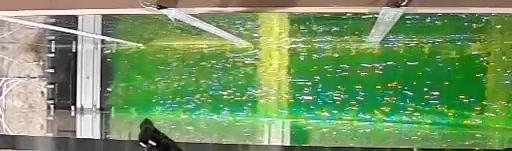}
\caption{$H=600\,{\rm mm}$; free surface before impact (top view) (176.7, 350.0, $\pm 3.3$ ms.)}
\label{fig:H60_tongue_top_view_before_impact}
\end{figure}

From the top-view images, it may be assumed that the flow is basically two-dimensional.
Unlike in the experimental setup of Janosi et al. \cite{Janosi_2004}
who observed instability of the wave front at the solid-liquid interface at long distance horizontal
runs, the longitudinal dimension of our tank is relatively short (about $3.33 H$ for $H=300\,{\rm mm}$
case and $1.67 H$ for $H=600\,{\rm mm}$).
For the $H=300\,{\rm mm}$ filling height the front of the downstream wave traveling along the horizontal bed does
not display significant instabilities prior the impact on downstream wall;
the wave front stays perpendicular to the direction of the flow, Fig.~\ref{fig:H30_fstopview}.
For the $H=600\,{\rm mm}$ case the front develops a wave tongue at the initial instances of the
wave propagation; later it is reabsorbed and the front becomes perpendicular to the lateral wall, Fig.~\ref{fig:H60_tongue_top_view_before_impact}.
All these small instabilities could nonetheless contribute to the significant scattering found for pressure registers,
mainly at the bottom sensor (1) which receives the direct impact of the advancing front.
%
%
%
\subsubsection{Water level measurements}
\label{sss:water_level_meas}
Water level measurements have been also conducted by digital image processing of video images.
This analysis was performed with an accuracy of $1.25\,{\rm mm}$.
Due to the reduced number of available video samples, no repeatability analysis has been conducted. Notwithstanding this, it is expected that the flow kinematics at least in the wave propagation prior to impact is repeatable.
This is confirmed by measurements of the arrival time to the lateral wall, which can be inferred from sensor $1$ pressure register. This issue is later discussed in section \ref{sss:pressure_time_history}.

\begin{figure}
\centering
\includegraphics[width=0.495\textwidth]{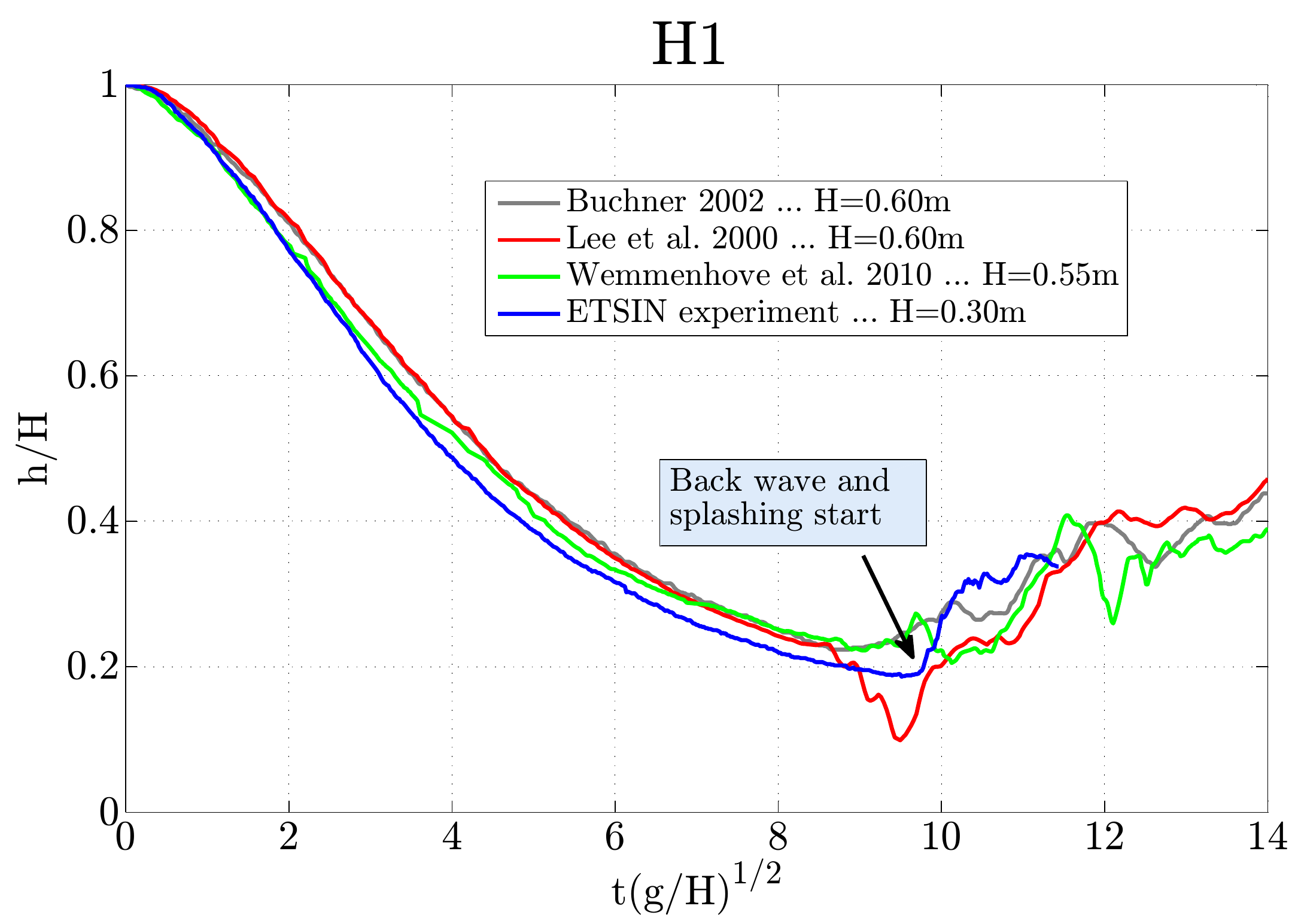}
\includegraphics[width=0.495\textwidth]{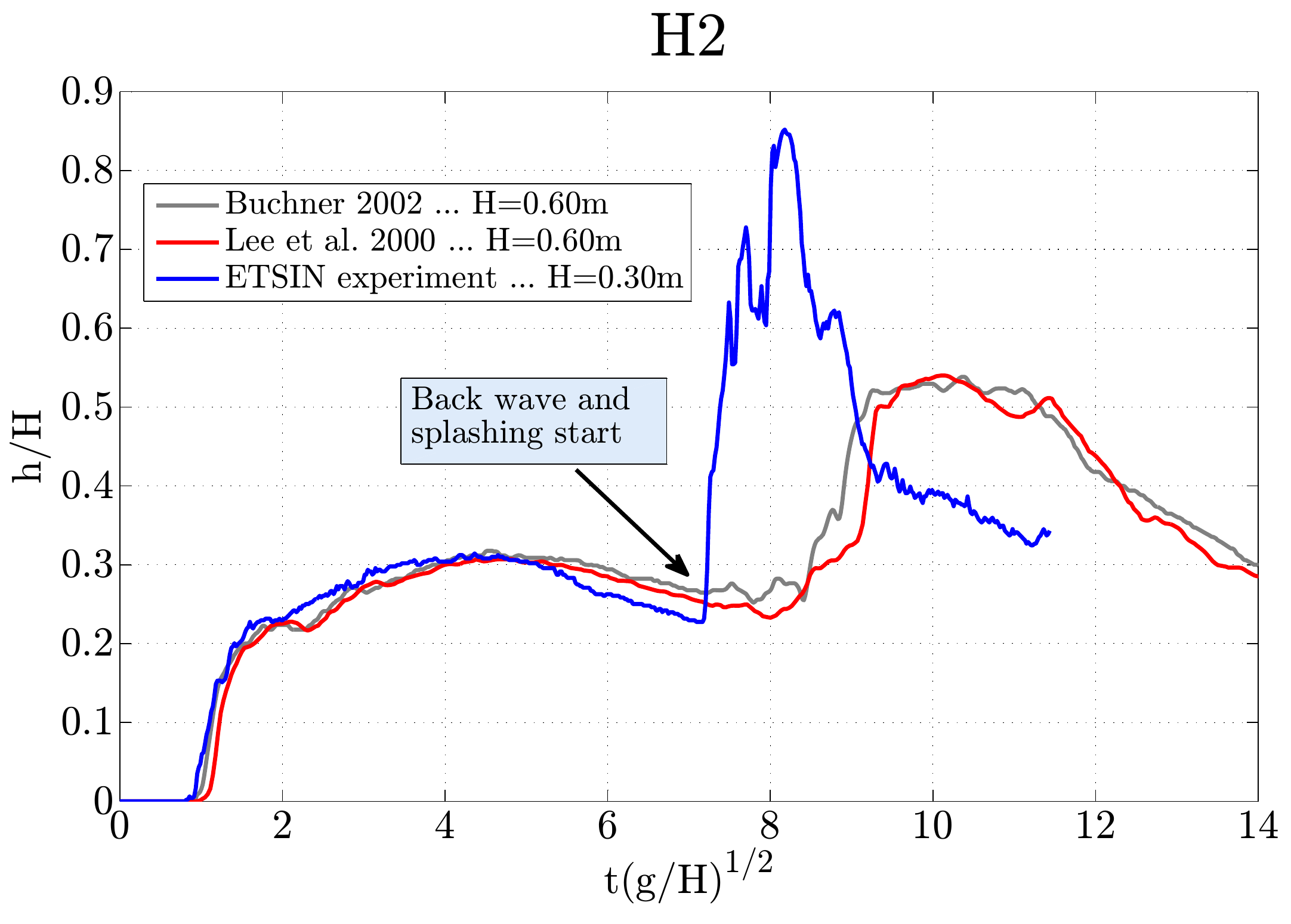}
\includegraphics[width=0.495\textwidth]{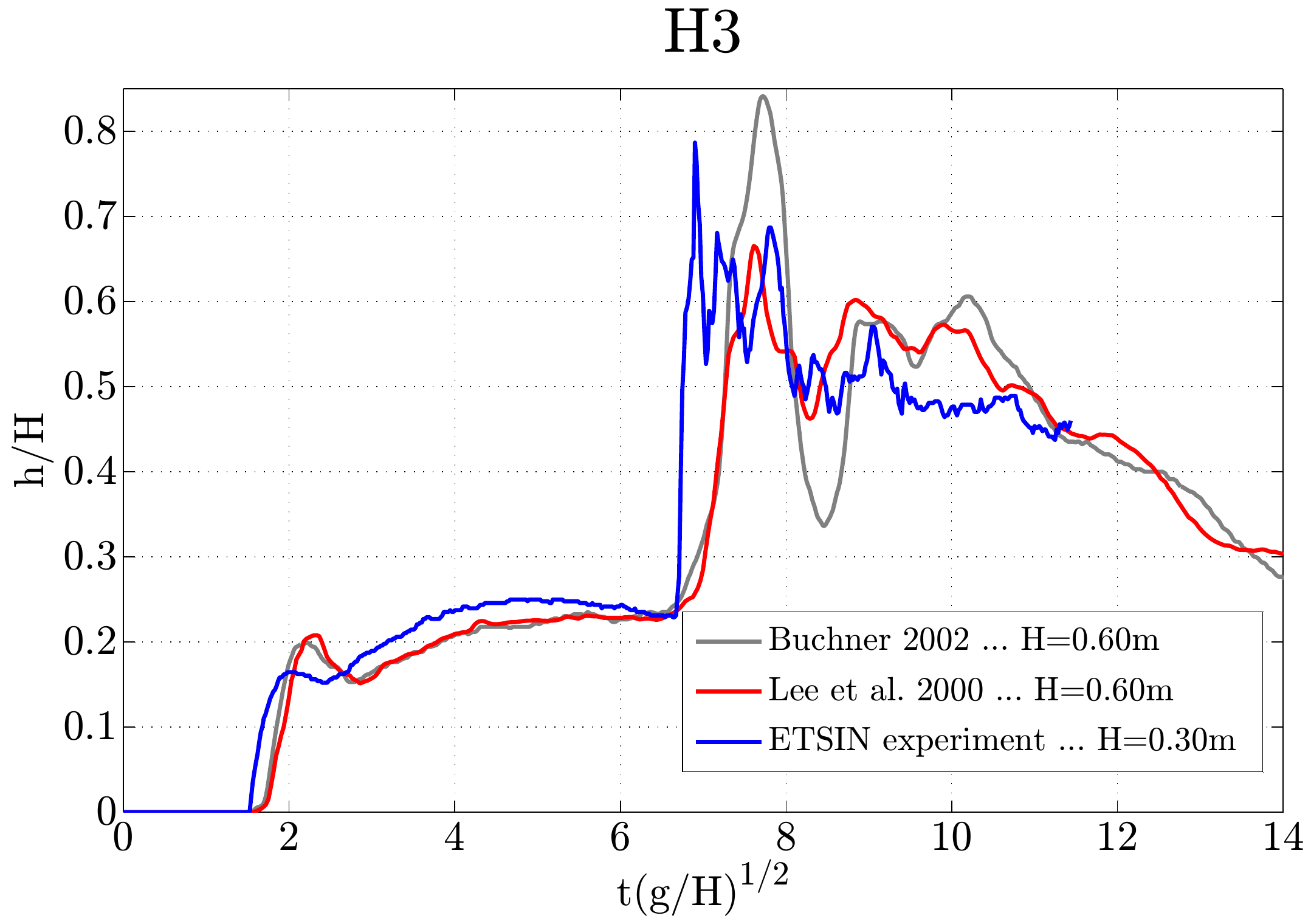}
\includegraphics[width=0.495\textwidth]{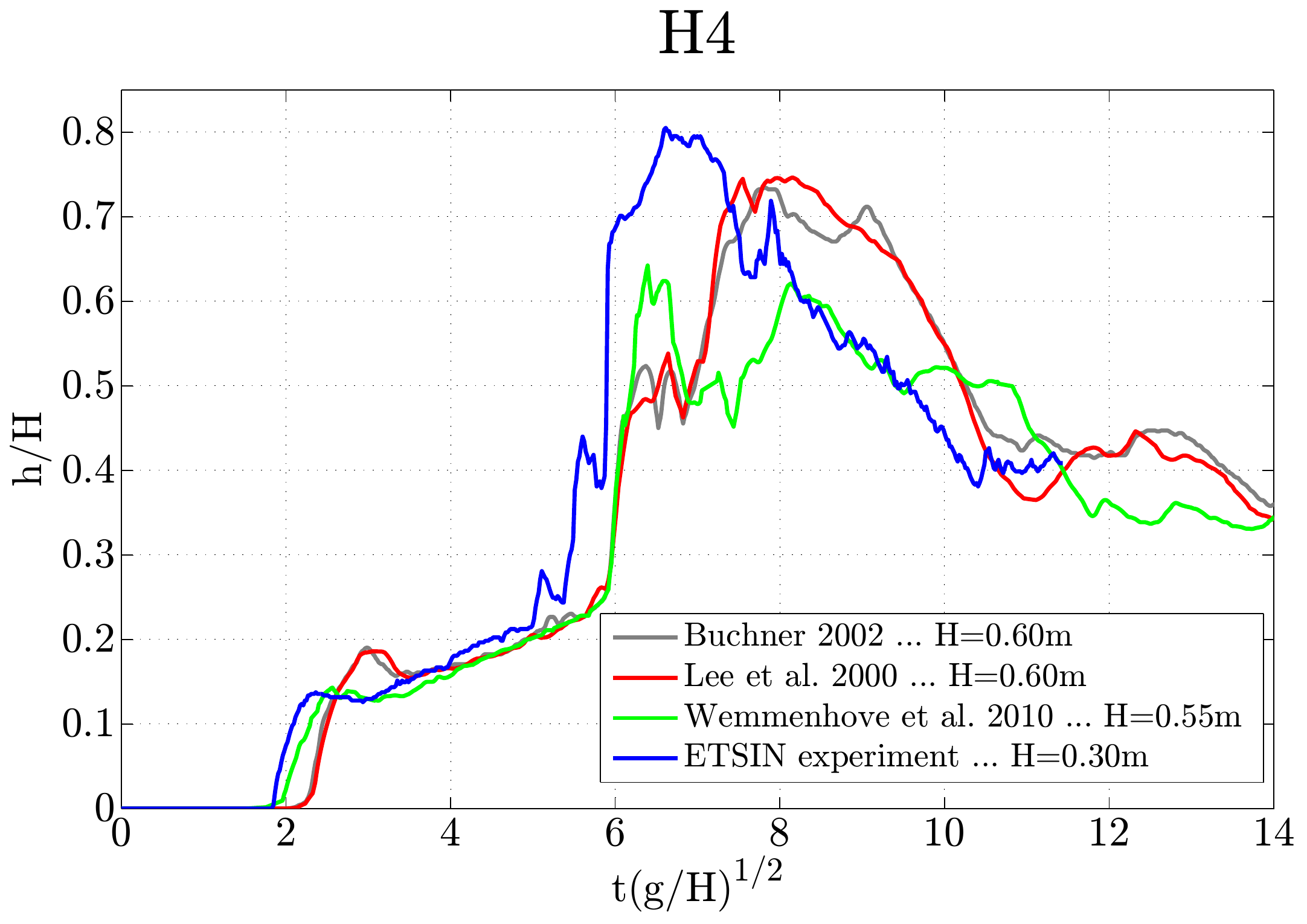}
\caption{Water level elevations at locations $H1$, $H2$, $H3$ and $H4$ for tests with $H=300\,{\rm mm}$ initial filling height compared to data from literature \cite{Buchner_2002PhD}\cite{Lee_2002}.}
\label{fig:WATER_LEVELS_300}
\end{figure}

The results of analysis of water levels at given locations $H1$, $H2$, $H3$ and $H4$ in time for $H=300\,{\rm mm}$ are provided in Fig.~\ref{fig:WATER_LEVELS_300}.
For simplicity, the downstream wave that is generated by the dam gate removal and travels along a dry bed is denoted as a primary wave in the following. The flow structures of this wave seem mostly laminar. When the primary wave impacts the vertical wall, a vertical run-up jet is created which then falls under gravity onto the underlying fluid. The back flow of the fluid causes a development of a plunging breaker which forms a combination of a bore propagation and a mixing layer when evolving on top of the laminar layer of fluid of the primary wave which is still advancing towards the wall, as plotted in Fig.~\ref{fig:H30_fs_evolution}. This back flow is denoted as a secondary wave in the paragraphs below.

When observing the plot for $H1$, Fig.~\ref{fig:WATER_LEVELS_300} (top-left), the decreasing part of the curve indicates the water discharge from the dam reservoir which agrees well with the data from literature, although being slightly more rapid. However, the increasing part of this curve indicates the backwards travelling secondary wave, which arrives earlier than in the related studies. Nevertheless the slope of the raise of the water level indicating also the shape of the secondary wave front agrees well in all the studies.

Plots for $H2$, $H3$ and $H4$ in Fig.~\ref{fig:WATER_LEVELS_300} provide information about the time between gate release and arrival of the primary wave to the given location, about the overall shape of the wave front, water level elevation and about the same characteristics for secondary wave. In general, it can be observed that the primary wave in our experimental runs propagates slightly faster downstream than in the related studies. A possible explanation of this effect might be the fact that there was a completely dry bed downstream the gate. Thus there was no thin layer of fluid which would resist the downstream flow and cause a delay of the primary wave, \cite{Janosi_2004}\cite{Stansby_1998}.
There is a good agreement in the slope of the water level elevation with other studies. However, the initial hump (which is found in studies by Buchner \cite{Buchner_2002PhD} and Lee et al. \cite{Lee_2002}) is missing. This may indicate that the horizontal bed in studies \cite{Buchner_2002PhD} and \cite{Lee_2002} was not completely dry.

After the hump, the character of the water level elevation curves agrees well in all compared studies,
however the major difference appears in time of arrival and elevation of the secondary wave. For $H3$ and $H4$, the secondary wave elevation agrees reasonably well, but results at location of $H2$ show a significant discrepancy. This is partially caused by complex free surface structures of the secondary flow which affected the optical images, but it also rises a question of the secondary flow repeatibility. Since this paper is focused on dynamics of the primary wave, this question remains opened for future studies.

\section{Results - Impact Pressure}

\subsection{Filling height $300\,{\rm mm}$}
\label{sss:h30_pressurepeaksanalysis}
\begin{figure}
\centering
\includegraphics[width=0.9\textwidth]{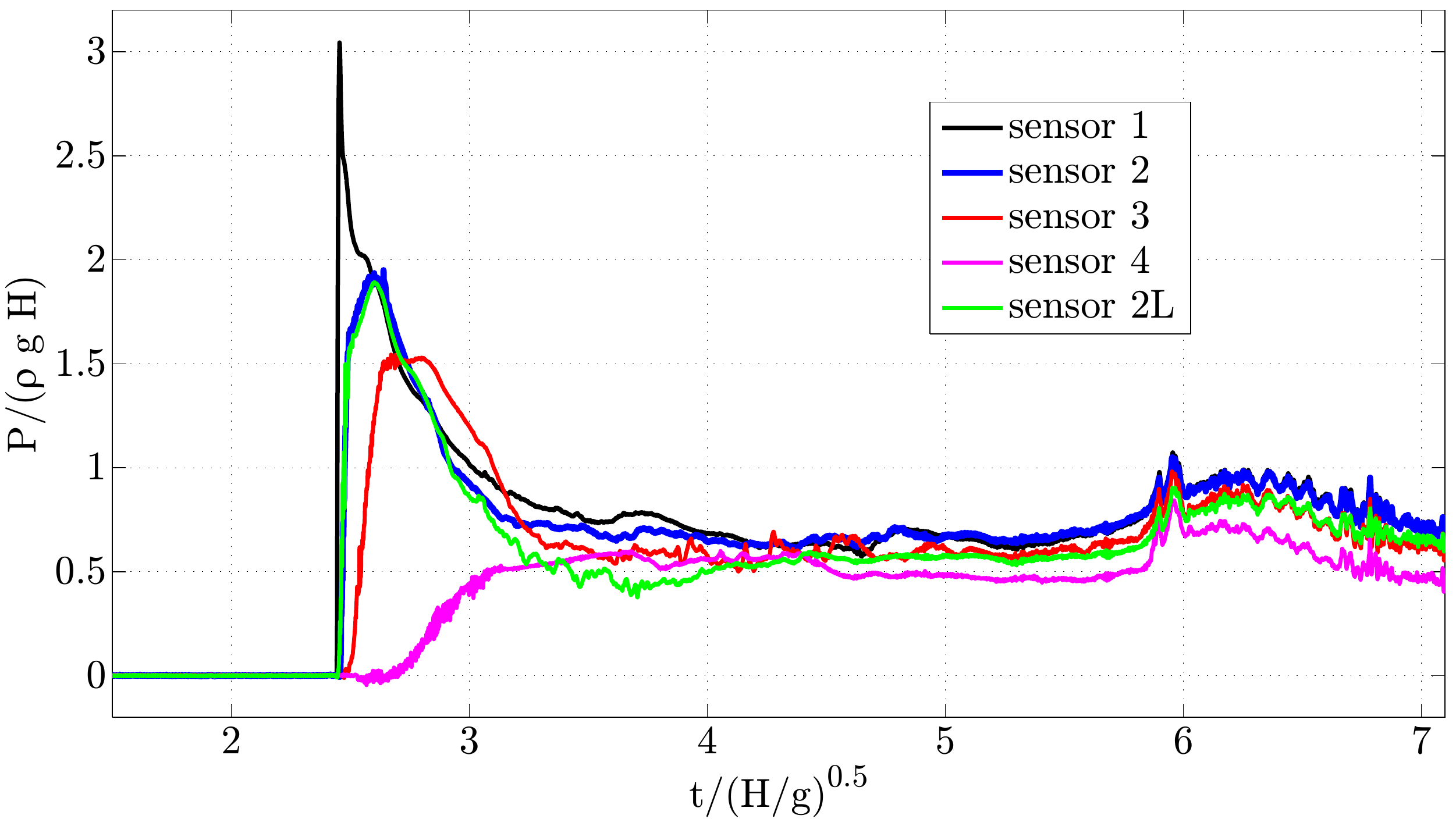}
\caption{$H=300\,{\rm mm}$; typical impact event pressure signals from all five pressure sensors.}
\label{fig:peak_event_5_sensors}
\end{figure}

The impact pressure was measured with five sensors at the vertical wall at the end of the downstream flume, as described in sections \ref{sec:exp_setup} and \ref{sec:test_matrix}. The positions of the pressure sensors at the wall are displayed in Fig.~\ref{fig:dam_scheme}.
The statistical analysis of the pressure peaks, rise times and the occurrence time, i.e. the time between the opening of the dam gate and the occurrence of the impact, was carried out based on data from $100$ test runs.
A typical dam break impact event signal as registered by the five sensors can be seen in Fig.~\ref{fig:peak_event_5_sensors}.
The recorded pressure $P$ is non-dimensionalized with regards to the hydrostatic pressure at the bottom of the reservoir and denoted as $P^*$.

It can be noticed that the highest peak is recorded by sensor number $1$ which is the sensor receiving the full impact, whilst the pressure of the other sensors is given by the run up of the flow. We can also observe that sensor number $4$,
i.e. the sensor located at the highest position, does not record a pure impact event, see Fig.~\ref{fig:H30_frontview}, and actually
the maximum for this sensor is obtained later in time, when the water falls back after running along the wall.
Sensor $2L$ is placed at the same height level as the sensor $2$ but off-centered by $37.5\,{\rm mm}$ (see Fig.~\ref{fig:dam_scheme}) so that the three-dimensionality of the impact event in terms of pressure can be explored.

\begin{figure}
\centering
\includegraphics[width=0.7\textwidth]{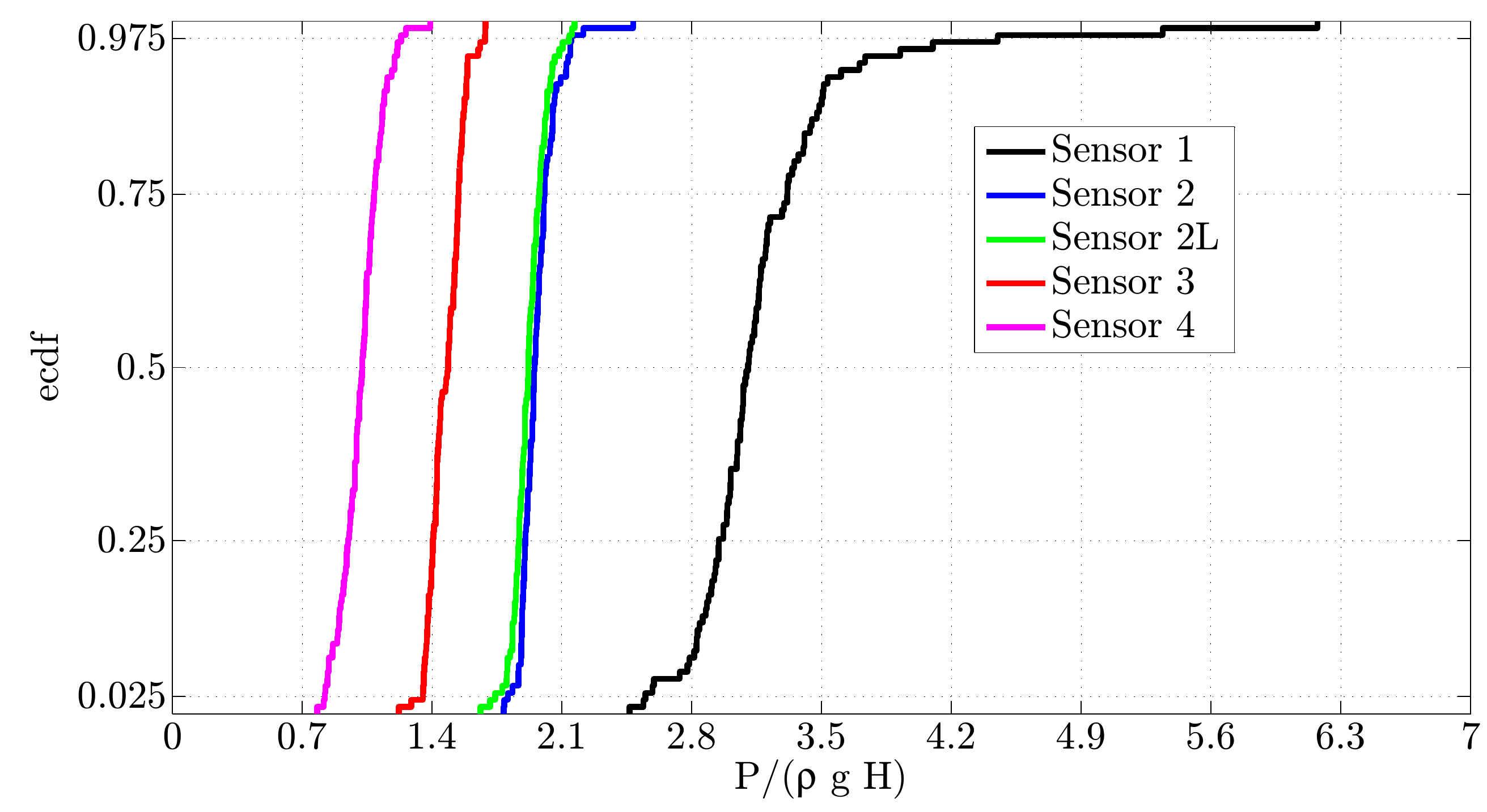}
\caption{$H=300\,{\rm mm}$; empirical cumulative distribution function (ecdf) of peak pressure for all five pressure sensors.}
\label{fig:H30_ensemble_peak_sensor1}
\end{figure}

Since the experiment has been repeated $100$ times, it is relevant to assess the dispersion of the recorded data.
Empirical cumulative distribution functions for the pressure peaks recorded by the five sensors are presented in
Fig.~\ref{fig:H30_ensemble_peak_sensor1}.
The dispersion of pressure peaks occurring in sensor $1$ registers is larger than the dispersion in peak values captured at other sensors where the impact is not so intense. The pressure peak values decay clearly with the distance from the bottom.
A summary of median values and confidence intervals can be seen in Tab.~\ref{tbl:h30stats}.

As can be observed in Fig.~\ref{fig:H30_ensemble_peak_sensor1}, sensor $1$ pressure median is approximately $3$ times the
static pressure at the bottom of the dam and the $97.5\%$ percentile goes up to approximately $4.5$ times the static pressure.
The advancing front can be assimilated to a jet impinging on a flat plate, for which an analytical solution of the
pressure and velocity fields is available \cite{Taylor_66_obliqueimpact}\cite{molteni_colagrossi_2009}.
The stationary value of $P/\rho\;v^2$ for an impinging jet is $0.5$ at the center and decays radially
towards the edge of the jet.
Considering the values of the bursting wave front velocity discussed in section \ref{sss:wavefront},
the ratio $P/\rho\;v^2$
obtained for present dam-break experiments are significantly larger (about $1.25$ for the median value and about $1.85$ for the $97.5\%$ percentile). This may be relevant for design aspects when considering transient loads in these types of flows.

\begin{figure}
\centering
\includegraphics[width=0.7\textwidth]{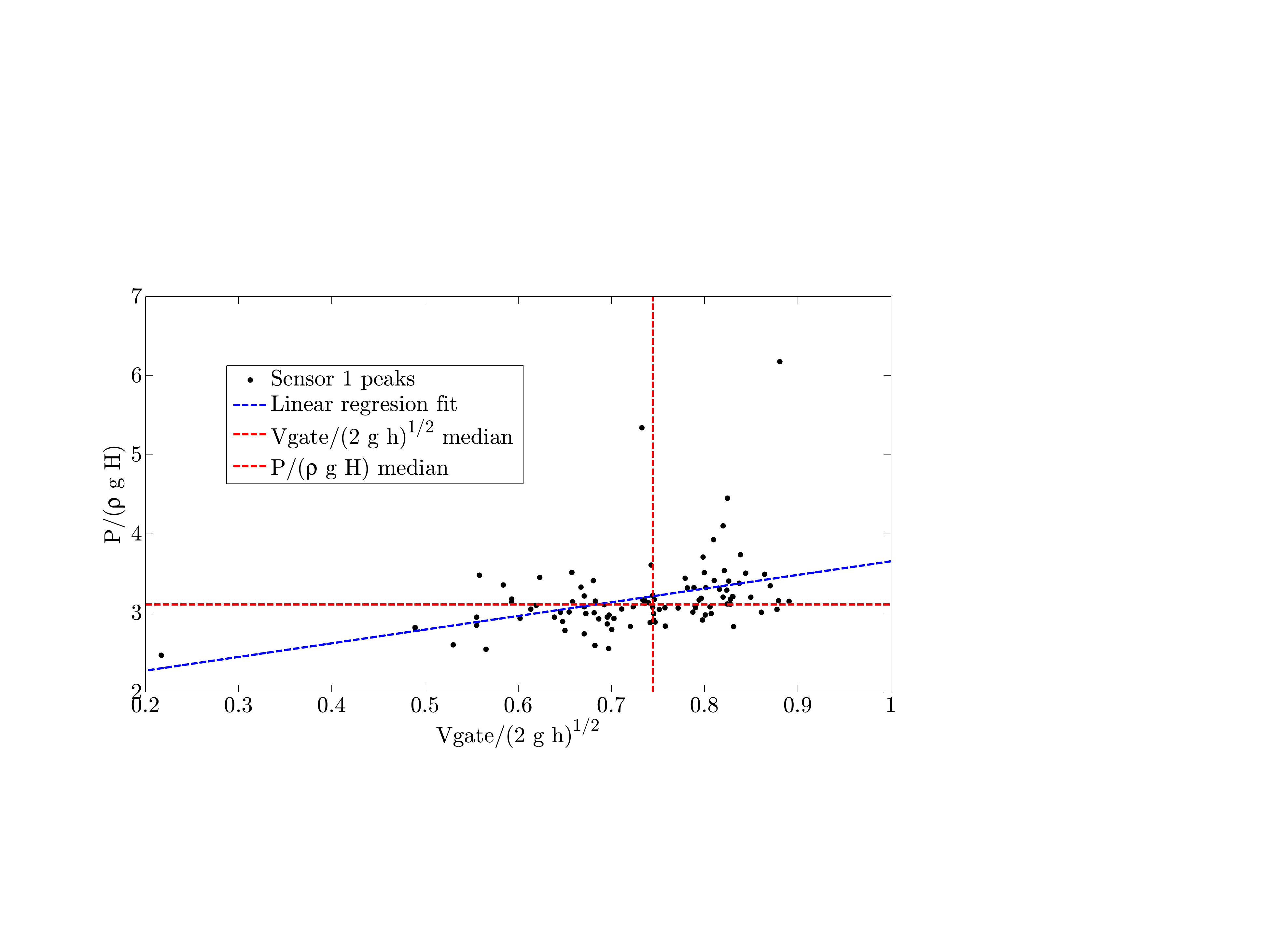}
\caption{$H=300\,{\rm mm}$; sensor $1$ peak pressure and gate velocity correlation.}
\label{fig:h30_p1_vgate_correlation}
\end{figure}

In order to investigate the influence of exogenous factors in the variability of this value, the correlation of the pressure peak value of sensor $1$ with the velocity of the gate has been explored, see Fig.~\ref{fig:h30_p1_vgate_correlation}.
It has been found that such correlation is not significant enough, considering the precision of the velocity measurements,
to state that the gate velocity is a relevant exogenous factor of variability in our experiments.
\begin{table*}[ht]
  \centering
  \begin{tabular}{|c|c|c|c|c|}
    \hline
Sensor	&	z [mm]	&	Median [mb]	&	97.5 Percentile	&	2.5 Percentile	\\ \hline
1	&	3	&	91.44	&	130.98	&	75.06	\\ \hline
2	&	15	&	57.49	&	63.31	&	53.26	\\ \hline
3	&	30	&	43.76	&	49.62	&	39.79	\\ \hline
4	&	80	&	30.15	&	36.32	&	24.18	\\ \hline
2L	&	15	&	56.3	&	63.00	&	51.23	\\ \hline
  \end{tabular}
  \caption{$H=300\,{\rm mm}$; summary of statistical data as recorded by pressure sensors for $100$ tests.}
  \label{tbl:h30stats}
\end{table*}
%
\subsubsection{Pressure time history analysis}
\label{sss:pressure_time_history}
\begin{figure}
\centering
\includegraphics[width=0.95\textwidth]{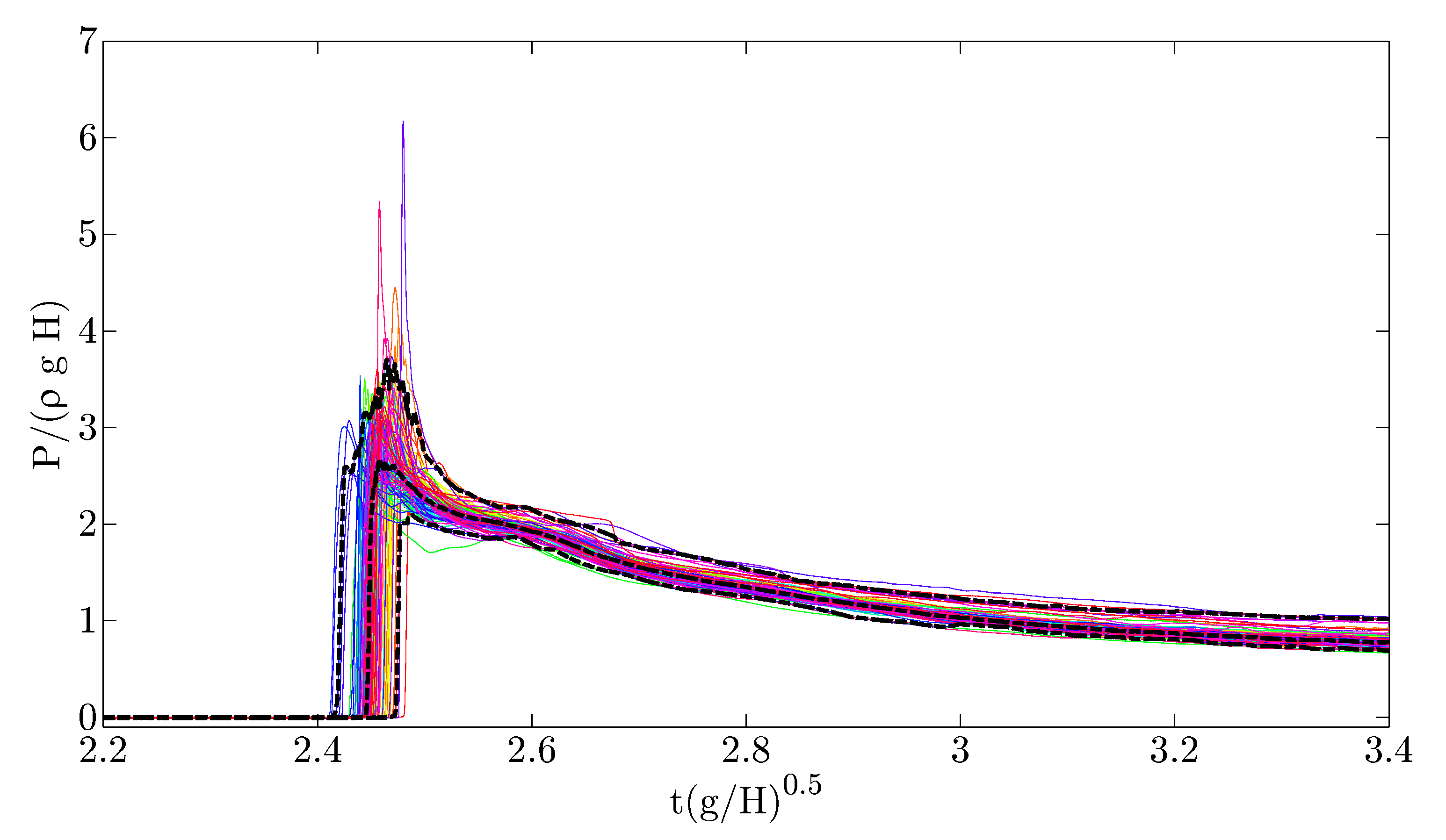}
\caption{$H=300\,{\rm mm}$; sensor $1$ pressure time histories for $100$ tests. The central black mark indicates the median, the lower and upper additional black markers represent the estimated $2.5\%$ and $97.5\%$ percentile levels respectively.}
\label{fig:h30_sensor1_pressure_th}
\end{figure}

In Fig.~\ref{fig:h30_sensor1_pressure_th}, the time histories of the $100$ tests for sensor $1$ are presented along with their median and the $2.5\%$ and $97.5\%$ percentile levels. For every instant a median and the corresponding confidence intervals are obtained. These are much smaller for the tail of the signal than for the impact event.

Variations in the occurrence time, as denoted by various instants in which the steep rise of pressure values is initiated, fall within a range of about $12\,{\rm ms}$ 
for the whole set of $100$ tests. This corresponds to $t^*=0.07$ in non-dimensional terms.
For comparison, this value is more than $5$ times smaller than the median of the gate removal duration, see section \ref{sss:gate_removal}.
This means that the variations in the occurrence time are less than $3\%$ of its median value.
Based on this observation a good repeatability of the wave front propagation is expected.

Since the confidence intervals are provided not only for the peak value but for the whole time history registered by the sensor, the data displayed in Fig.~\ref{fig:h30_sensor1_pressure_th} provide a relevant information that can be used for computational fluid dynamics (CFD) codes validation.
\subsubsection{Rise and decay times analysis}
\label{sss:risetime30}
\begin{figure}
\centering
\includegraphics[width=0.65\textwidth]{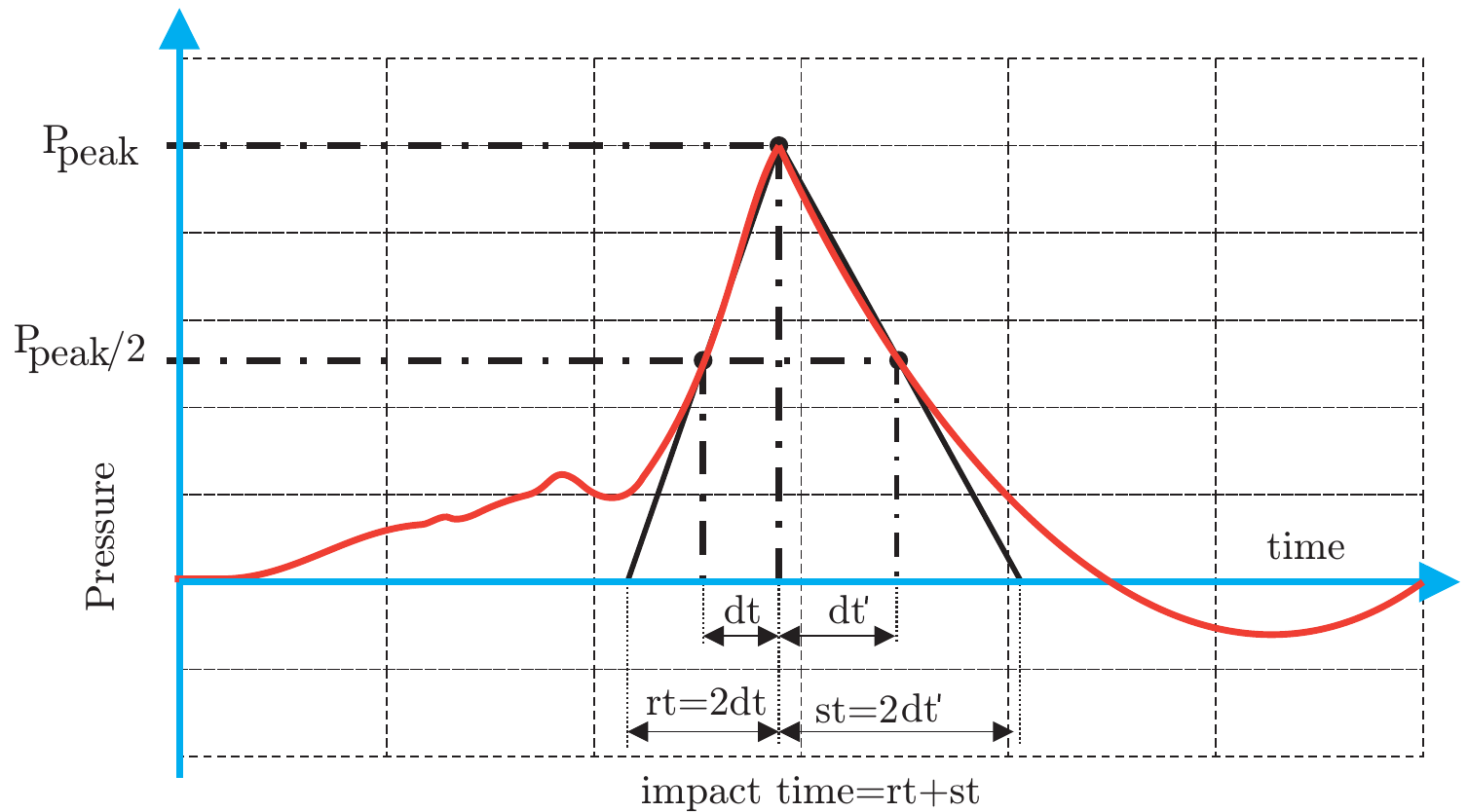}
\caption{Rise time definition}
\label{fig:rise_time_def}
\end{figure}

An analysis of the rise time and the decay time has also been carried out for sensor $1$. The rise time is important in sloshing phenomena because the structural response is not only dependent on
the peak pressure value but also on the duration of the event \cite{graczyk_moan_oe2008}\cite{wang_etal_dynamicstructcoef_isope2012}. The definition of the rise time used in this study is taken from ISOPE 2012 benchmark \cite{Loysel_Chollet_Gervaise_etal_gtt_isope2012}.
The rise time $rt$ is defined equal to twice the duration between the instants when the pressure signal rises to its half-maximal and maximal value (see fig \ref{fig:rise_time_def}).
The decay time $st$ is defined analogously by looking at the signal after the impact and the summation of both is defined as the impact time or impact duration.
The latter will be useful to model the impulse of the impact event.

\begin{figure}
\centering
\includegraphics[width=0.7\textwidth]{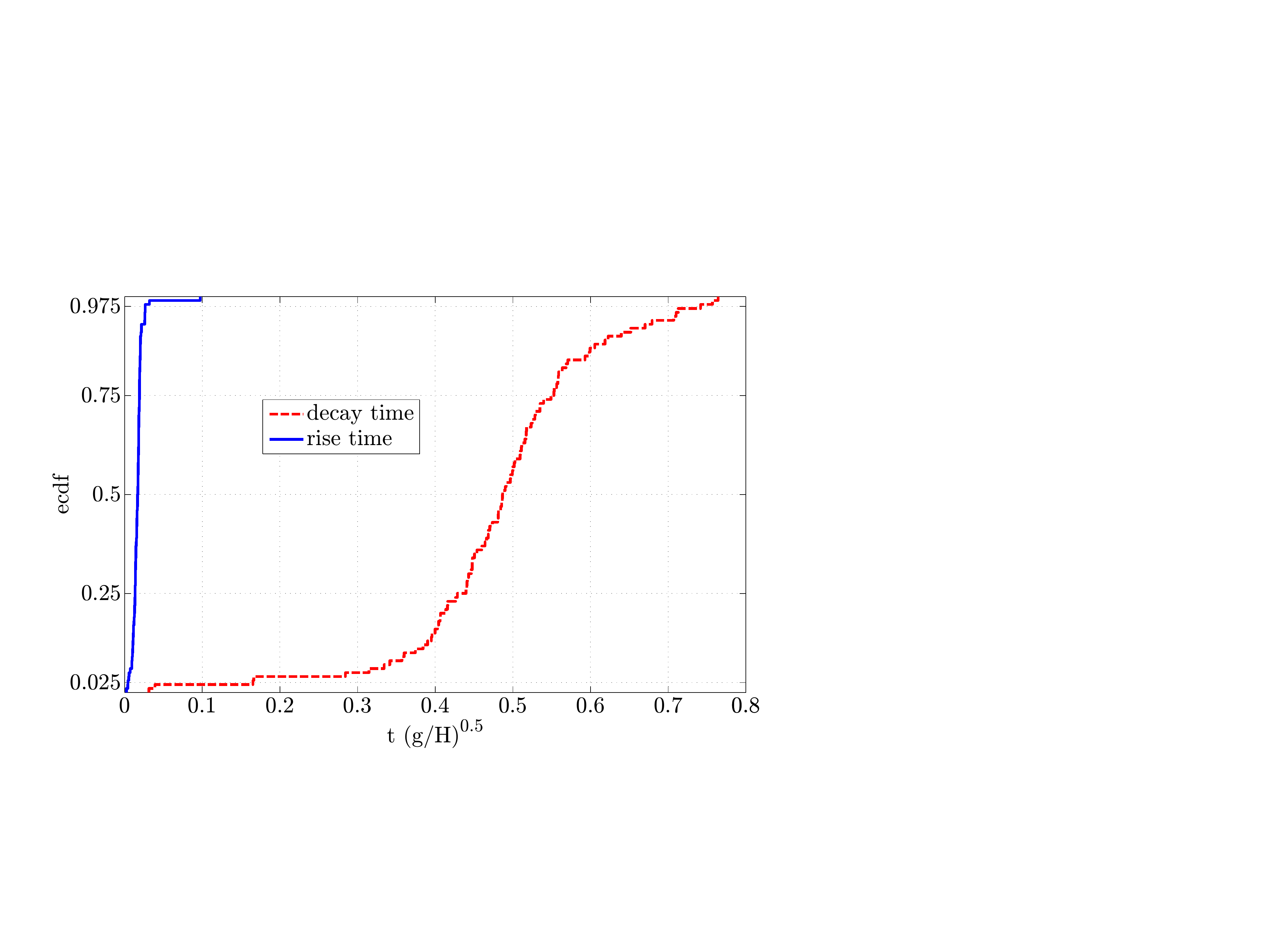}
\caption{$H=300\,{\rm mm}$; empirical cumulative distribution function (ecdf) of rise and decay times as registered at sensor $1$.}
\label{fig:H30_ensemble_rise_time sensor1}
\end{figure}

In Fig.~\ref{fig:H30_ensemble_rise_time sensor1}, the empirical cumulative distribution functions (ecdfs) of rise and decay times for the pressure time history from sensor $1$ are presented. The $95\%$ confidence interval in the rise time ranges from $1.5\,{\rm ms}$ to $4.5\,{\rm ms}$.
Thus the applied sampling rate of $20\,{\rm kHz}$ is considered sufficient in order to resolve well the peak pressures and the overall pressure time history.
The decay times are an order of magnitude larger than the rise times and are also presented in Fig.~\ref{fig:H30_ensemble_rise_time sensor1}. This corresponds to data from the sample pressure register presented in Fig.~\ref{fig:peak_event_5_sensors}.

\begin{figure}
\centering
\includegraphics[width=0.7\textwidth]{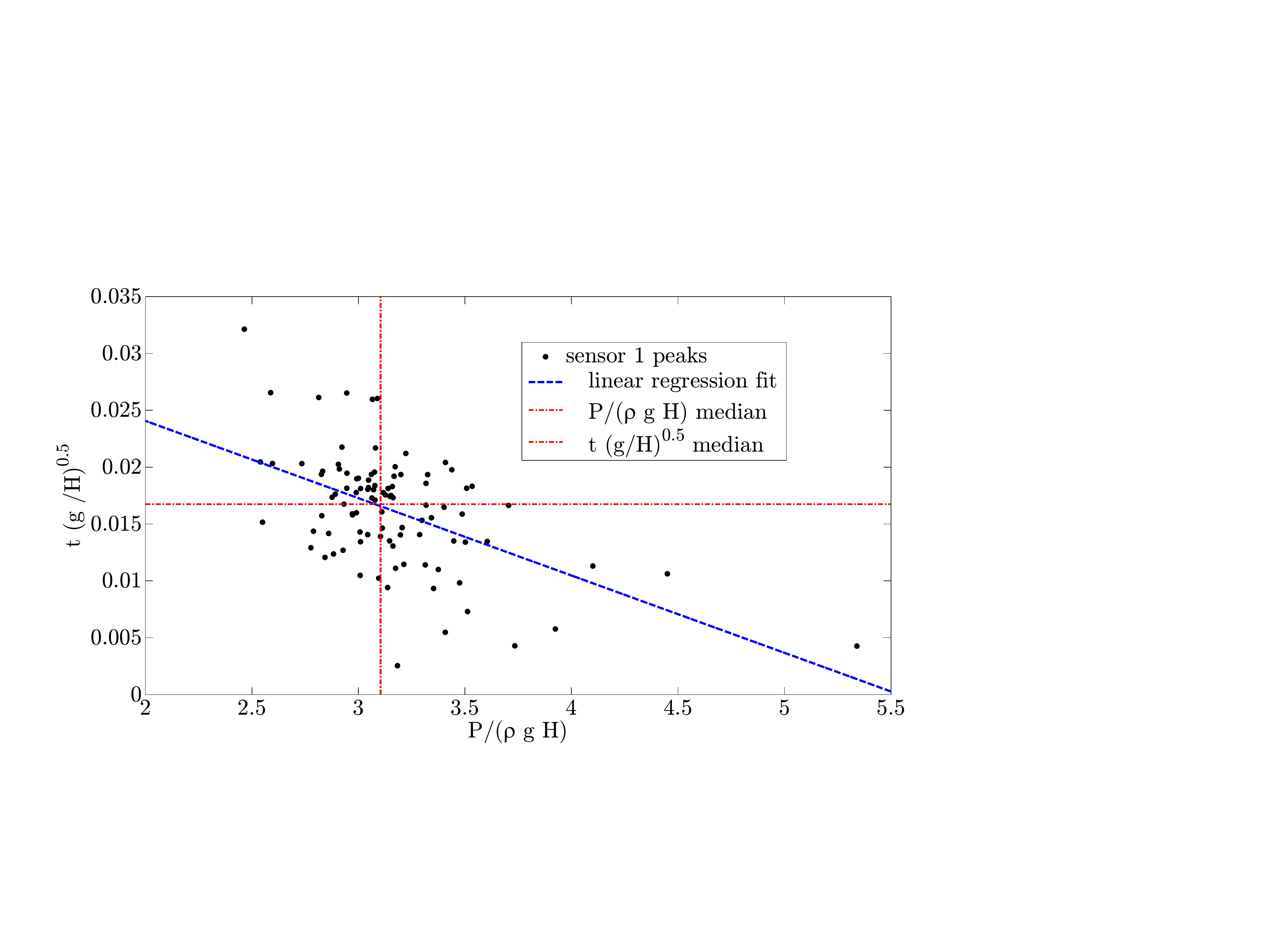}
\caption{$H=300\,{\rm mm}$; joint distribution of rise time vs pressure peak as registered at sensor $1$.}
\label{fig:h30_risetime_p1_joint}
\end{figure}

It is interesting to observe both the peak pressure and the rise time as a realization of the same random
phenomenon, and thus a joint distribution of the rise time and the peak pressure values is shown in Fig.~\ref{fig:h30_risetime_p1_joint}.
It can be appreciated that the rise time and the pressure peaks present a negative correlation (linear correlation factor equal to $-0.385$).
This anti-correlation had already been described by e.g. \cite{graczyk_moan_oe2008} who observed substantially shorter rise times for the highest pressure impact cases in the sloshing context. This applies also within this study as can be seen in Fig.~\ref{fig:h30_risetime_p1_joint}.

%
\subsubsection{Impulse}
\label{sss:impulse30}
It is interesting to analyze the impacts from the perspective of the pressure impulse.
The pressure impulse is the integral of the pressure over the
duration of the impact \cite{peregrine2003}, i.e.
\begin{equation}
\label{eq:impulse}
I(\mathbf{x})=\int_{\mathrm{Impact\;time}} P(\mathbf{x},t)dt
\end{equation}
The pressure impulse is a relevant physical variable when analyzing the impact of a wave on a structure as its values have more consistent magnitudes than the values of the peak pressure \cite{delorme_colagrossi_etal09}.

The pressure impulse for each test is now taken as the area of pressure curve during the impact time (Fig.~\ref{fig:rise_time_def}) and it is compared with the approximation obtained by the area of the triangle defined by the pressure peak and the impact time. Modeling of impact events by triangular pressure curves is found in the literature in the context of the structural finite element analysis of sloshing loads \cite{Graczyk_Moan_Wu_saos2007}.

\begin{figure}
\centering
\includegraphics[width=0.7\textwidth]{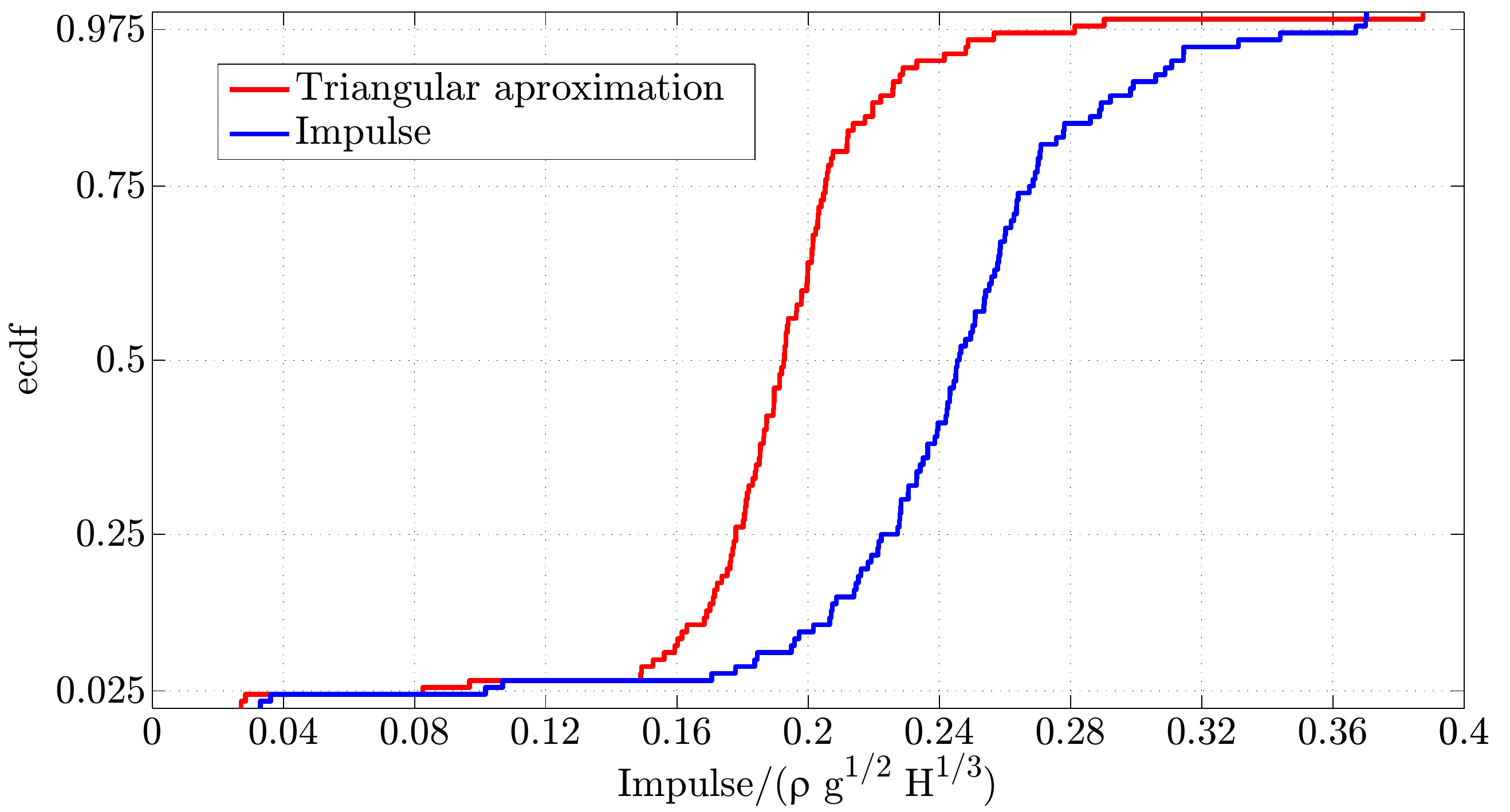}
\caption{$H=300\,{\rm mm}$; the empirical cumulative distribution function (ecdf) of pressure impulse and of triangular approximation for sensor $1$.}
\label{fig:h30_impulse_p1}
\end{figure}

The empirical cumulative distribution functions (ecdfs) of the impulse and of the triangular approximation, both non-dimensionalized with respect to typical time and pressure, are presented in Fig.~\ref{fig:h30_impulse_p1} for the $100$ tests.
The median of the pressure impulse is $5.15\,{\rm mb\cdot s}$ and the median of the triangular approximation is $4.04\,{\rm mb\cdot s}$.  In average, the triangular approximation neglects part of the area below the pressure peak curve causing an error of order of 25\%.

If we set a triangular approximation based on the sum of the typical (median) rise plus decay time at sensor $1$ (Fig.~\ref{fig:H30_ensemble_rise_time sensor1}) and on the typical (median) sensor $1$ peak pressure (Fig.~\ref{fig:H30_ensemble_peak_sensor1}), its value is $(0.003+0.085)\cdot 91.44/2=4.02$ $mb\cdot s$ which is almost identical to the median of all the triangular approximations for each separate test presented above. This suggest that the pressure impulse can be estimated by analyzing impact times and pressure peaks independently while considering a safety margin of the order of 25\%.
\subsubsection{Three-dimensionality}
\label{sss:h30_2d}
\begin{figure}
\centering
\includegraphics[width=0.95\textwidth]{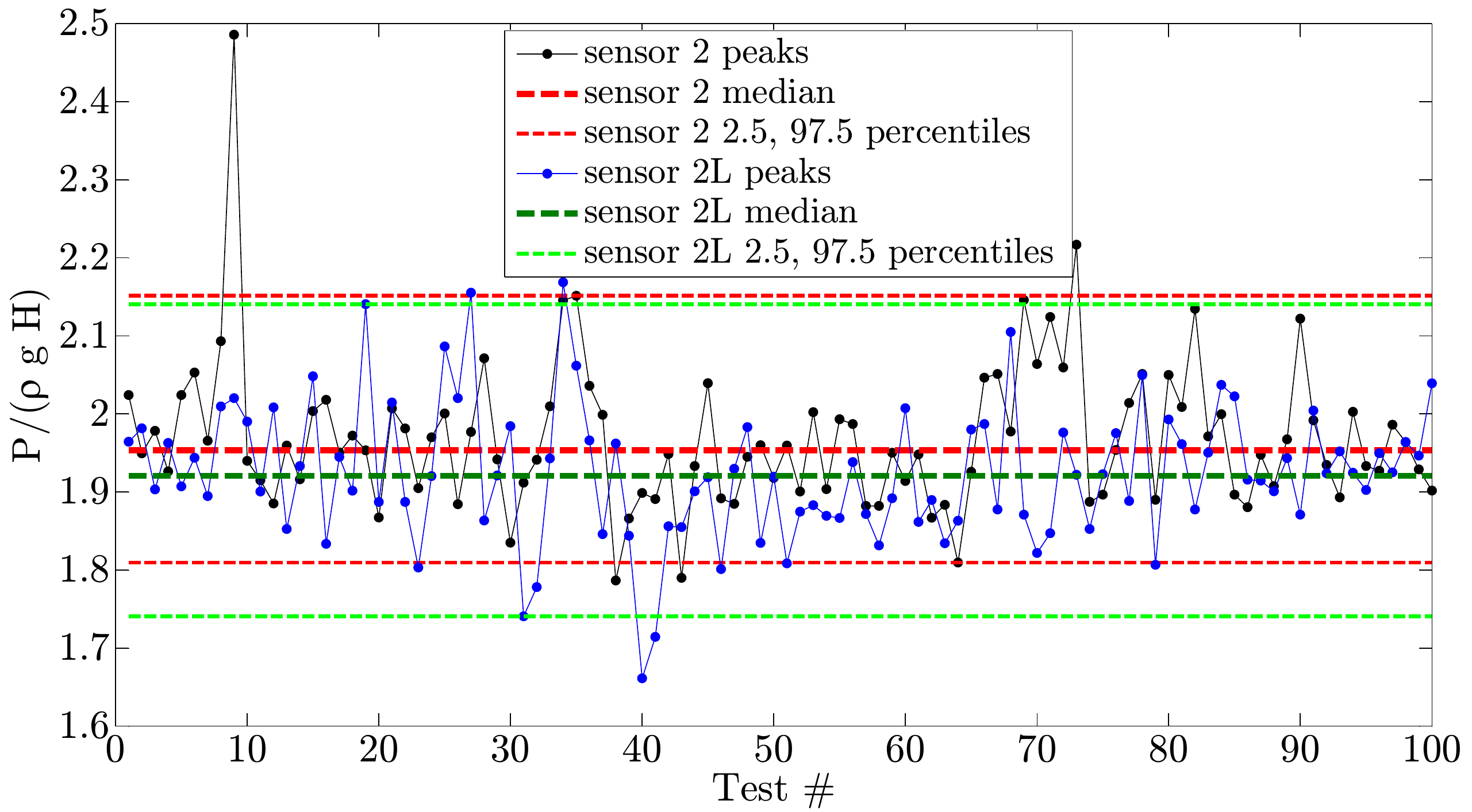}
\caption{$H=300\,{\rm mm}$; peak pressure values at sensors $2$ and $2L$ for 100 tests.}
\label{fig:H30_ensemble_peak_sensors_2_2L}
\end{figure}

\begin{figure}
\centering
\includegraphics[width=0.24\textwidth]{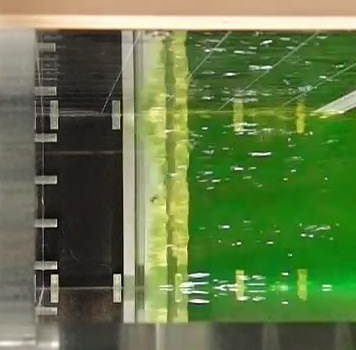}
\includegraphics[width=0.24\textwidth]{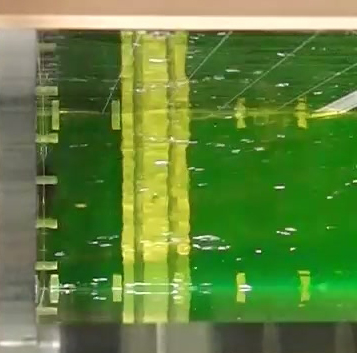}
\includegraphics[width=0.24\textwidth]{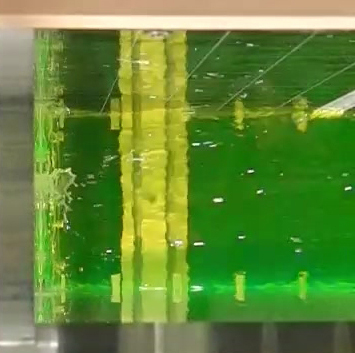}
\includegraphics[width=0.24\textwidth]{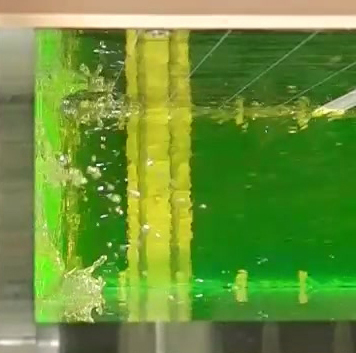}
\caption{$H=300\,{\rm mm}$; free surface during impact and run-up (top view) (476.6, 506.6, 540.0, 603.3, $\pm 3.3$ ms). See supplementary materials for a complete movie.}
\label{fig:H30_fstopviewwall}
\end{figure}

\begin{figure}
\centering
\includegraphics[width=0.65\textwidth]{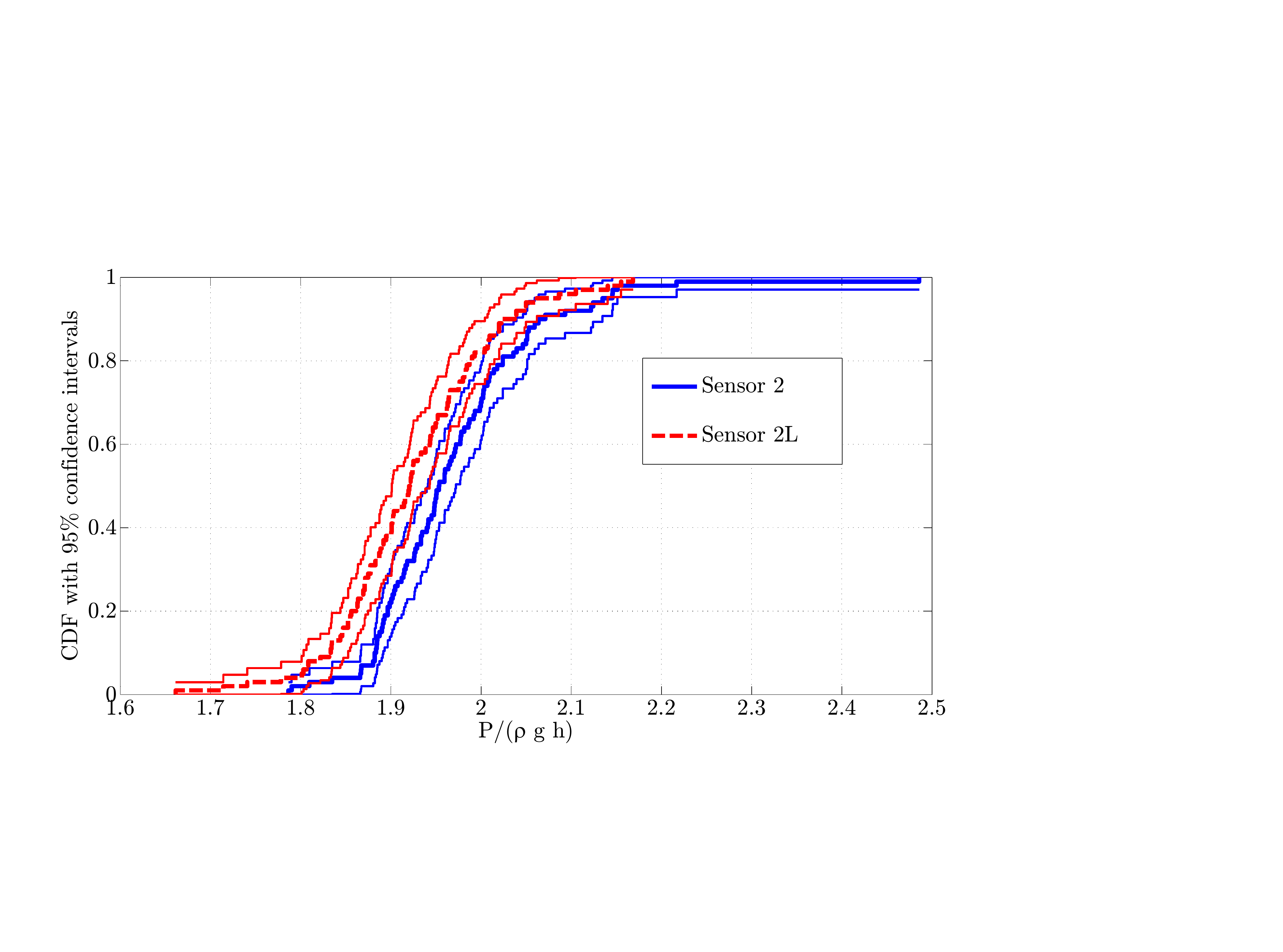}
\caption{$H=300\,{\rm mm}$; Kolmogorov-Smirnoff equal distribution hypothesis contrast test of equal distributions for peak pressures at sensors $2$ and $2L$.}
\label{fig:H30_sensors_2_2L_ks2}
\end{figure}

Sensors number $2$ and number $2L$, which are the two sensors situated at the same height above the horizontal bed in order to check the three-dimensionality of the problem, register pressure signals that are similar but exhibit some differences. The median peak pressure for sensor number $2L$ (the off-centered one) is slightly lower than for sensor $2$ and so is the $2.5\%$ and $97.5\%$ percentile, see Tab.~\ref{tbl:h30stats} and Fig.~\ref{fig:H30_ensemble_peak_sensors_2_2L}.
The maximum recorded peak pressure is significantly larger for the centered sensor than
for the displaced one. From the top view of Figs. \ref{fig:H30_fstopview} and \ref{fig:H30_fstopviewwall}, it seems that the front is aligned with the wall before impact but as the run-up progresses three-dimensional structures are evolved and the run-up is no longer homogeneous.
The hypothesis of these two distributions being the same is rejected at $0.05$ significance level using a
Kolmogorov-Smirnoff equal distribution hypothesis contrast test (see Fig.~\ref{fig:H30_sensors_2_2L_ks2}).
%
\subsubsection{Peaks correlation}
\label{sss:h30_peaks_correlation}
\begin{figure}
\centering
\includegraphics[width=0.49\textwidth]{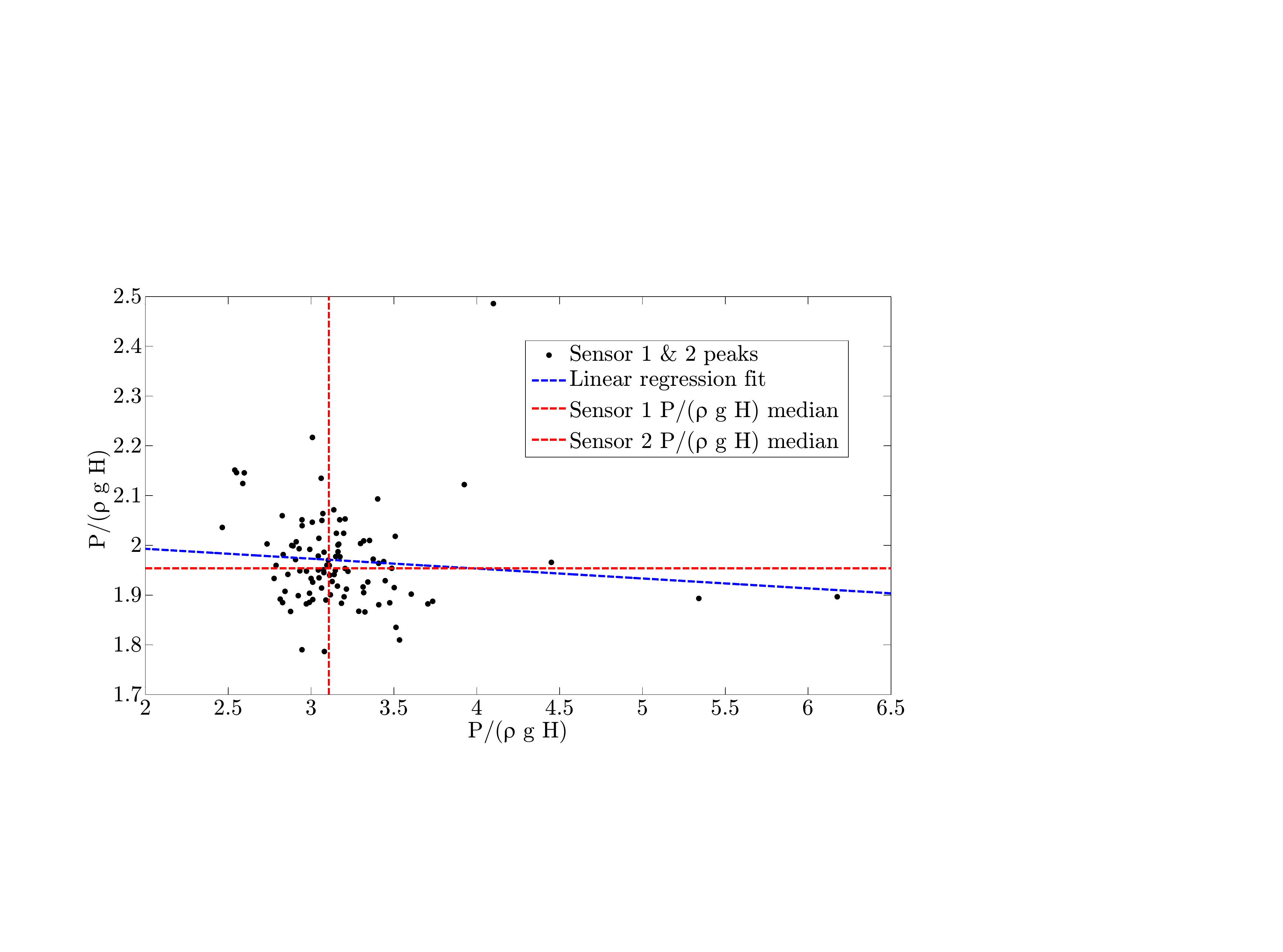}
\includegraphics[width=0.49\textwidth]{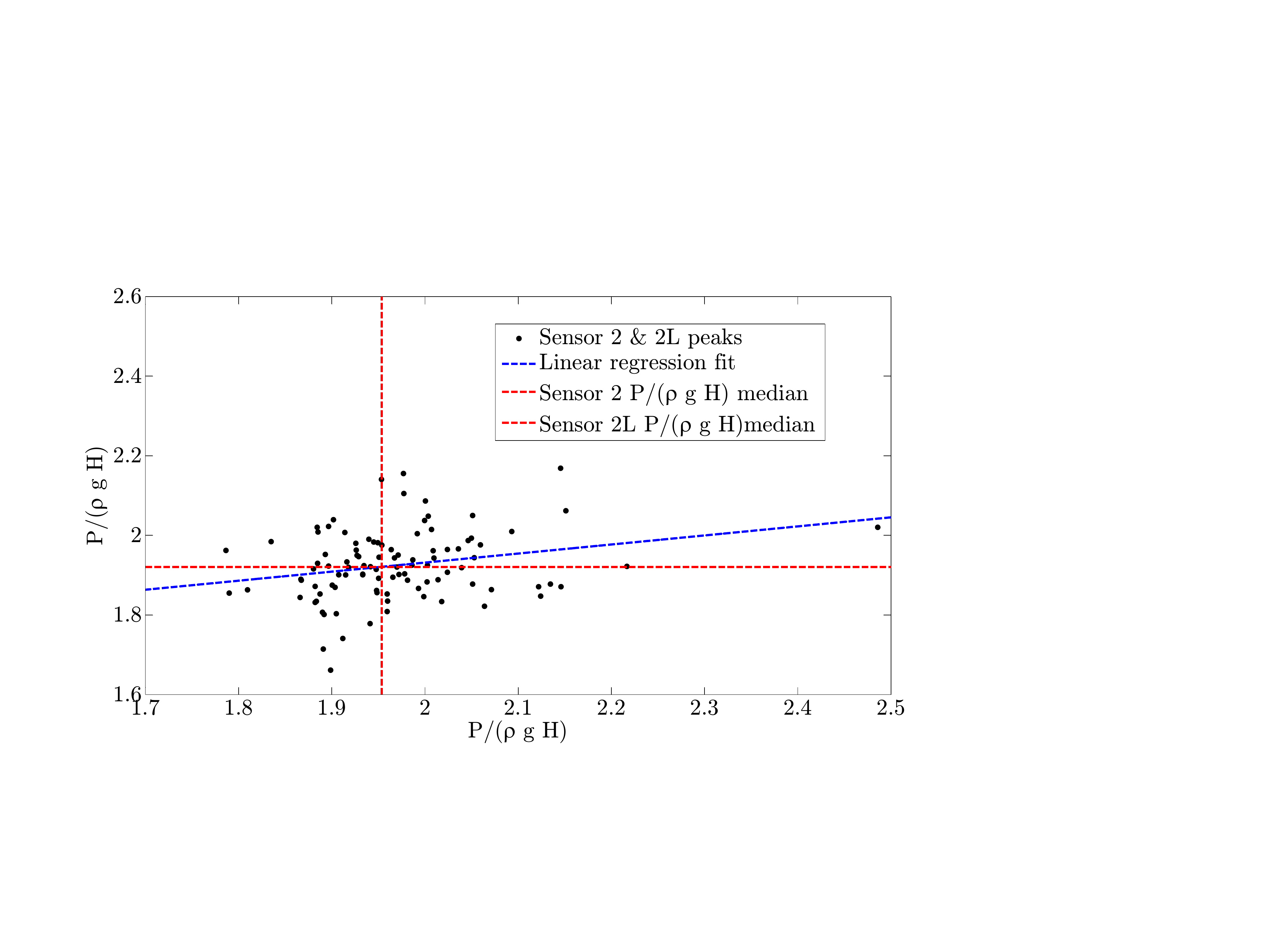}
\caption{$H=300\,{\rm mm}$; correlation between pressure peaks at sensor $1$ and $2$ (left) and between pressure peaks at sensor $2$ and $2L$ (right).}
\label{fig:H30_sensorscorrelation}
\end{figure}

It is relevant to analyze the correlations between the pressure peaks obtained from different
sensors. The linear correlation factor for all cases is presented in Tab.~\ref{tbl_h30_correlations}.
The correlation between the recorded pressure peaks is not very strong
which is contradictory to expectation that more energetic impacts would show consistently in all peaks.
This lack of correlation suggests that the pressure peaks are extremely localized.
Data for sensor $1$ vs sensor $2$, and sensor $2$ vs sensor $2L$ are presented in figure
\ref{fig:H30_sensorscorrelation}. The correlation between sensor $1$ and sensor $2$ is small
and they are in principle slightly anti-correlated.
But as can be appreciated in Tab.~\ref{tbl_h30_correlations}, the correlation grows when comparing sensor $1$ to sensor $3$ and $4$.
This trend would need further analysis but it represents a relevant piece of information. On the other hand, the correlation between sensor $2$ and sensor $2L$ is the largest found in the table. All this may indicate that variations in
the vertical direction
are more intricate than in the transversal direction,
which was an aim the experimental setup was built with.
\begin{table*}[ht]
  \centering
\begin{tabular}{|c|c|c|c|c|c|}
  \hline
Sensor	&	1	&	2	&	2L	&	3	&	4	\\ \hline
1	&	---	&	-0.0707	&	0.0347	&	0.1439	&	0.2056	\\ \hline
2	&	s	&	---	&	0.2515	&	0.1833	&	0.012	\\ \hline
2L	&	s	&	s	&	---	&	0.2052	&	0.0937	\\ \hline
3	&	s	&	s	&	s	&	---	&	0.1362	\\ \hline
4	&	s	&	s	&	s	&	s	&	---	\\ \hline
\end{tabular}
   \caption{Correlation between pressure peaks at all five sensors for $H=300\;mm$ filling height
   (\emph{s} stands for symmetric).}
  \label{tbl_h30_correlations}
\end{table*}
%
\subsubsection{Comparisons with published data}
A series of comparisons with existing data from the literature can be now undertaken.

\begin{enumerate}
\item Zhou et al. \cite{lee_zhou_cao_2002_jfe_dambreak}\cite{Zhou_1999} published a series of results for a dam break experiment.
The present setup for $H=300\,{\rm mm}$ is a half scale setup of Zhou et al. who conducted measurements in three different positions at the impacting wall, the lowest one being equivalent to sensor $4$ of the present analysis (see Fig.~\ref{fig:dam_scheme}).
The force transducers of Zhou et al. were much larger than the ones used in this study. Zhou used $90\,{\rm mm}$ diameter transducer while the present experimental campaign is performed with $4.2\,{\rm mm}$ diameter transducer, see section \ref{sss:presssensors} for details. Therefore it is necessary to be cautious in the comparisons presented in Fig.~\ref{fig:lee_p2_compared_graph}. A series of comments comparing these data follow:
\begin{enumerate}
\item The order of magnitude of the registered pressure is similar ($P^*\sim 0.8$)
\item The matching in the occurrence time ($t^* \sim 2.8$) is good.
\item The pressure in their experimental register rises rapidly in time to a larger value than any of the registers reported here (approximately $P^*\sim 0.7$ at $t^*\sim 3.0$). However only a single measurement is documented in \cite{lee_zhou_cao_2002_jfe_dambreak}\cite{Zhou_1999}.
Furthermore, the sensor $4$ does not register a pure impact event in our campaign but in their case, the steepness of the curve is larger. This could be related to the fact that lower edge of the $90\,{\, mm}$ wide sensor is proportionally lower than the edge of sensor $4$. Thus it may be loaded by the more energetic part of the flow and register a significantly larger impact.
\item After the initial impact peak, Zhou's register drops to and lays close to the $2.5\%$ percentile of the present experimental campaign registers. However, Zhou's sensor averages the pressure over a larger area in which the pressure decays with increasing elevation. Since only a single measurement of Zhou et al. is available, this agreement is satisfactory.
\item In both studies, a mild rise in pressure is recorded at time $t^* \sim 6$ which about corresponds to the moment when the plunging breaker hits the underlying fluid.
\item Overall comparison of the two studies is quite consistent.
\end{enumerate}
The experimental data of Zhou et al. \cite{lee_zhou_cao_2002_jfe_dambreak}\cite{Zhou_1999} have been used for validation of numerical codes in a series of papers. Colagrossi et al.\cite{colagrossi2003} and Asai et al. \cite{asai_etal_2012_dambreak_isph} reported a good agreement for a Smoothed Particle Hydrodynamics (SPH) solver and Park et al. \cite{Park2012176} and Abdolmaleki et al. \cite{Abdolmaleki_etal_dambreak2004} provided similar results for RANS solutions. However, they do not register the initial peak in sensor $4$, which is consistent with the present experimental campaign.
\begin{figure}
\centering
\includegraphics[width=0.85\textwidth]{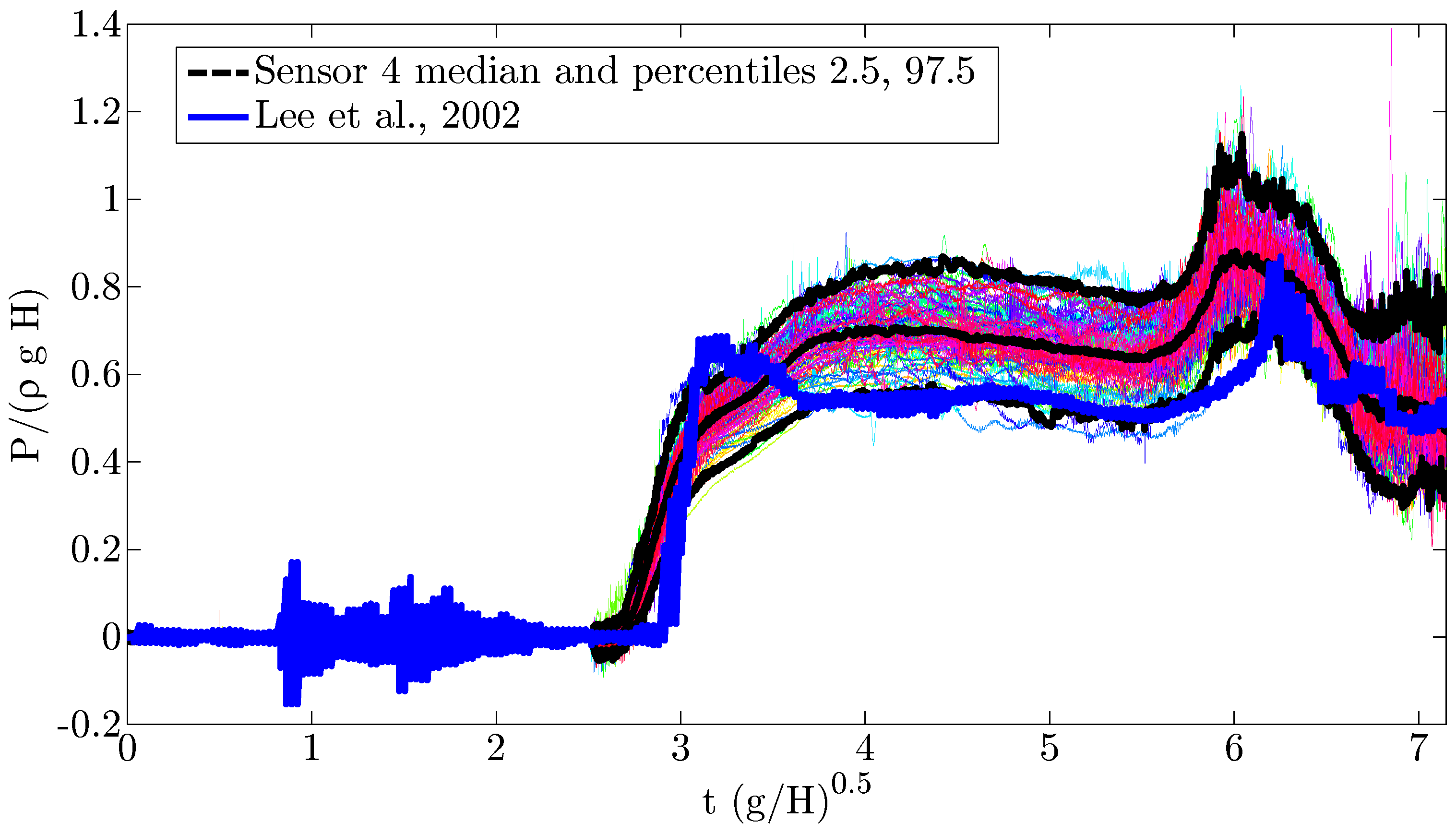}
\caption{$H=300\,{\rm mm}$; Sensor 4 pressure time history. Comparison with \cite{lee_zhou_cao_2002_jfe_dambreak}}
\label{fig:lee_p2_compared_graph}
\end{figure}
\item Wemmenhove et al. \cite{Wemmenhove_2010} performed measurements with the same experimental setup as Zhou et al. \cite{lee_zhou_cao_2002_jfe_dambreak,Zhou_1999} but using a different sensor type at different position.
Two recorded time histories of pressure are available. The sensor position corresponds to location of sensor $2$ in this paper, but the sensor diameter is not documented. A comparison with present data is shown in figure \ref{fig:Wemmenhove_sensor2_compared_graph}.
A series of comments comparing these data follow:
\begin{enumerate}
\item Wemmenhove et al.\cite{Wemmenhove_2010} pressure peak values are in fair agreement with the median pressure at sensor $2$ of the present experimental campaign.
\item There are significant differences in the pressure registers after the peak. Unfortunately, there is no explanation for this behavior.
\item For one of the measurements the occurrence time falls within the range of the present campaign but that is not the case for the other one. The dispersion in the occurrence time is difficult to explain according to the present campaign.
\item Some unexpected negative values are measured just prior to the peak in the present campaign. They are related to spurious electronic wetting effects of the sensor membrane.
\end{enumerate}
\begin{figure}
\centering
\includegraphics[width=0.85\textwidth]{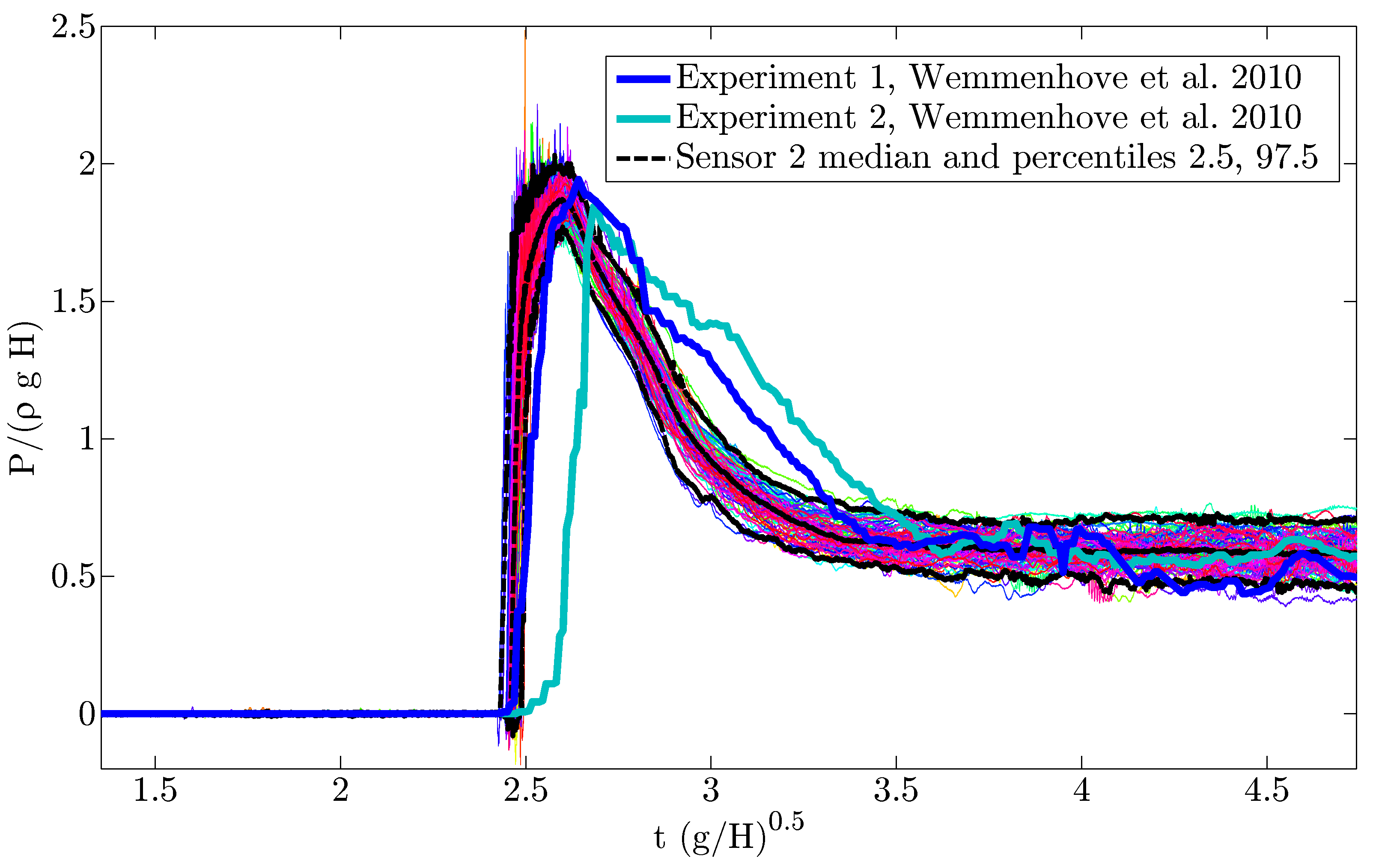}
\caption{$H=300\,{\rm mm}$; Sensor 2 pressure time history. Comparison with \cite{Wemmenhove_2010}}
\label{fig:Wemmenhove_sensor2_compared_graph}
\end{figure}
\item Kleefsman et al. \cite{kleefsman_etal_jcp_2005} performed measurements with the same experimental setup like Zhou et al. \cite{lee_zhou_cao_2002_jfe_dambreak}\cite{Zhou_1999} but considering a cuboidal obstacle placed on the horizontal bed downstream the dam gate. Its position is upstream of the lateral vertical wall, but there is a pressure positioned at the same height as sensor $3$ of the current study. Thus, some qualitative comparisons can be performed, see figure \ref{fig:kleefsman_sensor3_compared_graph}.
  However, a description of the sensor type is missing in \cite{kleefsman_etal_jcp_2005}.
  A series of comments comparing these data follow:
\begin{enumerate}
\item The magnitude of the peak pressure recorded by Kleefsman et al. \cite{kleefsman_etal_jcp_2005} is similar to the median of peak pressure at sensor $3$ in this study.
\item Occurrence time of the pressure at the obstacle in \cite{kleefsman_etal_jcp_2005} is smaller since the obstacle is placed closer to the dam gate than the vertical wall at the end of the flume. If the occurence time at an obstacle is predicted using the current data from sensor $3$, the difference in the predicted value and in occurence time recorded by Kleefsman is about $6\%$.
\item The difference in the curve profile after the impact may be explained by the fact, that the obstacle in \cite{kleefsman_etal_jcp_2005} does not fill the entire width of the downstream flume and a complex three-dimensional flow runs over and around the obstacle towards the downstream end of the flume.
\end{enumerate}
\begin{figure}
\centering
\includegraphics[width=0.85\textwidth]{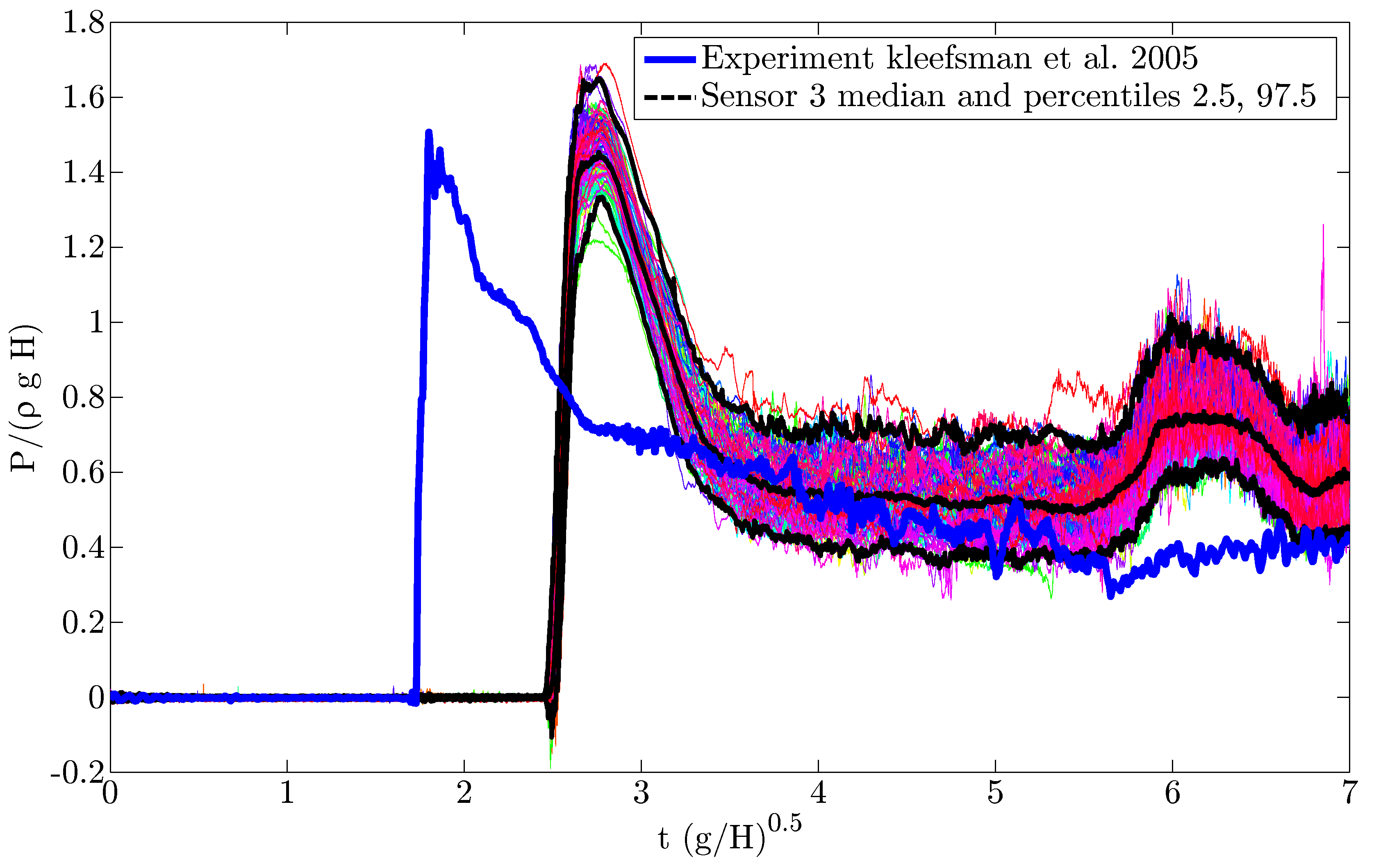}
\caption{$H=300\,{\rm mm}$; Sensor 3 pressure time history. Comparison with \cite{kleefsman_etal_jcp_2005}}
\label{fig:kleefsman_sensor3_compared_graph}
\end{figure}
\end{enumerate}

\subsection{Filling height $600\,{\rm mm}$}
The impact pressure for $600\,{\rm mm}$ was measured with the same setup of five pressure sensors at the vertical wall downstream the dam gate as in the previous sections (Fig.~\ref{fig:dam_scheme}).
The statistical analysis of the pressure peaks, rise times and the occurrence time is carried out for data from $100$ test runs.

\begin{figure}
\centering
\includegraphics[width=0.9\textwidth]{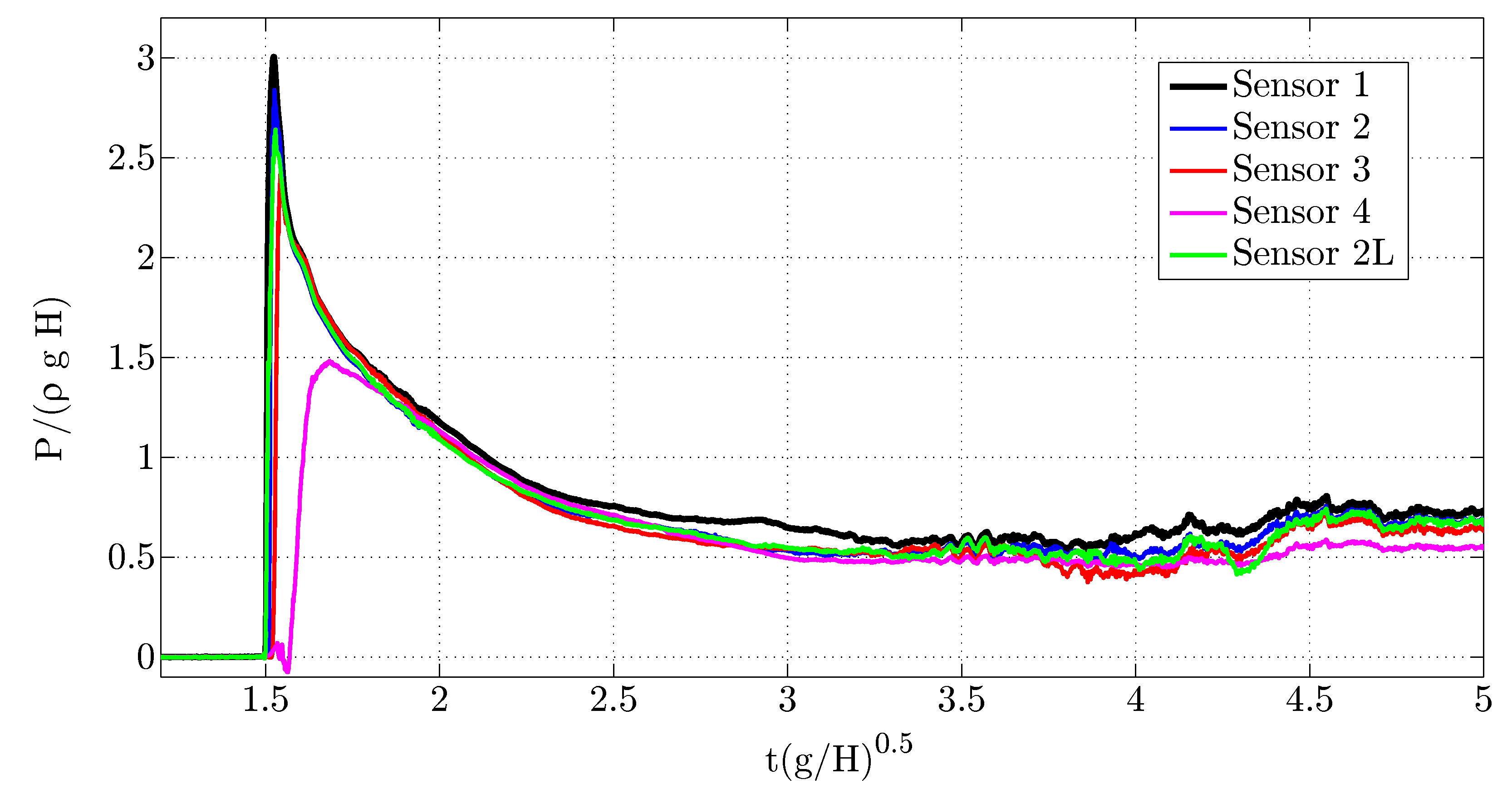}
\caption{$H=600\,{\rm mm}$;  typical impact event pressure signals from all five pressure sensors.}
\label{fig:h60_peak_event_5_sensors}
\end{figure}

\begin{figure}
\centering
\includegraphics[width=0.7\textwidth]{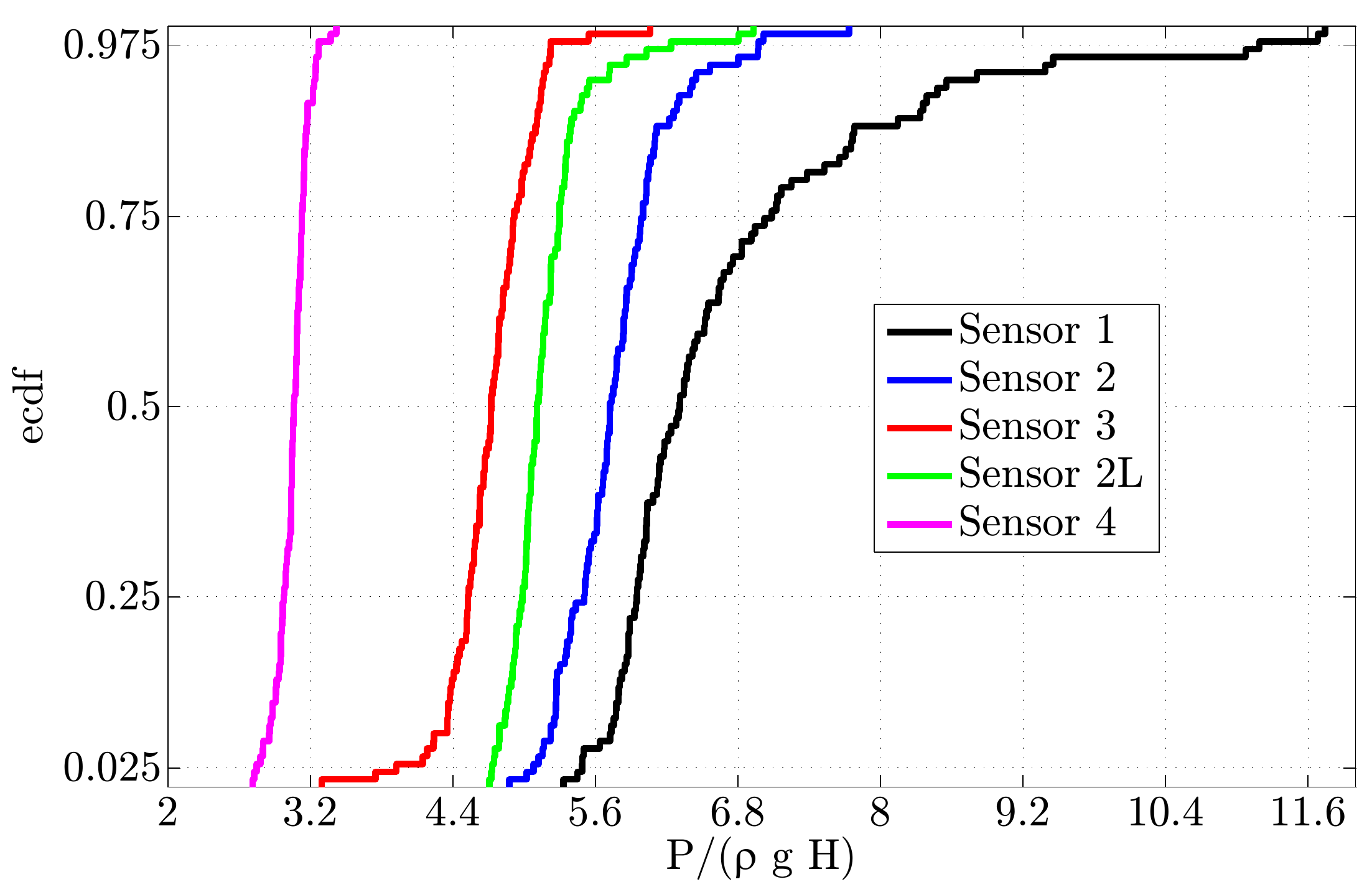}
\caption{$H=600\,{\rm mm}$; empirical cumulative distribution function (ecdf) of peak pressure for all five pressure sensors.}
\label{fig:H60_ensemble_peak_sensor1}
\end{figure}

A typical dam break impact event signal as registered by the $5$ sensors can be seen in Fig.~\ref{fig:h60_peak_event_5_sensors}. As expected, the highest pressure peak is recorded by sensor $1$, but the registered values for sensors $1-3$ lay significantly closer to each other than in the $H=300\,{\rm mm}$ case presented in Fig.~\ref{fig:peak_event_5_sensors}.
The empirical cumulative distribution functions (ecdf) of the impact pressures for all five sensors are presented in Fig.~\ref{fig:H60_ensemble_peak_sensor1}.
As for $H=300\,{\rm mm}$, the width of the confidence interval of pressure peaks recorded by sensor $1$ is larger than for the other sensors, where the impact is not so intense.
A summary of these values can be seen in Tab.~\ref{tbl:h60stats}.
The medians for sensors $1-3$ and their distribution are consistent with data from the representative test presented in Fig.~\ref{fig:h60_peak_event_5_sensors}.

Similarly to section \ref{sss:h30_pressurepeaksanalysis}, the correlation of the peak pressures at sensor $1$ with the gate velocity has been analyzed and no significant correlation has been found.
A thorough comparison of results for $300\,{\rm mm}$ and $600\,{\rm mm}$ filling height tests is presented in section \ref{ss:30_60}.

\begin{table*}[ht]
  \centering
  \begin{tabular}{|c|c|c|c|c|}
    \hline
Sensor	&	z [mm]	&	Median [mb]	&	97.5 Percentile	&	2.5 Percentile	\\ \hline
1	&	3	&	185.7	&	343.8	&	161.5	\\ \hline
2	&	15	&	168.4	&	205.2	&	149.4	\\ \hline
3	&	30	&	138.9	&	153.7	&	115.4	\\ \hline
4	&	80	&	90.09	&	96.1	&	80.8	\\ \hline
2L	&	15	&	150.2	&	183.5	&	138.9	\\ \hline
  \end{tabular}
  \caption{$H=600\,{\rm mm}$; summary of statistical data as recorded by pressure sensors for $100$ tests.}
  \label{tbl:h60stats}
\end{table*}
%
\subsubsection{Pressure time history analysis}
\begin{figure}
\centering
\includegraphics[width=0.9\textwidth]{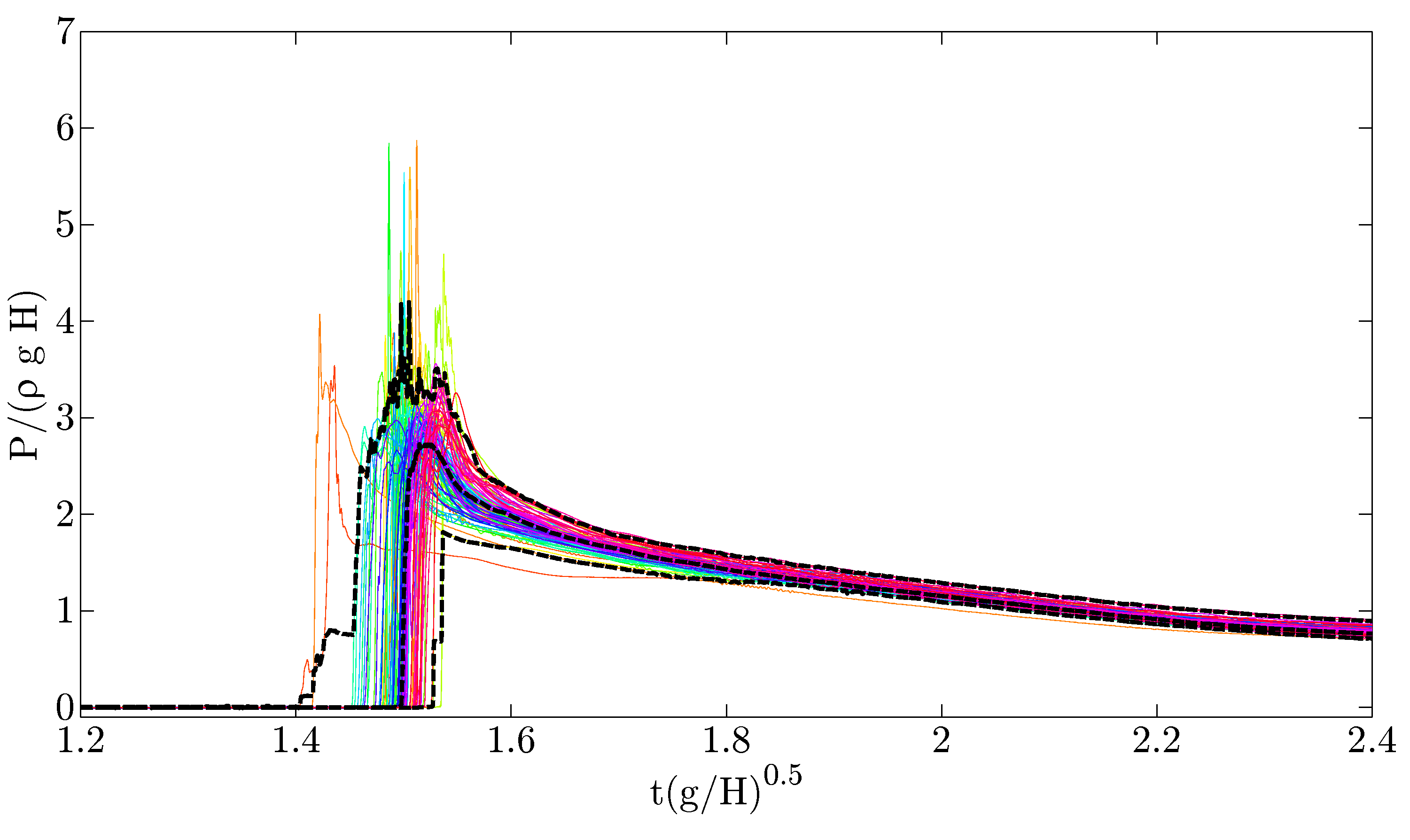}
\caption{$H=600\,{\rm mm}$; sensor 1 pressure time histories for 100 tests. The central black mark indicates the median, the lower and upper additional black markers represent the estimated 2.5\% and 97.5\% percentile levels respectively.}
\label{fig:h60_sensor1_pressure_th}
\end{figure}

In Fig.~\ref{fig:h60_sensor1_pressure_th}, the time histories of all $100$ tests for sensor $1$ are presented along with their median and the 2.5\% and 97.5\% percentile levels. For every instant a median and the corresponding confidence intervals are obtained.
There are some differences when compared with the time histories presented for $H=300\,{\rm mm}$ case in Fig.~\ref{fig:h30_sensor1_pressure_th}, but the overall behavior and major trends remain similar.
%
\subsubsection{Rise and decay times analysis}
\begin{figure}
\centering
\includegraphics[width=0.7\textwidth]{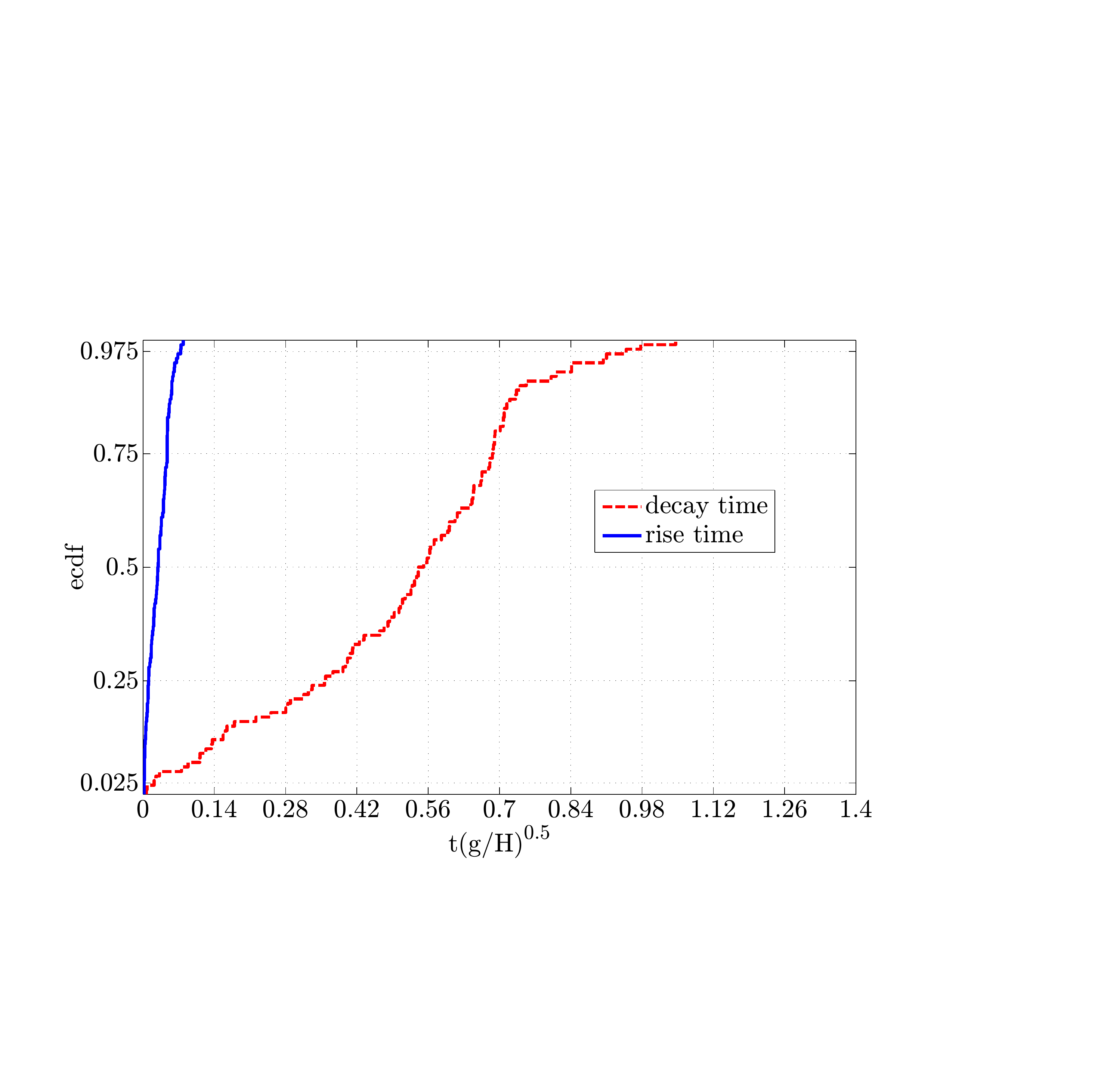}
\caption{$H=600\,{\rm mm}$; empirical cumulative distribution function (ecdf) of rise and decay times as registered at sensor $1$.}
\label{fig:H60_ensemble_rise_time sensor1}
\end{figure}

An analysis of the rise time has been carried out analogously to the one performed in section \ref{sss:risetime30}.
In figure  \ref{fig:H60_ensemble_rise_time sensor1}, the empirical cumulative distribution functions of the rise time and the decay time for the pressure time history in sensor $1$ for the $100$ tests are displayed.
The rise time ranges from $0.5\,{\rm mm}$ to $6\,{\rm mm}$ and similarly to $H=300\,{\rm mm}$ filling height case the decay time is an order of magnitude larger than the rise time.

\begin{figure}
\centering
\includegraphics[width=0.7\textwidth]{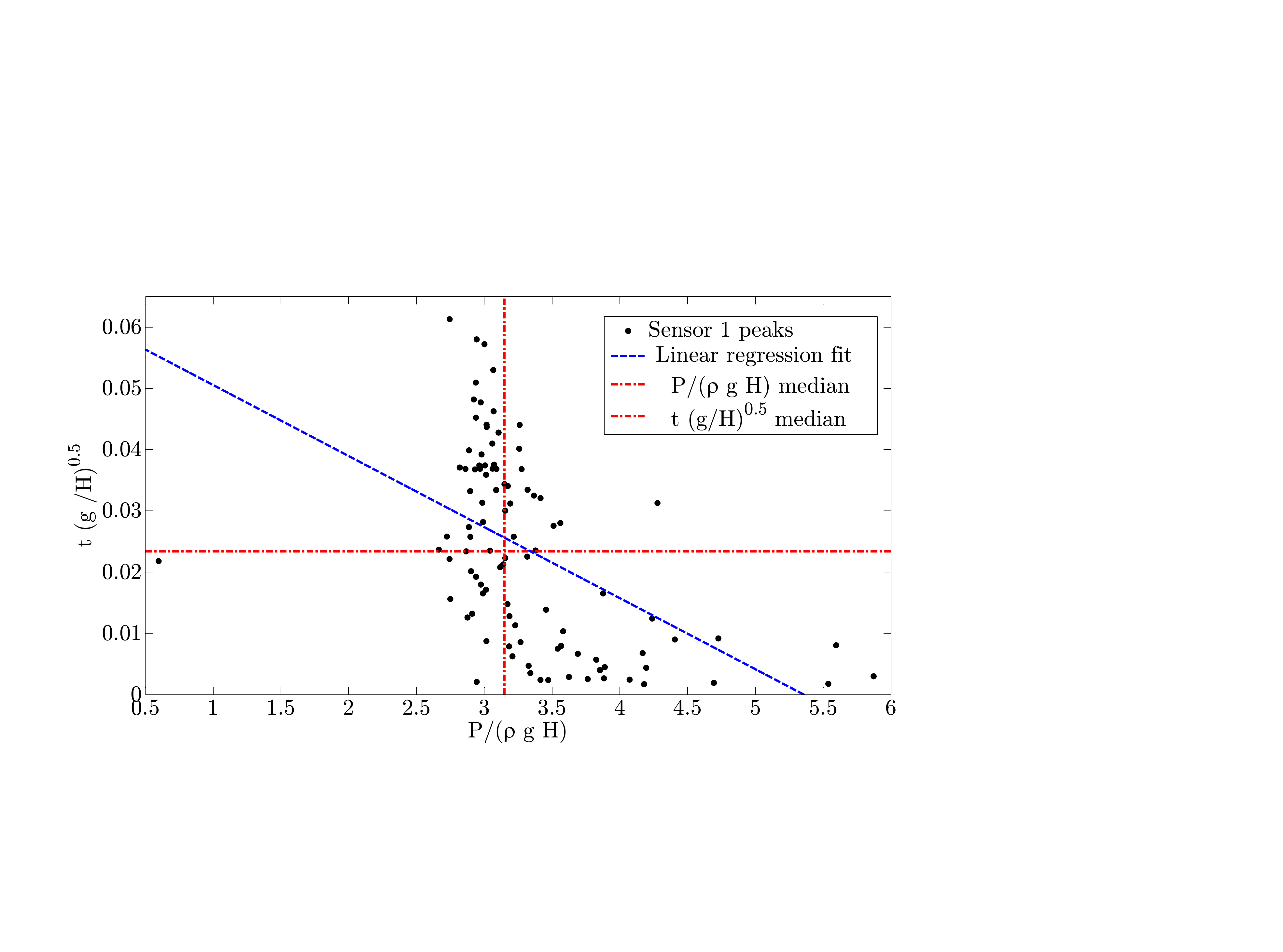}
\caption{$H=600\,{\rm mm}$; joint distribution of rise time vs pressure peak as registered at sensor $1$.}
\label{fig:h60_risetime_p1_joint}
\end{figure}

It is again interesting to observe both the peak pressure and the rise time as a realization of the same random phenomenon and a joint distribution of the rise times and the peak pressure values is shown in Fig.~\ref{fig:h60_risetime_p1_joint}.
It can be appreciated that the rise time and the pressure peaks present a negative correlation. The linear correlation factor is equal to $-0.376$ which closely resembles the correlation found for $H=300\,{\rm mm}$ case presented in section \ref{sss:risetime30}.
%
\subsubsection{Impulse}
\begin{figure}
\centering
\includegraphics[width=0.7\textwidth]{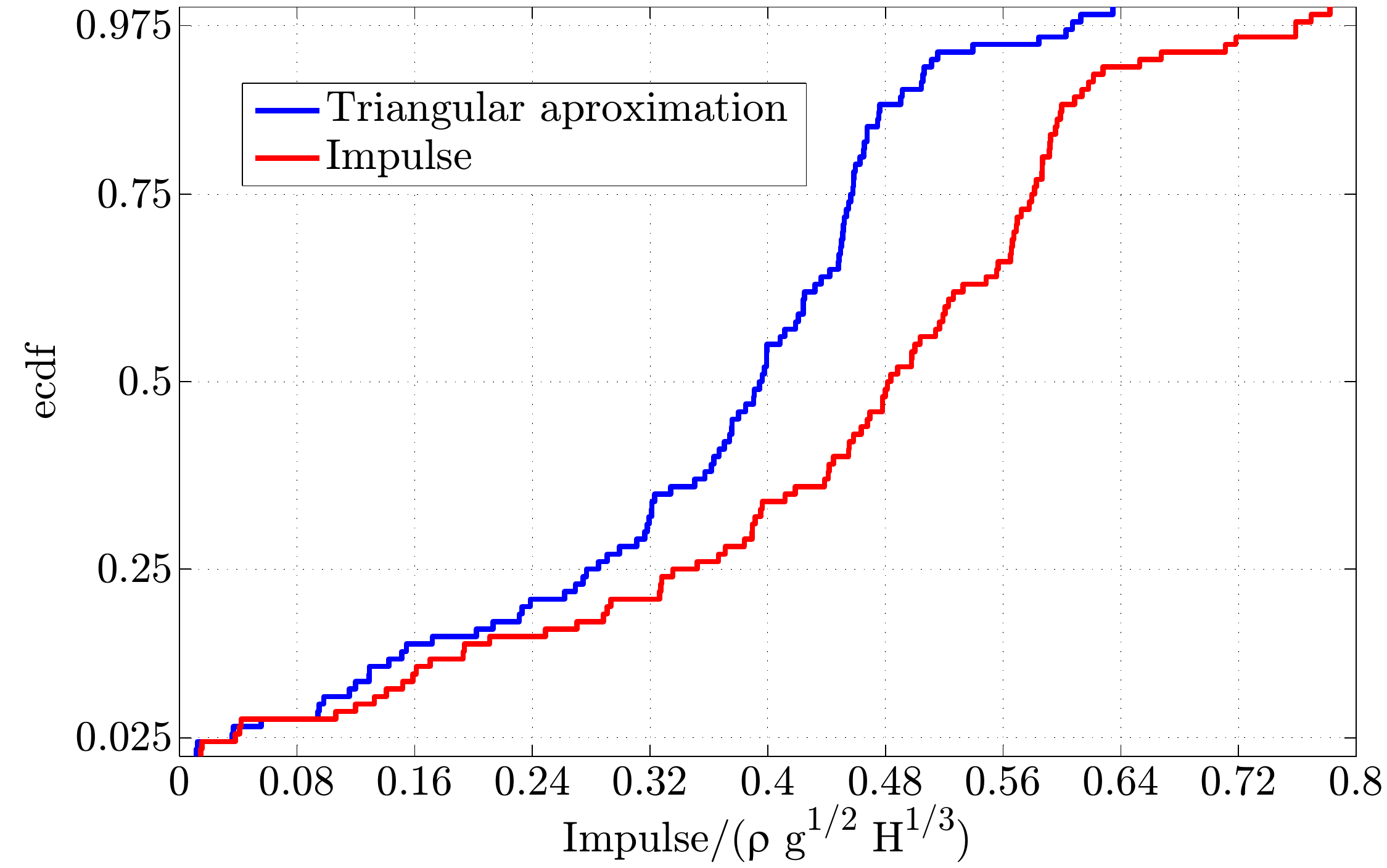}
\caption{$H=600\,{\rm mm}$;  the empirical cumulative distribution function (ecdf) of pressure impulse and of triangular approximation for sensor $1$.}
\label{fig:h60_impulse_p1}
\end{figure}

The impact events for $H=600\,{\rm mm}$ are also analyzed from the perspective of the pressure impulse.
The empirical cumulative distribution functions of the pressure impulse and its triangular approximation for the $100$ tests are presented in Fig.~\ref{fig:h60_impulse_p1}.
The spread of the distribution is larger than for $H=300\,{\rm mm}$ case which is coherent with the larger variability found for the other statistics for this case.

The median of the pressure impulse is $12.74\,{\rm mb\,s}$ and the median of the triangular approximation is $10.44\,{\rm mb\,s}$.
As for $H=300\,{\rm mm}$, if we set a triangular approximation based on the sum of the typical (median) rise plus decay time at sensor $1$ (Fig.~\ref{fig:H60_ensemble_rise_time sensor1}) and on the typical (median) sensor $1$ peak pressure (Fig.~\ref{fig:H60_ensemble_peak_sensor1}), its value is $(0.007+0.104)\cdot 185.69 /2=10.30$ $mb\cdot s$ which is a close value to the median of all the triangular approximations for each separate test.
This statistical consistency is very interesting considering that impact loads are usually modeled for structural finite element computations as an equivalent triangular load whose characteristics are defined by the averages of the peak pressure and impact time \citep{Graczyk_Moan_Wu_saos2007}.
\subsubsection{Three-dimensionality}
\label{h60_sss_2d}
\begin{figure}
\centering
\includegraphics[width=0.9\textwidth]{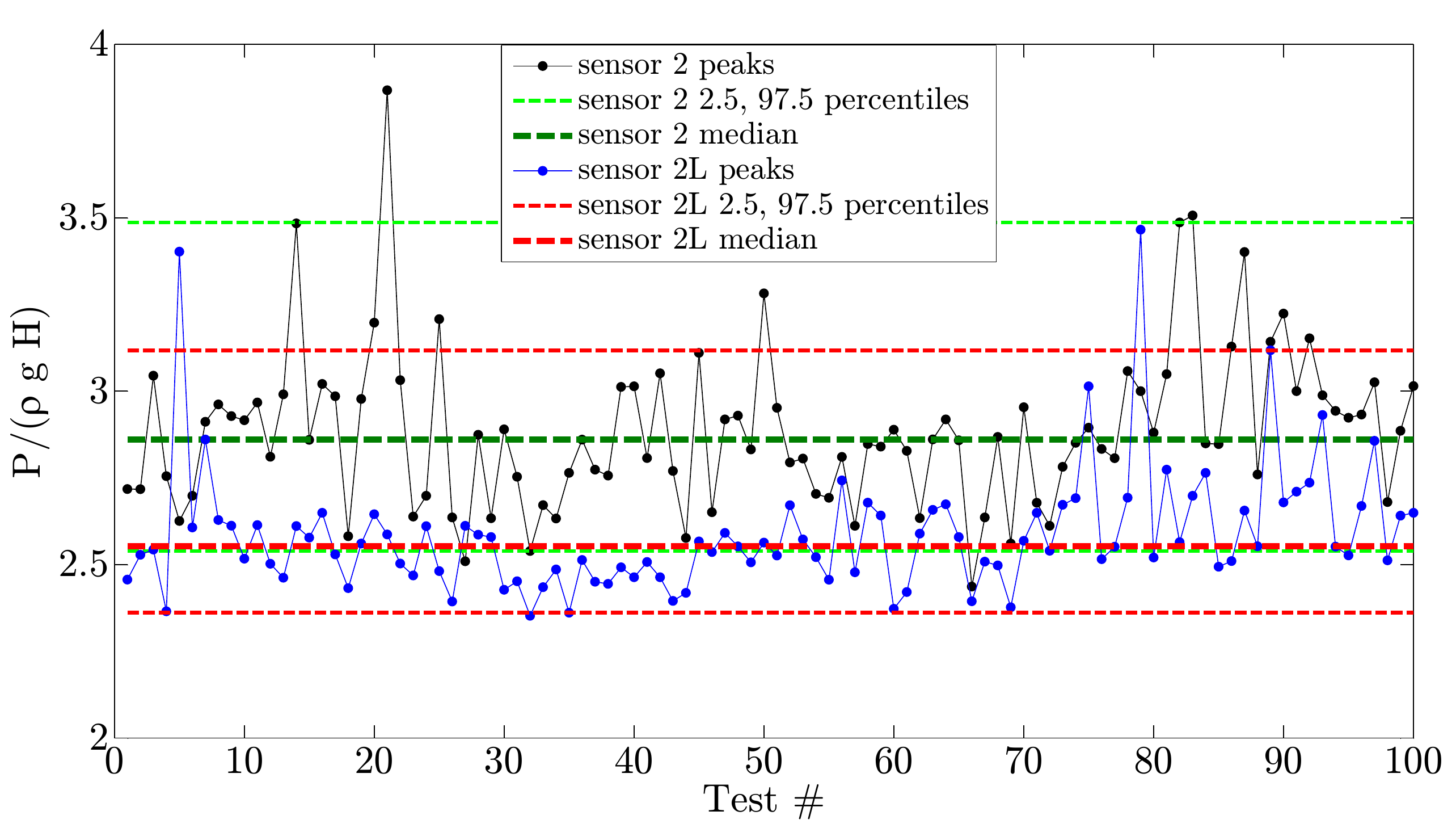}
\caption{$H=600\,{\rm mm}$; peak pressure values at sensors $2$ and $2L$ for 100 tests.}
\label{fig:H60_ensemble_peak_sensors_2_2L}
\end{figure}

\begin{figure}
\centering
\includegraphics[width=0.65\textwidth]{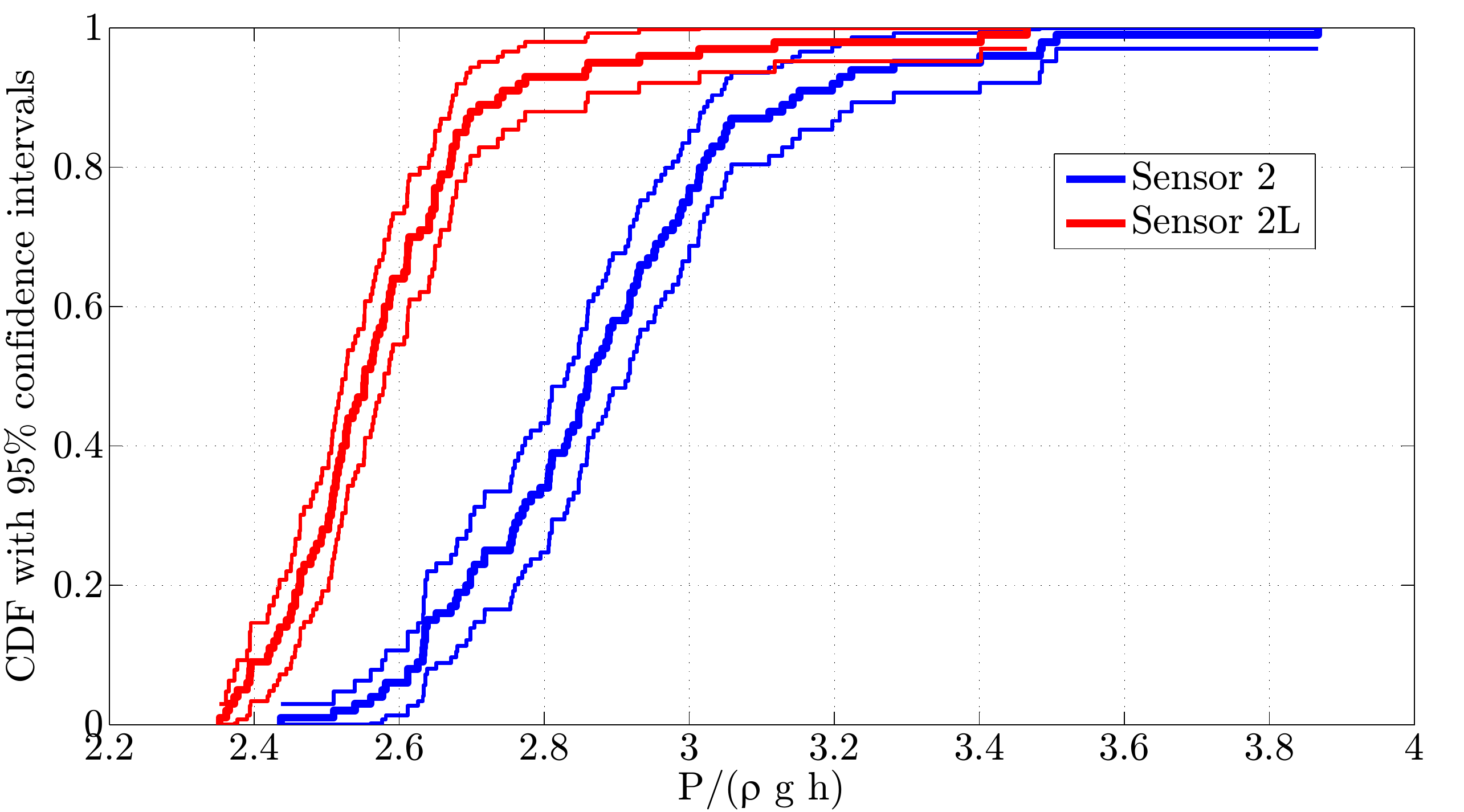}
\caption{$H=600\,{\rm mm}$; Kolmogorov-Smirnoff equal distribution hypothesis contrast test of equal distributions for peak pressures at sensors $2$ and $2L$.}
\label{fig:H60_sensors_2_2L_ks2}
\end{figure}

The three-dimensionality of the problem is checked using the registers of sensors $2$ and $2L$. The recorded pressure curves at these two sensors are close to each other but exhibit several differences. The median pressure for sensor number $2L$ is around $10\%$ lower than the median pressure at sensor $2$ (Tab.~\ref{tbl:h60stats}). The same applies when comparing values of the $2.5\%$ percentile (Fig.~\ref{fig:H60_ensemble_peak_sensors_2_2L}). The maximum pressure peak is significantly larger for the centered sensor than for the displaced one. Based on $0.05$ significance level of Kolmogorov-Smirnoff equal distribution hypothesis contrast test, the distributions for sensor $2$ and $2L$ cannot be consider equal, see Fig.~\ref{fig:H60_sensors_2_2L_ks2}. The differences are larger than for the case $H=300\,{\rm mm}$ discussed in section \ref{sss:h30_2d}.

Looking at the flow details, from the top view of Fig.~\ref{fig:H60_tongue_top_view_before_impact}, it seems the front is initially perturbed but then becomes aligned with the wall before the impact.
The increase in difference between the statistics of the sensors $2$ and $2L$ for  $H=600\,{\rm mm}$, compared to  $H=300\,{\rm mm}$ could be explained by the increase in Reynolds number between both cases and the correspondent increase in turbulence intensity as the run-up takes place on the downstream wall (see supplementary materials corresponding to figures \ref{fig:H30_fstopview}, \ref{fig:H60_tongue_top_view_before_impact}).
%
\subsubsection{Peaks correlation}
\label{sss:h30_peaks_correlation}
Analysis of the correlations between the pressure peaks obtained from the different sensors has been carried out. The linear correlation factor for all cases is presented in Tab.~\ref{tbl_h60_correlations}.
The correlation between sensor $1$ and sensor $2$ is much stronger
than for $H=300\,{\rm mm}$ and this positive correlation remains also for other sensors when compared to sensor $1$. On the other hand, the correlation between sensor $2$ and sensor $2L$
is still positive as for $H=300\,{\rm mm}$ but in the case of $H=600\,{\rm mm}$ its value is smaller than for the correlation of sensors located vertically above each other. This supports the idea that variations in the transversal direction are larger for this filling height, as pointed in section \ref{h60_sss_2d}.

\begin{table*}[ht]
  \centering
\begin{tabular}{|c|c|c|c|c|c|}
  \hline
Sensor	&	1	&	2	&	2L	&	3	&	4	\\ \hline
1	&	---	&	0.2581	&	0.0539	&	0.1185	&	0.1387	\\ \hline
2	&	s	&	---	&	0.1418	&	0.3555	&	0.1856	\\ \hline
2L	&	s	&	s	&	---	&	0.0828	&	0.0567	\\ \hline
3	&	s	&	s	&	s	&	---	&	0.4275	\\ \hline
4	&	s	&	s	&	s	&	s	&	---	\\ \hline
  \hline
\end{tabular}
  \caption{Correlation between pressure peaks for $H=600\;mm$, \emph{s} stands for symmetric}
  \label{tbl_h60_correlations}
\end{table*}

\subsection{Discussion on 300mm and 600mm experiments}
\label{ss:30_60}
\begin{figure}
\centering
\includegraphics[width=0.9\textwidth]{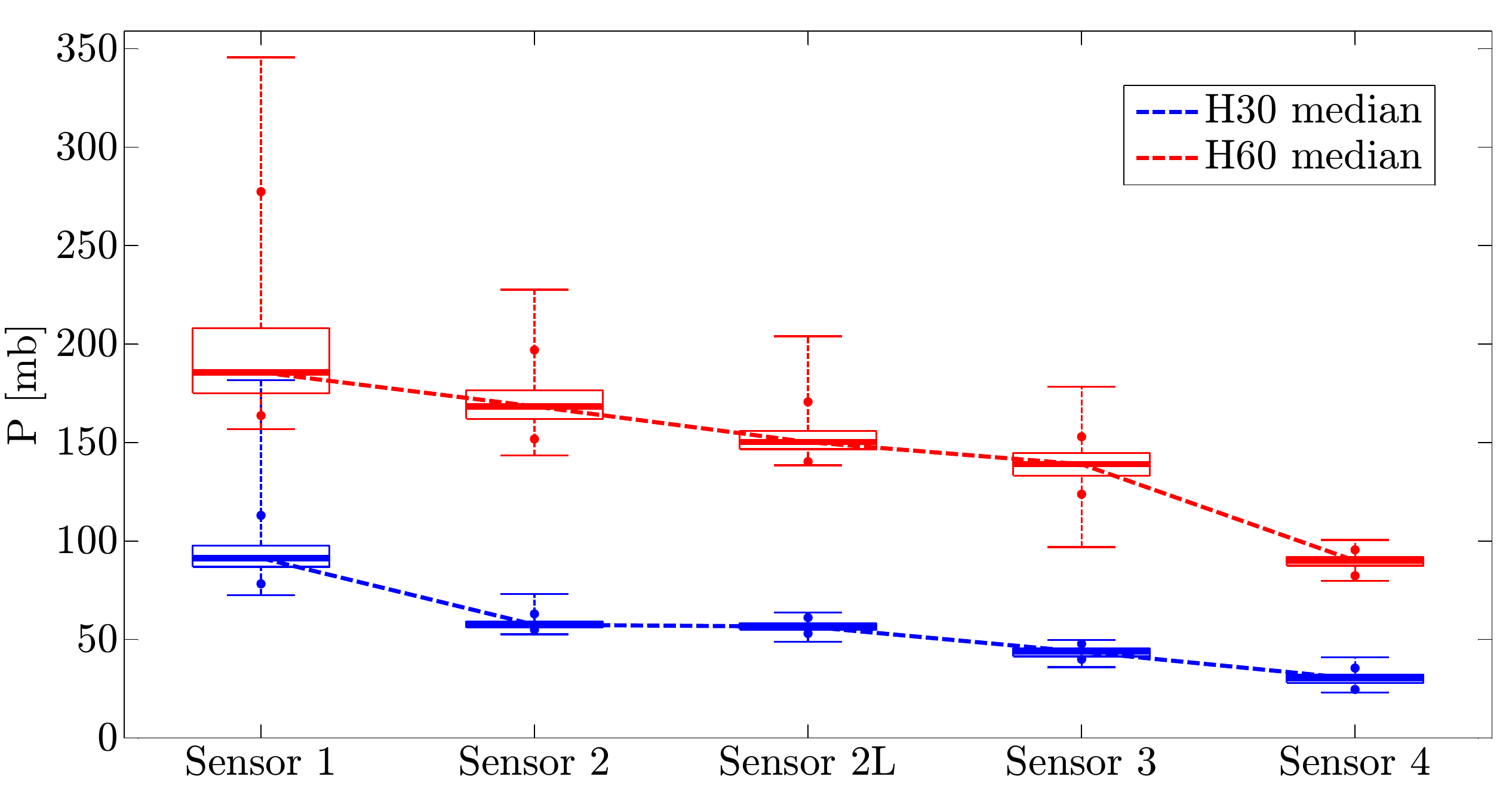}
\caption{H=300mm, H=600mm: comparison of median and 95\% confidence interval for peak pressure and all five sensors.}
\label{fig:H30_H60_median}
\end{figure}

\begin{figure}
\centering
\includegraphics[width=0.495\textwidth]{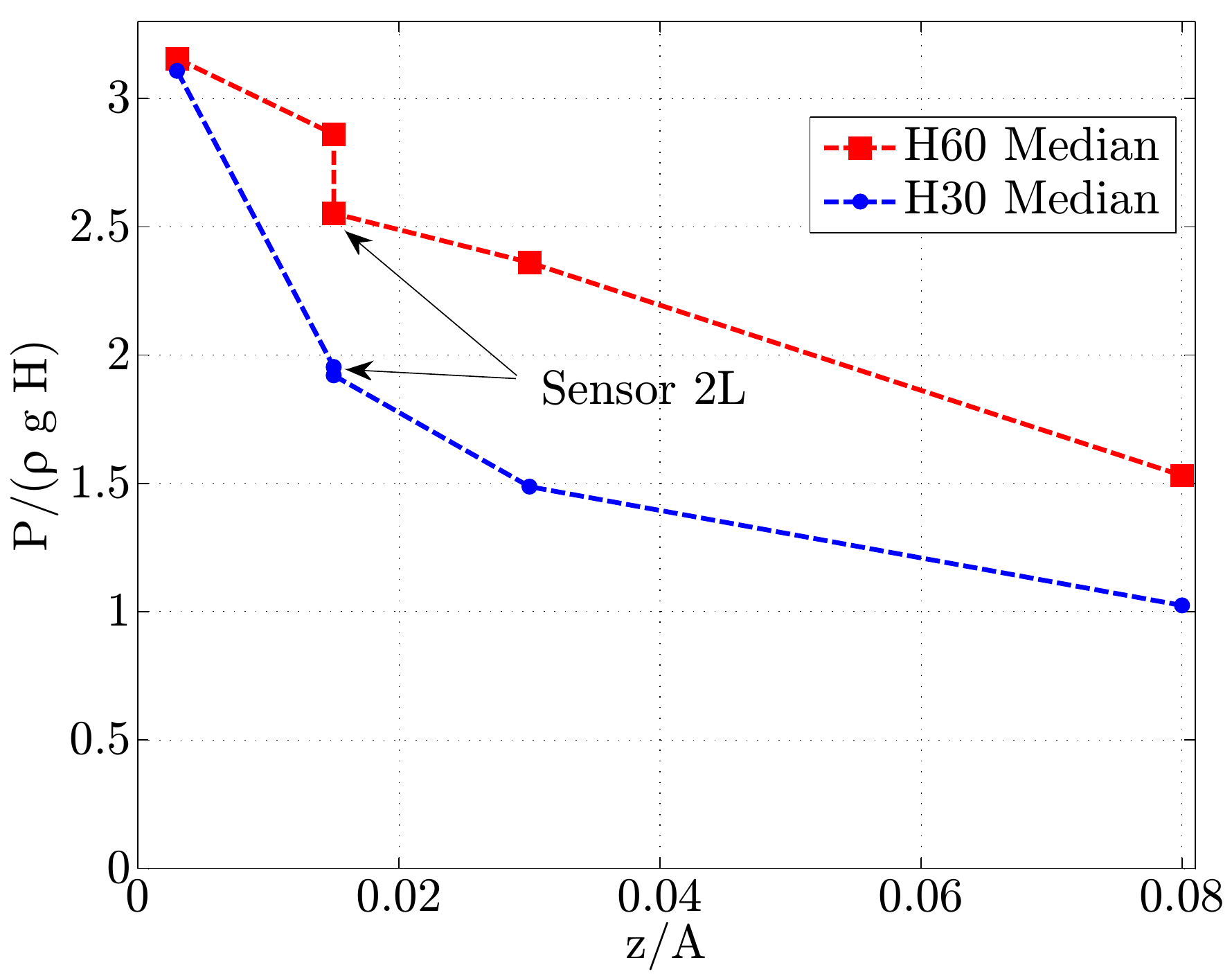}
\includegraphics[width=0.495\textwidth]{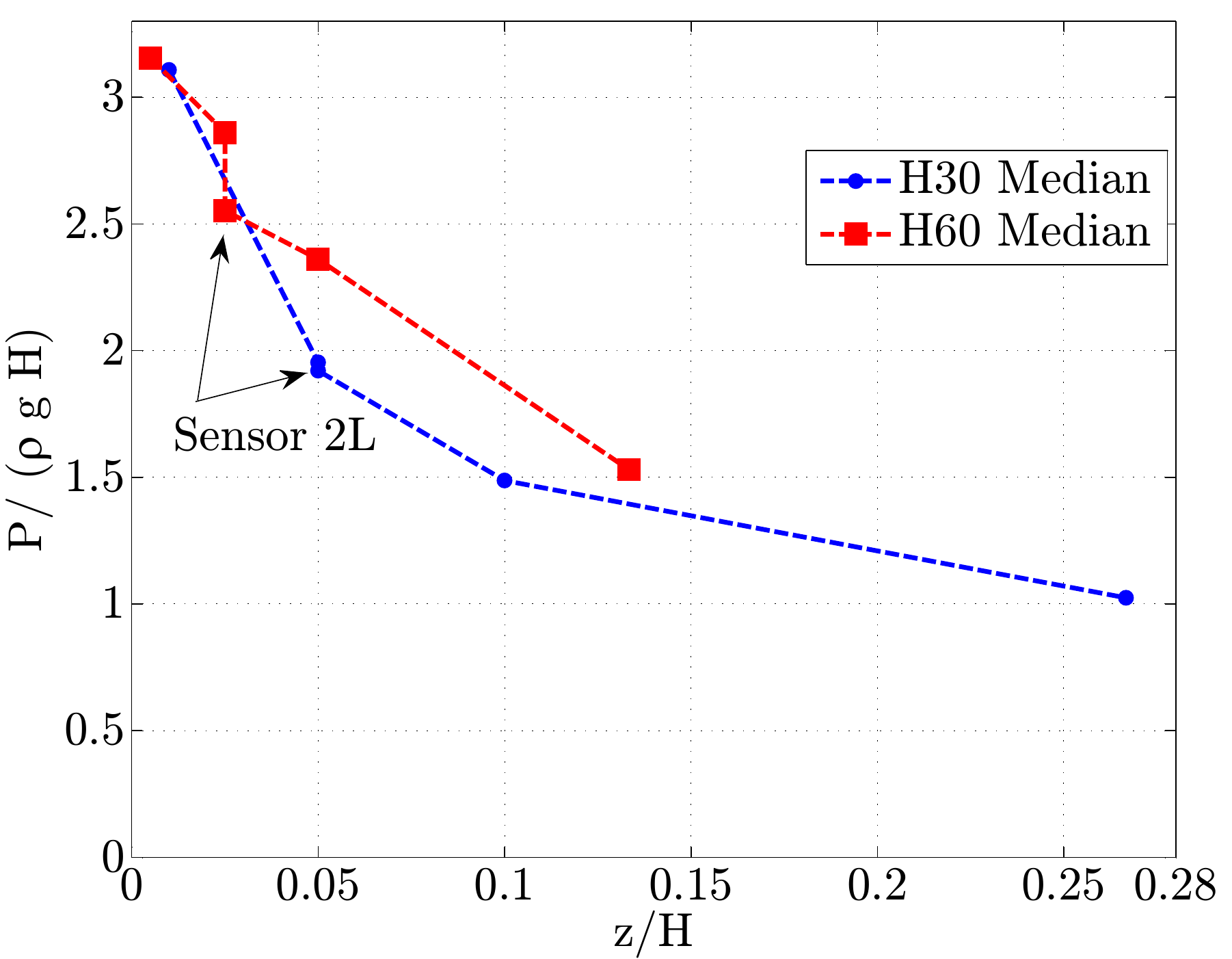}
\caption{H=300mm, H=600mm: comparison of median of peak pressure for all sensors as a function of the sensor height related to the tank length (left) and to still water height (right).}
\label{fig:libor_pmedian_h}
\end{figure}

A comparison between the two sets of experiments, i.e. between the tests with $H=300\,{\rm mm}$ and  $H=600\,{\rm mm}$ filling height, is presented in Fig.~\ref{fig:H30_H60_median}.
The median of peak pressures and relevant confidence intervals for all sensors are provided.
The box plot reports the following quantities:
\begin{enumerate}
  \item The lower and upper limit of each box indicate the 25\% and 75\% percentile levels, i.e. the lower and upper quartile respectively.
  \item The central mark indicates the 50\% percentile level, i.e. the median.
  \item The whiskers indicate the observed minimum and the observed maximum.
  \item The lower and upper additional markers represent the estimated 2.5\% and 97.5\% percentile levels respectively.
\end{enumerate}
It can be observed that:
\begin{enumerate}
\item The median as well as the length of the confidence intervals is larger for $H=\,{\rm 600mm}$ test case.
\item The dependence of the median for sensor $1$ on the filling height $H$ is basically linear. This is consistent with the assumption of pressure being proportional to the square of the velocity of the impacting flow, as discussed in slamming context in \cite{Greco2012148}.
\item The linear dependence found for pressure $1$ is not valid for the other sensors.
Nevertheless larger pressures are found for the case with $H=\,{\rm 600mm}$ (this is further discussed later).
\item It is noticeable that the two-dimensionality of the studied flow is lost for the case with $H=\,{\rm 600mm}$. Although three-dimensional effects can be experienced for both filling heights, the deviation between sensor $2$ and $2L$ is comparatively negligible for the $H=300\,{\rm mm}$ case.
\end{enumerate}
%
%
The non-dimensional median of the pressure peaks,  plotted as a function of sensor location,  
is presented in Fig.~\ref{fig:libor_pmedian_h} for the $H=300\,{\rm mm}$ and  $H=600\,{\rm mm}$ cases.
Due to the linear dependence with $H$ found for sensor $1$, the curves collapse for this sensor. 
That is not the case for the other sensors for which the larger $H$ experiments provide a larger non-dimensional pressure. This can be related with the fact that the distance to the wall ($A/H$) is comparatively shorter for the
case with $H=600\,{\rm mm}$ (see fig. \ref{fig:dam_scheme} for a definition of these parameters). 
Actually the data for the two cases become much closer when the sensor height is
made non-dimensional also with $H$, Fig.~\ref{fig:libor_pmedian_h} (right).

This set of comparative data provides a good benchmark for computational models and numerical simulations of this flow since it comprises a complete set of statistically analyzed data (including both the average values and the confidence intervals).

\section{Conclusions}
The experimental study on the dam break flow over the dry horizontal bed have been have been performed for two different initial filling heights in the dam reservoir.
The kinematics of the flow has been described in detail and compared to data from literature.
A reasonable agreement in free surface evolution has been achieved for both the wave height as well as the wave front velocity.  The analysis of water front highlighted the onset of transversal variability, which is much larger for the high filling cases.

The dynamics of the flow, in particular the pressure loads at the downstream wall, has been investigated.
A set of a hundred tests have been carried out for each filling height in order to provide statistically relevant data for the loads at the downstream wall.
In total, five pressure sensors were applied. An array of four sensors have been placed along the centerline of the tank wall at different heights and the fifth sensor have been offset in horizontal direction.
One of the sensors has been placed at the lowest feasible level above the bed considering its diameter.
For this sensor, the studied flow event can be compared to a transient impinging jet. Results for the
pressures obtained with this sensor when the wave front touches the wall 
have been confronted with the analytical solution for a steady impinging jet finding
that the representative values are significantly larger, which may have implications when designing
walls to protect from these types of flows. 

The values of the pressure peaks at the downstream wall present a scattering which has not been reflected in the literature yet. Exogenous dependence of this scattering on gate rise velocity has been explored without finding a significant correlation. The scattering in the pressure peaks becomes substantially larger for the high filling cases.

The rise time, the decay time and the pressure impulse at the sensor, which is placed just above the bed, has been investigated.
Consistency has been found between the medians of the impact duration time, peak pressure and pressure impulse.
Significantly larger values for decay times than for the rise times have been documented, with relatively larger values for the high filling cases.

Pressure registers of an offset sensor have been explored. It has been found that there are significant statistical differences between the pressure loads in the center and in this horizontally offset sensor and that these differences increase with the filling height.

The dependence of the pressure values on the filling level has been found to be linear in statistical terms for the sensor placed just above the bed, the one with the highest impact pressures.
That is not the case for the other sensors for which the tendency is not exactly linear. Proportionally larger pressures are found for the high filling case.

The time histories during the whole impact events have been also presented and discussed. With the low filling level, the setup becomes a half scaled setup of the reference work \cite{lee_zhou_cao_2002_jfe_dambreak}. When comparing data between the two studies, the pressure register of Lee et al. \cite{lee_zhou_cao_2002_jfe_dambreak} lays outside the 95\% confidence interval during a large part of the impact event. This may be attributed to the fact that Lee et al. \cite{lee_zhou_cao_2002_jfe_dambreak} used sensors of significantly larger diameter and ignored the stochastic nature of the phenomenon when presenting data from a single experiment. Comparisons with other works from the literature have been carried out finding overall good agreement.

The results presented in this paper are aimed to provide a wider perspective to the dam break problem that would be useful for validation of existing and future computational models and simulation results. Details on both kinematics and dynamics of the dam break flow evolution are rendered and can serve as a basis for a thorough assessment of computational model performance. In order to enable a comprehensive analysis, the empirical distributions and corresponding confidence intervals existing (and usually neglected) in experimental works are provided. All the materials (experimental videos, pressure registers etc.) are made available as a part of the paper so that they can serve as basis for further analysis by third parties or used as validation data for numerical simulations.

A substantial amount of future work threads arise naturally from the present work. First a repeatability analysis of the kinematics is desirable, aiming at looking for sources of variability of
the pressure loads at the downstream wall. Understanding the physical mechanisms during the impact event, which comprise bubble entrapment, generation of vortical structures, corner waves etc., is of interest. Attention can also be paid to the source of different values of measured magnitudes for the different filling heights. As a next step, fluid-structure interaction may be studied either at the downstream wall or at obstacles incorporated into the setup.

\begin{ack}
This work was supported by the European Regional Development
Fund (ERDF), project NTIS - New Technologies for Information Society, European Centre of Excellence, CZ.1.05/1.1.00/02.0090 and by the Spanish Ministry for Science and Innovation under
grant TRA2010-16988 ``\textit{Ca\-rac\-te\-ri\-za\-ci\'on Num\'erica y
Experimental de las Cargas Fluido-Din\'amicas en el transporte de Gas Licuado}''.
We thank Dr. Gabriele Bulian, Prof. Francisco Molleda, Javier Gonz\'alez Nieto and Francisco Javier Dom\'{i}nguez
for assisting in different issues during the experimental campaign, data analysis and manuscript writing.
\end{ack}

%

\end{document}